\documentclass[twocolumn,prx,aps,eqsecnum,amsmath,amssymb,showpacs,floatfix,superscriptaddress]{revtex4-1}
\usepackage[latin9]{inputenc}
\usepackage{array}
\usepackage{multirow}
\usepackage{amsmath}
\usepackage{graphicx}

\makeatletter

\providecommand{\tabularnewline}{\\}

\usepackage{dcolumn}
\usepackage{bm}
\usepackage{times}
\usepackage{color}\usepackage{color}

\makeatother

\begin{document}

\title{Fingerprints of spin-orbital polarons and of their disorder\\
in the photoemission spectra of doped Mott insulators with orbital
degeneracy}

\author{Adolfo Avella}

\affiliation{Dipartimento di Fisica ``E.R. Caianiello'', Università degli Studi
di Salerno, I-84084 Fisciano (SA), Italy}

\affiliation{CNR-SPIN, UoS di Salerno, I-84084 Fisciano (SA), Italy}

\affiliation{Unità CNISM di Salerno, Università degli Studi di Salerno, I-84084
Fisciano (SA), Italy}

\author{Andrzej M. Ole\'{s}}

\affiliation{Marian Smoluchowski Institute of Physics, Jagiellonian University,
Prof. S. \L ojasiewicza 11, PL-30348 Kraków, Poland }

\affiliation{Max-Planck-Institut für Festkörperforschung, Heisenbergstrasse 1,
D-70569 Stuttgart, Germany }

\author{Peter Horsch}

\affiliation{Max-Planck-Institut für Festkörperforschung, Heisenbergstrasse 1,
D-70569 Stuttgart, Germany }

\date{28 July, 2017}
\begin{abstract}
We explore the effects of disordered charged defects on the electronic
excitations observed in the photoemission spectra of doped transition
metal oxides in the Mott insulating regime by the example of the
$R_{1-x}$Ca$_{x}$VO$_{3}$ perovskites, where $R=$ La, $\dots$, Lu.
A fundamental characteristic of these vanadium $d^{2}$ compounds with
partly filled
$t_{2g}$ valence orbitals is the persistence of spin and orbital order
up to high doping, in contrast to the loss of magnetic order in high-$T_c$
cuprates at low defect concentration. We demonstrate that the disordered
electronic structure of doped Mott-Hubbard insulators can be obtained
with high precision within the unrestricted Hartree-Fock approximation.
In particular:
(i) the atomic multiplet excitations in the inverse photoemission spectra
and the various defect-related states and satellites are well reproduced,
(ii) a robust Mott gap survives up to large doping, and
(iii) we show that the defect states inside the Mott gap develop a soft
gap at the Fermi energy.
The soft defect states gap, that separates the highest occupied from
the lowest unoccupied states, can be characterized by a shape and
a scale parameter extracted from a Weibull statistical sampling of the
density of states near the chemical potential. These parameters provide
a criterion and a comprehensive schematization for the insulator-metal
transition in disordered systems. We demonstrate that charge defects
trigger small spin-orbital polarons, with their \emph{internal} kinetic
energy responsible for the opening of the soft defect states gap. This
\emph{kinetic} gap survives disorder fluctuations of defects and is
amplified by the long-range $e$-$e$ interactions, whereas in the atomic
limit we observe a Coulomb singularity. The small size of spin-orbital
polarons is inferred by an analysis of the inverse participation ratio
which explains the origin of the robustness of spin and orbital order.
Using realistic
parameters for the perovskite vanadate system La$_{1-x}$Ca$_{x}$VO$_{3}$,
we show that its soft gap is well reproduced as well as the marginal
doping dependence of the position of the chemical potential relative
to the center of the lower Hubbard band. The present theory uncovers
also the reasons why the $d^{1}\rightarrow d^{0}$ satellite excitations,
which directly probe the effect of the random defect fields on the
polaron state, are not well resolved in the available experimental
photoemission spectra for La$_{1-x}$Ca$_{x}$VO$_{3}$.
\end{abstract}

\maketitle

\section{Introduction\label{sec:intro}}

In this work, we deal with transition metal oxides that are in their
intrinsic state Mott insulators as a result of strong electron repulsions
and not due to band structure effects as in semiconductors
\cite{Ima98,Hwa12}. Mott insulating materials typically display different
realizations of quantum magnetism and some give rise to rare quantum spin
liquid states \cite{Bal10,Sac11,Sav17,Zho17}. Doping Mott insulators can
have striking consequences. For example, doping the two-dimensional (2D)
antiferromagnetic (AF) Mott insulator La$_{2}$CuO$_{4}$ with Sr, Ba or Ca
gives rise to high-$T_{c}$ superconductivity \cite{Lee06,All09,Ber11,Ave14},
with an insulator to superconductor transition and the disappearance
of AF order at very low doping \cite{Ell89,Kha93,Kas98}. Manganites
are paradigmatic examples of systems characterized by spin, orbital
and charge degrees of freedom that are controlled by spin-orbital
superexchange interactions \cite{Kug82,Tok00,Ole05,Kha05,Hor07,Ole12}.
Doping the Mott insulator LaMnO$_{3}$, for instance, leads to a variety
of insulating spin-, orbital-, and charge- ordered phases \cite{Kil99,Fei99,Feh04,Dag04,Gec05,Kha11,Sna16}, as well as
to a metallic regime that displays a colossal magnetoresistance
\cite{Dag01,Dag05,Tok06}.

A very interesting class of materials where the orbital degree of
freedom plays an important role are the $3d$-electron systems with
active $t_{2g}$ degrees of freedom that drive orbital fluctuations
\cite{Kha05,Pav07}. In the titanium perovskites, they play a prominent
role in the spin-orbital order or may even trigger a disordered state
\cite{Kha00,Moc02,Kha03,Ish04,Che08,Ulr15}, while superconductivity
was discovered at SrTiO$_{3}$ interfaces \cite{Sch64,Koz09,Man11,Gab13}.
We shall focus here on the vanadium perovskites
$R$VO$_{3}$, where $R$=La,$\dots$,Lu, which reveal temperature-induced
magnetization reversals \cite{Ren98} as a result of the coupling
of spin and orbital degrees of freedom of the $t_{2g}$ valence electrons
\cite{Ren00,Kha01,Hor03,Ulr03,Sir03,Sir08}. This class of compounds
has interesting phase diagrams with two complementary spin-orbital
ordered phases and a pure orbital-ordered phase
\cite{Miy03,Miy06,Fuj10,Miy07,Sag07,Ree11,Lin17}.
Remarkably, the $C$-type spin and $G$-type orbital ordered phase
in La$_{1-x}$Sr$_{x}$VO$_{3}$, Pr$_{1-x}$Ca$_{x}$VO$_{3}$ and
Y$_{1-x}$Ca$_{x}$VO$_{3}$ is robust up to high-doping concentrations
and also the metal-insulator transition takes place only at quite
substantial doping \cite{Miy00,Fuj05,Fuj08,Ree16}. This striking
robustness of the spin-orbital ordered state against doping as compared
to the fragile AF state in the cuprates is one of the motivations
for our work.

The combination of spin and orbital degrees of freedom triggers
spin-orbital (SO) polarons \cite{Dag06,Woh09,Mona,Bie17}. We shall
discuss in this work why small SO polarons in the insulating regime of
cubic vanadates \cite{Ave15,Ree16} are much more strongly bound to charge
defects than spin polarons in high-$T_{c}$ materials \cite{Che91,Che09}.
The reduced mobility of doped holes or electrons inside the SO polarons
implies a weaker screening of defect potentials by the doped charge
carriers and thus provides an explanation for the shift of the insulator
to metal transition towards high doping concentrations in the vanadates.
Here, we shall support these conclusions by a careful analysis of the
doping dependence of the density of states, i.e., relevant for
photoemission (PES) and inverse photoemission (IPES). We explore the
localization of defect states wave functions, the defect states gap inside
the Mott-Hubbard (MH) gap, and finally the reduction of spin and orbital
order and its relation to the many-body SO polaron wavefunction and their
binding to defects.

The cubic vanadates represent a class of compounds with quantum fluctuating
orbitals and spins, even in the absence of doping. In contrast to
the manganites, the cubic vanadates have very small Jahn-Teller interactions,
a consequence of the $t_{2g}$ nature of their valence electrons.
Therefore, orbital occupations are not rigid even in the ordered phases,
but fluctuate \cite{Kha01,Hor03,Hor08}. Several peculiar features
can be traced back to orbital quantum fluctuations, such as ferromagnetism
driven by orbital singlet fluctuations \cite{Kha01} and orbital Peierls
dimerization \cite{Ulr03,Hor03,Sir03,Sir08}. Indeed, the joint spin and
orbital fluctuations are particularly strong in the ordered
$C$-type AF and $G$-type alternating orbital (AO), i.e., $C$-AF/$G$-AO
phase, realized in these compounds when they are doped \cite{Hor11}.
Furthermore, it was shown that charged defects tend to enhance orbital
Peierls dimerization \cite{Hor11}. An important motivation for the
investigation of the cubic vanadates is a large experimental data
base for these systems, which includes the phase diagrams of many
undoped compounds \cite{Miy03,Miy06,Fuj10} and their pressure dependence
\cite{Biz12}, as well as the doping dependence of the optical conductivity
for several systems \cite{Fuj08}. In this work, our focus is on the
doping dependence of the PES
spectra of the vanadium perovskites \cite{Mai98,Mai00,Pen99}.

Solving this problem involves, for example, answering questions like:
(i)~What is the nature of defect states in a strongly correlated
system, i.e., to what extent are such defect states different from
those in usual semiconductors or band insulators \cite{Gan05,Dra07}?
(ii)~What happens to the MH gap in the presence of defects?
(iii)~Which are the different features of defect states in
MH insulators with orbital degeneracy within $t_{2g}$ transition
metal oxides as compared to those in doped high-$T_{c}$ superconductors?
(iv)~Which methods, among those capable to yield reliable results
for MH insulators, can be efficiently extended to take into account
defects and disorder? To answer these questions is a
formidable challenge, as defects greatly influence the subtle interplay
of spin, orbital, lattice and charge degrees of freedom.

\begin{figure}[b!]
\includegraphics[width=0.95\columnwidth]{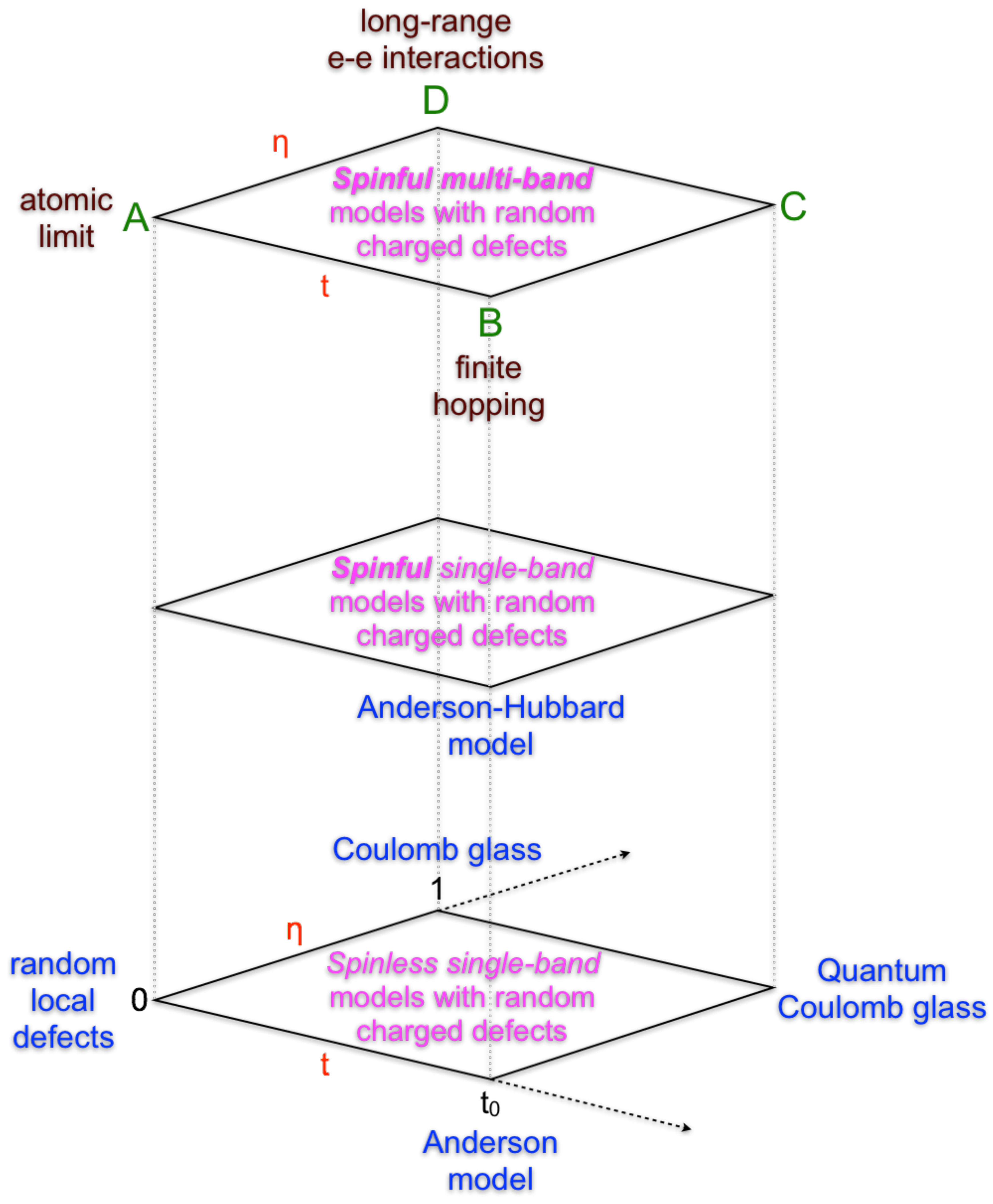} \caption{
Schematic hierarchy of typical models for systems with disorder:
On the lowest floor, generic models with a single orbital per site
and no spin are shown. These models are further distinguished depending
on whether they involve long-range $e$-$e$ interactions ($\eta$-scale)
and/or kinetic energy ($t$-scale). By adding spin to charge carriers
one finds single orbital Hubbard-type models shown on the second floor.
On the third top floor, reside models with spin and orbital degrees
of freedom of interest in this paper, designed for doped spin-orbital
systems. The labels A, B, C, and D refer to models explained and
investigated in following Sections.\label{fig:tower}}
\end{figure}

The calculation of the electronic structure of a disordered system
with charged defects and long-range $e$-$e$ interactions is a difficult
optimization problem even for the simplest models for the defect states
in the gap of semiconductors, e.g., the Coulomb glass (CG) model
\cite{Gan05}.
The effect of disorder and the resulting localization of electron
states has a long history \cite{And58,Lee85}. The reason for the
calculational complexity of the insulating phase is the absence of
metallic screening, and therefore the energy and occupation of a defect
state depends on that of far distant random defects because of the
long-range Coulomb interaction \cite{Lee85}. In contrast, in the
metallic state, because of the presence of a Fermi surface, perturbative
diagrammatic techniques can be applied to deal with multiple scattering
corrections leading to a Coulomb anomaly in the density of states
(DOS) $N(\omega)$ \cite{Alt79}. For the insulating phase, it was
argued by Pollak \cite{Pol70} that $e$-$e$ interactions should
lead to a depletion of $N(\omega)$ at the chemical potential in systems
with disorder. For the CG model, Efros \textit{et al.}
\cite{Efr75,Efr76,Efr11} could show, using some simplifying criteria,
that the Coulomb interaction generates at a soft gap at the chemical
potential in $N(\omega)\propto|\omega|^{\nu}$, with an exponent $\nu=d-1$
determined by the spatial dimension $d$. This is called the Coulomb
gap \cite{Efr75}. Recently, Shinoaka and Imada \cite{Shi09,Shi10}
reported unconventional soft gaps for models with only short-range
interactions. Furthermore, a disorder-induced Coulomb or zero-bias
anomaly was observed by Epperlein \textit{et al.} \cite{Epp97} for
the quantum CG model and by Yun Song \textit{et al.} \cite{Son09}
for an extended Anderson-Hubbard model.

We base our study on a generic spinful three-band model for the $t_{2g}$
electrons on the vanadium ions \cite{Ave15} that provides a complete
description of the magnetic and orbital ordered phases. The model
includes the local Hubbard-Hund interactions and thereby describes
the atomic multiplet structure of the V ions. Moreover, the model
contains the long-range $e$-$e$ interactions, and thereby it includes
the dielectric screening 
of the $t_{2g}$ electrons, and it contains the Coulomb potentials
of the random defects. Finally, there is the flavor conserving kinetic
energy of $t_{2g}$ electrons, and moreover crystal field and Jahn-Teller
interactions. All these interactions together yield spin-orbital superexchange
and pure orbital interactions that determine the different spin- and
orbital ordered phases \cite{Kha01,Hor08}. This model can be simplified
by the elimination of degrees of freedom and by the removal of terms,
such as the kinetic energy or the Hubbard interaction for instance,
to obtain simpler models that have been used in the study of disorder.
For instance, one may arrive at the CG \cite{Efr75} or
at the quantum CG \cite{Pol70,Epp97} models after elimination
of spin and orbital degrees of freedom, or at the Anderson-Hubbard
model \cite{Shi09} after elimination of the orbital degrees of freedom
(see Fig.~\ref{fig:tower}).

It requires highly efficient methods like the Hartree, the Hartree-Fock
\cite{Miz00,Bar00,Che09,Sch11,Hor11}, or the density functional method
\cite{Sol08,Car06,Pav12} to analyze systems with defects and disorder.
Besides disorder, the calculations of the electronic structure and
excitations of the spinful multiband models with charged defects are
complicated by the fact that our systems are Mott insulators. We have
already demonstrated that the Hartree-Fock method is also capable to
describe faithfully the atomic multiplet structure in the spin-orbital
degenerate case provided the system has a broken spin and orbital
symmetry and one can apply the unrestricted Hartree Fock (uHF) version
of the method \cite{Hor11,Ave13}. In presence of local off-diagonal terms,
the Hubbard-Hund interaction has to be expressed in rotational invariant
form in spin and orbital space.
The uHF method \cite{Miz96,Miz97,Wen10}, is known to be reliable for
systems with spontaneously broken symmetries as, for instance,
multiband models for manganites \cite{Miz00,Sol09}, or
the iron-based superconductors \cite{Kub09,Flo11,Luo13,Dag14} or
clusters of transition metals with magnetic ground states \cite{Bar00}.
The uHF method designed and used in this paper is capable to treat
simultaneously phenomena that arise at distinct energy scales: the
high energy scale of $\sim1$ eV related to the (on-site and intersite)
Coulomb interactions in proximity of the defects and the low energy
scale of $\sim0.1$ eV that is characteristic of the orbital physics
and controls the electronic transport in doped materials. Accordingly,
the method is able to address the onset of the orbital order and its
remarkable robustness in doped La$_{1-x}$Sr$_{x}$VO$_{3}$ and
Y$_{1-x}$Ca$_{x}$VO$_{3}$.

\begin{figure}[t!]
\includegraphics[width=1\columnwidth]{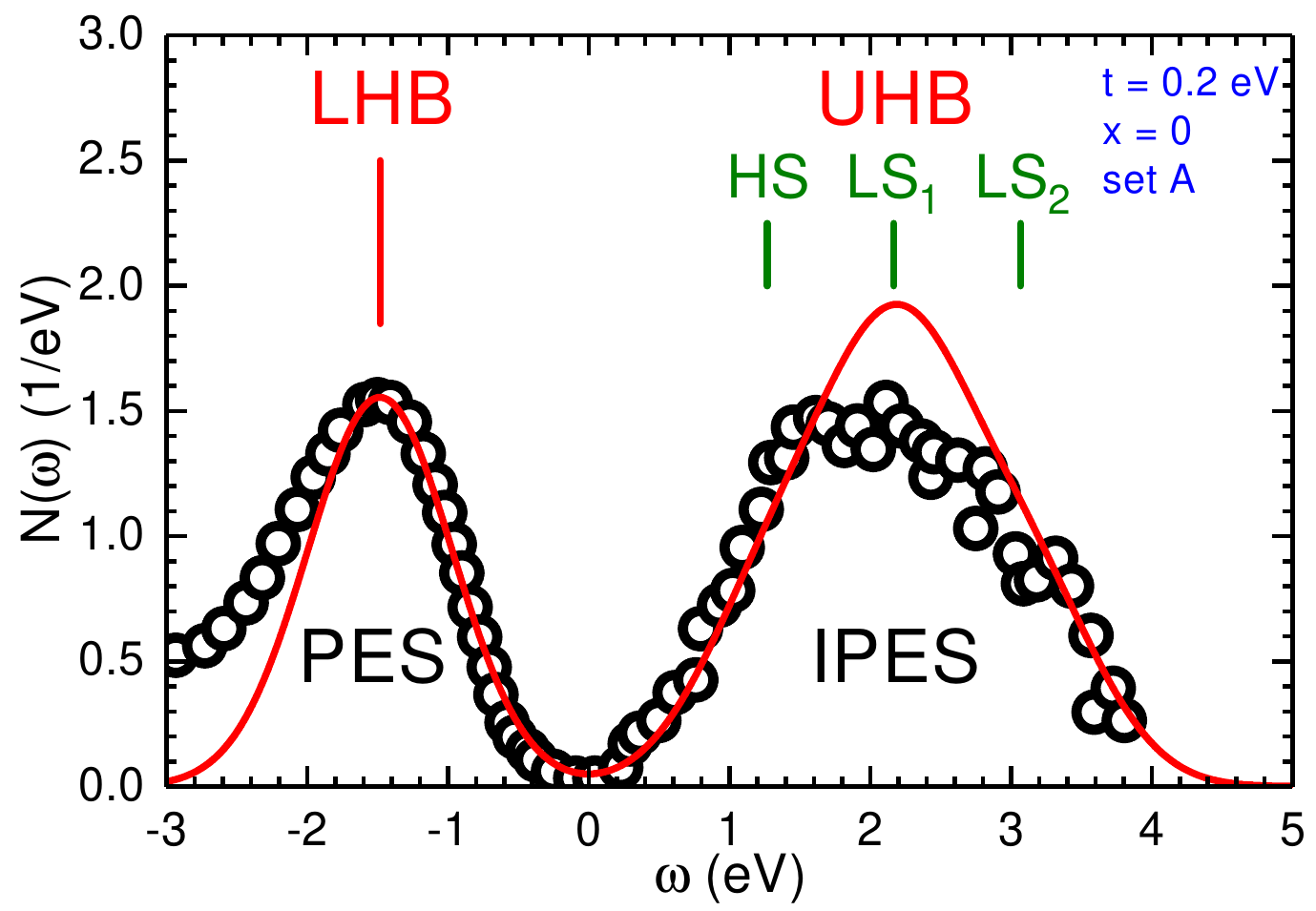} \caption{
Photoemission (PES) and inverse photoemission (IPES) spectra for undoped
LaVO$_{3}$ in the vicinity of the chemical potential, which is located
at $\omega=0$. Experimental data (black hollow circles) were obtained by
Maiti and Sarma \cite{Mai00}; the solid (red) curve was obtained
from the theory reported in Sec.~\ref{sec:dop}, with the three peaks
of the UHB marked as HS, LS$_{1}$, and LS$_{2}$ standing for the
high-spin and the two low-spin excitations, respectively. \label{fig:maiti}}
\end{figure}

Figure~\ref{fig:maiti} displays PES and IPES data of Maiti and Sarma
\cite{Mai00} for the undoped reference compound LaVO$_{3}$. The data
provides evidence that the system is a Mott insulator. One recognizes
the LHB in PES and one also sees the UHB in IPES, where the theoretically
expected multiplet bands have some correspondence with features in the
experiment. The fact that the Hubbard gap in the vicinity of $\omega=0$
appears as a soft gap suggests that the sample contained possibly some
defects at the surface \cite{Che09}. The experimental data is compared
with the uHF calculation, where the DOS $N(\omega)$ is broadened with
an extra linewidth of $\gamma=0.5$ eV. This has the consequence that
the multiplets of the UHB are no longer resolved in the theoretical
DOS $N(\omega)$.

A surprising feature of the electronic structure of cubic vanadates
is the persistence of the MH gap up to high doping concentrations
\cite{Hor11}. This has been demonstrated most clearly by optical
conductivity experiments \cite{Fuj08} in La$_{1-x}$Sr$_{x}$VO$_{3}$
and Y$_{1-x}$Ca$_{x}$VO$_{3}$ for doping concentrations up to $x=0.1$
and 0.17, respectively. Our uHF calculations of the statistically
averaged DOS $N(\omega)$ reproduces this robustness of the MH bands
beyond doping concentrations of 50\%, despite the fact that substantial
spectral weight is transferred from the Hubbard bands to defect states.
The latter appear as satellites and as states inside the MH gap. We
shall discuss the energetics of these states, the sum rules and the
spectral weight transfers.

Our main focus is on the defect states that appear in the MH gap and
that themselves form a \textit{defect states (DS) gap} at the chemical
potential. The DS gap depends both on the kinetic energy parameter
$t$ and the $e$-$e$ interactions strength which we can tune with
a coupling constant $\eta\in[0,1]$, where $\eta=0$ corresponds to
the absence of $e$-$e$ interactions and $\eta=1$ to the estimated
typical strength for these interactions in vanadates. We perform a
statistical analysis of the power law behavior of the DS gap in the
DOS $N(\omega)\propto|\omega|^{\nu}$ in the vicinity of of the chemical
potential. The exponent $\nu$ is nonuniversal and depends on $t$
and $\eta$. It has been shown that the kinetic energy of the $t_{2g}$
electrons plays a fundamental role for the formation of the DS gap
in the cubic vanadates \cite{Ave15}. This mechanism is distinct from
the Coulomb gap mechanism. We furthermore show that, with increasing
$\eta$, $e$-$e$ interactions of Coulomb type do increase the DS
gap. Yet in the atomic limit, that is without kinetic energy ($t=0$),
we found that $e$-$e$ interactions alone are not strong enough in
the cubic vanadates to open a Coulomb gap.

We investigate the localization of defect states by means of the inverse
participation number (IPN) \cite{Bel70,Tho74,Weg80}. We have generalized
this concept here to the case with spin- and orbital- degeneracy.
We find that at moderate doping, all wave functions are well localized.
The states in the Hubbard bands are less localized than the defect
states inside the Mott gap that contribute to the soft gap. These
states are typically localized on 1-2 sites with tiny admixtures from
further neighbors. Interestingly, we observe a discontinuity in the
localization of the wave functions below and above the chemical potential
when $e$-$e$ interactions are switched on. The small participation
number ($<2$) for the doped hole states can be taken as an unambiguous
sign that holes are in small \textit{spin-orbital polaron states}
that are strongly bound to the charge defects, i.e., basically on
a single bond. This leads to a reduction of spatial symmetry \cite{Ave15}.
The strong localization of wave functions appears consistent with
experiments by Nguyen and Goodenough \cite{Ngu95} who analyzed the
magnetic properties of the La$_{1-x}$Ca$_{x}$VO$_{3}$ system, and
who concluded that carriers are in trapped small polaron states.
Moreover, in a recent combined x-ray and neutron diffraction study
of Pr$_{1-x}$Ca$_{x}$VO$_{3}$, Reehuis \textit{et. al.} \cite{Ree16}
found signatures for spin-orbital polarons from the change of spin
and orbital correlations close to the insulator-metal transition at
$x\simeq0.23$.

In PES experiments of gapped systems, the position of the chemical
potential $\mu$ is a subtle issue as it is determined by defects
\cite{Dam03,Dam16}. For the doped cubic vanadates, we find that $\mu$
lies in the center of the DS states gap that forms inside the MH gap.
We find that the distance of $\mu$ from the center of the LHB scales
with the binding strength of the Ca defect and is basically unchanged
by doping up to 50\%, consistent with the PES study of Maiti and Sarma
\cite{Mai98,Mai00} for La$_{1-x}$Ca$_{x}$VO$_{3}$. We interpret
this as manifestation of small polaron physics. Furthermore, we show
that $d^{1}\rightarrow d^{0}$ satellites in PES spectra can provide
a precise fingerprint of the state of spin-orbital polarons and the
strength and variance of the random defect fields acting on it.

The paper is organized as follows. In Sec.~\ref{sec:model}, we introduce
the triply degenerate Hubbard model for $t_{2g}$ electrons in the
doped perovskite vanadates, such as La$_{1-x}$Ca$_{x}$VO$_{3}$
with Ca$^{2+}$ charged defects replacing randomly some La$^{3+}$
ions. The model includes local (on-site) and long-range Coulomb interactions,
as well as the Coulomb potentials induced by Ca defects, which increase
the energies of the $t_{2g}$ electrons located close to the defects
taking also into account the contributions of more distant defects
at random positions. We treat the $e$-$e$ interactions in the uHF
approximation and consider two parameter sets for La$_{1-x}$Ca$_{x}$VO$_{3}$
motivated by the experimental data in Sec.~\ref{sec:hf}. Using this
approach and performing self-consistent calculations for realistic
$t_{2g}$ hopping integral, $t=0.2$ eV, we obtained the one-particle
DOSs presented in Sec.~\ref{sec:multi}. They are interpreted using the
generic structure of the PES and IPES spectra in a doped system with
orbital degeneracy derived from the atomic limit in Sec.~\ref{sec:atom}.
We analyze the multiplet structure and comment on the PES and IPES
data obtained for the undoped LaVO$_{3}$ \cite{Mai00}. The numerical
spectral weights near the atomic limit confirm the exact calculations,
as shown in Sec.~\ref{sec:atomw}. Furthermore, using the results
simulating the atomic limit (i.e., for a very low value of the hopping
integral $t=0.01$ eV) we extract the effects that arise due to finite
kinetic energy and emphasize the role played by active bonds in
Sec.~\ref{sec:bond}. While the excitations for the active bonds can be
resolved from the spectra, we also analyze the sum rules obeyed by the
structures seen in the PES. In Sec.~\ref{sec:gap},
we elucidate the kinetic mechanism of the gap in the present system
and perform the Weibull analysis \cite{Ave15}. The electronic states
that contribute to the spectral functions have various degrees of
localization, which is universal for various defect realizations,
as we show in detail in Sec.~\ref{sec:ipr}. Actually, more localized
states appear at the edges of Hubbard bands. In Sec.~\ref{sec:polarons},
we explain gradual changes of spin-orbital order with increasing doping
within the polaron picture. Finally, in Sec.~\ref{sec:pes}, we briefly
address the experimental results for PES and IPES in doped systems
and demonstrate that the large defect potential is responsible for the
overall scenario consistent with the experiments. The final discussion
and summary are presented in Sec.~\ref{sec:summa}. The technical
details of the present calculations are reported in two appendices:
in Appendix A one finds a general discussion of the algorithm for
a system with random defects, while in Appendix B it is explained
what one can learn by a perturbative analysis in terms of spin-orbital
polarons.



\section{Mott insulator with charged defects\label{sec:model}}


We shall describe the doped $R_{1-x}$Ca$_{x}$VO$_{3}$ compounds
(here~$R$ stands either for yttrium Y or for lanthanum La) by means
of a multiband Hubbard model
describing the subspace of $t_{2g}$-states at vanadium ions. Previous
studies \cite{Kha05,Kha01,Hor03,Kha04,Hor07} have shown that $t_{2g}$
orbitals are crucial for the description of the perovskite vanadates
as Mott insulators.
Each dopant \textendash{} calcium Ca$^{2+}$ ions randomly replacing
Y$^{3+}$ or La$^{3+}$ones in this specific case \textendash{} is
equivalent to an effective charged defect sourcing a long range Coulomb
potential. The degree of randomness of the dopants' locations is controlled,
in principle, by the annealing procedure used (or not used: full randomness
as in the case we analyze here) during the fabrication of the samples
and leads to a quite general disorder problem in the Mott insulating
regime. The sequence of terms in the following Hamiltonian is selected
to highlight the crucial role played by the charged defects and the
long-ranged $e$-$e$ interaction in the present system \cite{Ave15},
\begin{align}
{\cal H}_{t_{2g}} & =\sum_{i}h_{i}n_{i}+\frac{1}{2}\sum_{i\neq j}V_{ij}n_{i}n_{j}
\nonumber \\
& +{\cal H}_{{\rm kin}}+{\cal H}_{{\rm int}}+{\cal H}_{\mathrm{CF}}
+{\cal H}_{\mathrm{JT}},\label{H3band}
\end{align}
where $n_{i}=\sum_{\alpha\sigma}n_{i\alpha\sigma}$ and
$n_{i\alpha\sigma}=d_{i\alpha\sigma}^{\dagger}d_{i\alpha\sigma}^{}$
are the electron density operator and the partial electron density
operator for orbital $\alpha$ and spin $\sigma=\uparrow,\downarrow$,
respectively, at site $i$ of a cubic Bravais lattice.

If we would restrict the Hamiltonian~(\ref{H3band}) to the single-orbital
case, the first two terms would represent the CG model \cite{Efr75} with
randomly distributed local levels $\{h_{i}\}$ and long-range $e$-$e$
interactions of Coulomb type. A very important difference though between
the model we propose and those usually encountered
in the literature resides in the realization of the disorder: We use
a fully random distribution of charged defects leading to a realistic
distribution of energy levels that depends on the analyzed crystal
structure and on the level of doping, and is very structured, that
is, very far from the uniform or Gaussian distributions usually used
{[}see Fig.~(\ref{fig_w}){]}. The additional inclusion of a nearest neighbor
hopping term would lead to the quantum CG \cite{Pol70,Epp97}. Without
long-range $e$-$e$ interactions, but with a finite nearest neighbor
hopping term between random levels, one would instead get the Anderson
model, see Fig.~\ref{fig:tower}. The strongly correlated cousins
of these models involve in addition the local Hubbard interaction
between electrons with opposite spins that could lead to the opening
of a Mott gap. One representative model of this class is the Anderson-Hubbard
model \cite{Shi09} that belongs to the next level of sophistication
in the hierarchy of models with disorder, see Fig.~\ref{fig:tower}.

\begin{figure}[b!]
\includegraphics[width=0.75\columnwidth]{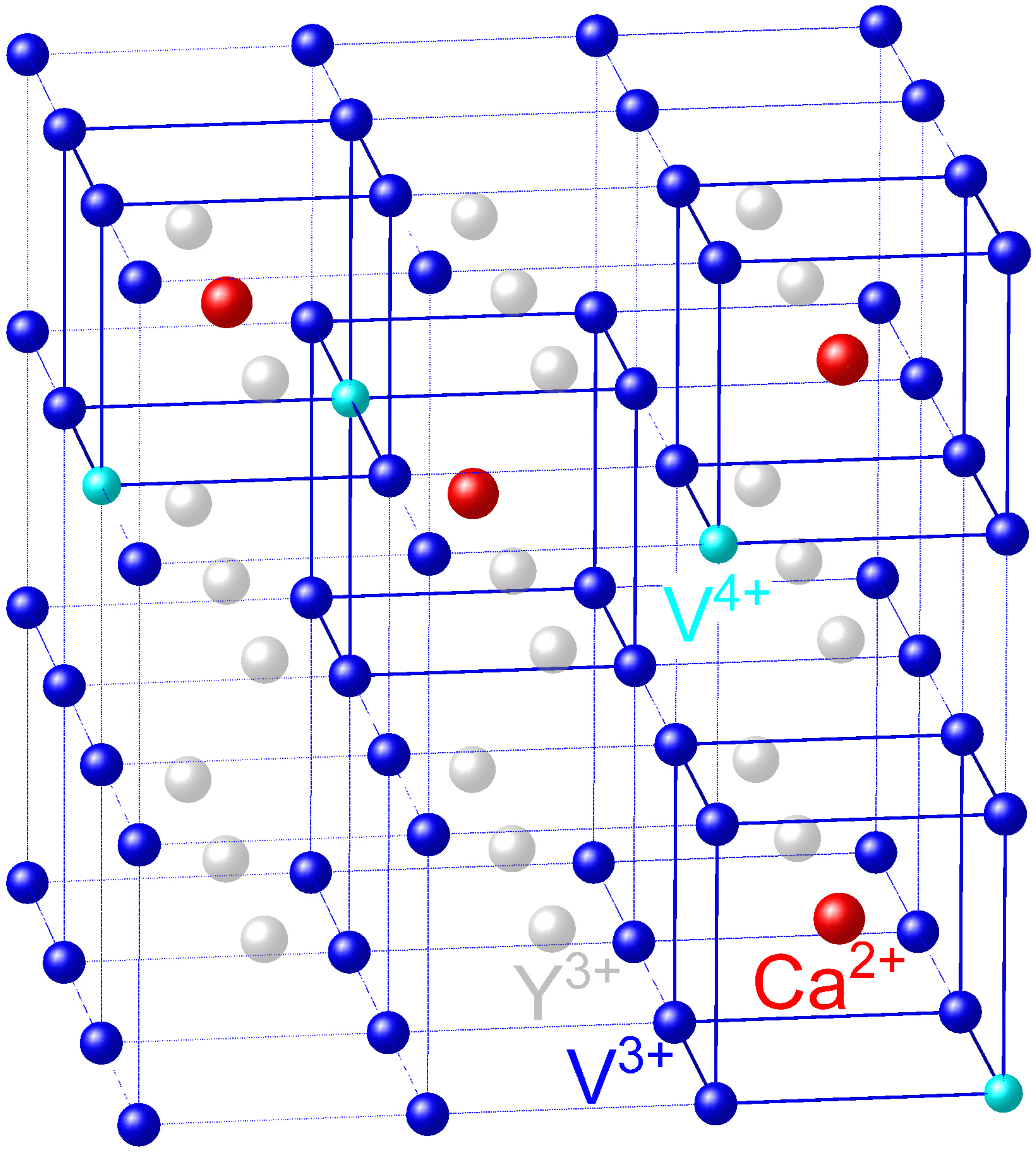} \caption{
Cubic vanadium lattice of the doped perovskite vanadate compound
$R_{1-x}$Ca$_{x}$VO$_{3}$ with random Ca$^{2+}$ defects replacing Y$^{3+}$
ions (O-ions are not shown). Doped holes occupy V$^{4+}$ ions in the cubic
V$^{3+}$ lattice and are attracted by the Ca$^{2+}$ ions on the cubes
surrounding the defects.\label{fig:defs}}
\end{figure}

As we shall see below in greater detail than what already announced
above, crucial to our analysis are the random fields,
\begin{equation}
h_{i}=\sum_{m}V_{im}^{{\rm D}},\label{hi}
\end{equation}
due to the electron-defect interaction $V_{im}^{{\rm D}}$, where
$m$ runs over the random defect sites that are located in the center
of the Vanadium cubes, see Fig.~\ref{fig:defs}, and the $e$-$e$
interaction $V_{ij}$. These interactions are screened by the dielectric
constant $\epsilon_{c}$ of the core electrons and depend on the distance
$r$ through the function,
\begin{equation}
v(r)=\frac{e^{2}}{\epsilon_{c}\,r}\equiv V_{D}\,\frac{d}{r}.\label{v}
\end{equation}
The contribution of the $t_{2g}$ states to the screening of the interactions
is not considered in the determination of $\epsilon_{c}$, but it
is included explicitly in the Hamiltonian~(\ref{H3band}). A typical
value for the core dielectric constant of the perovskite vanadates
is $\epsilon_{c}\simeq5$ \cite{Tsv04}. In the following, we shall
consider the potential energy $V_{{\rm D}}\equiv e^{2}/\epsilon_{c}\,d$
as an independent parameter, instead of~$\epsilon_{c}$. Here $d=\sqrt{3}a$
is the distance between the defect ion and its closest vanadium neighbors
and $a$ is the vanadium cubic lattice constant, i.e., the distance
between the closest vanadium ions (see Fig.~\ref{fig:defs}).


The screened interaction $v(r)$ (\ref{v}) defines both the $e$-$e$
interaction and the defect potential in case of defects with a net
charge of one as in the present case, i.e.,
\begin{equation}
V_{im}^{{\rm D}}=v(r_{im}),\quad V_{ij}=\eta v(r_{ij}),\label{pots}
\end{equation}
where $r_{im}$ and $r_{ij}$ represent the distances between a V
ion at site $i$ and a Ca defect at site $m$ ($r_{im}$) or another
V ion at site $j$ ($r_{ij}$), respectively. We have introduced the
parameter $\eta\in[0,1]$ in 
$V_{ij}$ in Eq.~(\ref{pots}) in order to to explore the importance
of $e$-$e$ interactions and the resulting dielectric screening through
the $t_{2g}$ electrons. In the physical case $\eta=1$, the system
is charge neutral and, as a result, the negatively charged defect
and the positively charged bound hole act overall as a dipole. In
the case of an insulator to metal transition, the screening due to
$t_{2g}$ electrons will switch from that of a semiconductor to that
of a metal, where defect charges are perfectly screened. For $\eta=0$
instead, the monopolar potential of the negatively charged defects
is unscreened as the polarization processes coming from the $t_{2g}$
electrons (and holes) are not active.

\begin{figure}[b!]
\includegraphics[width=0.7\columnwidth]{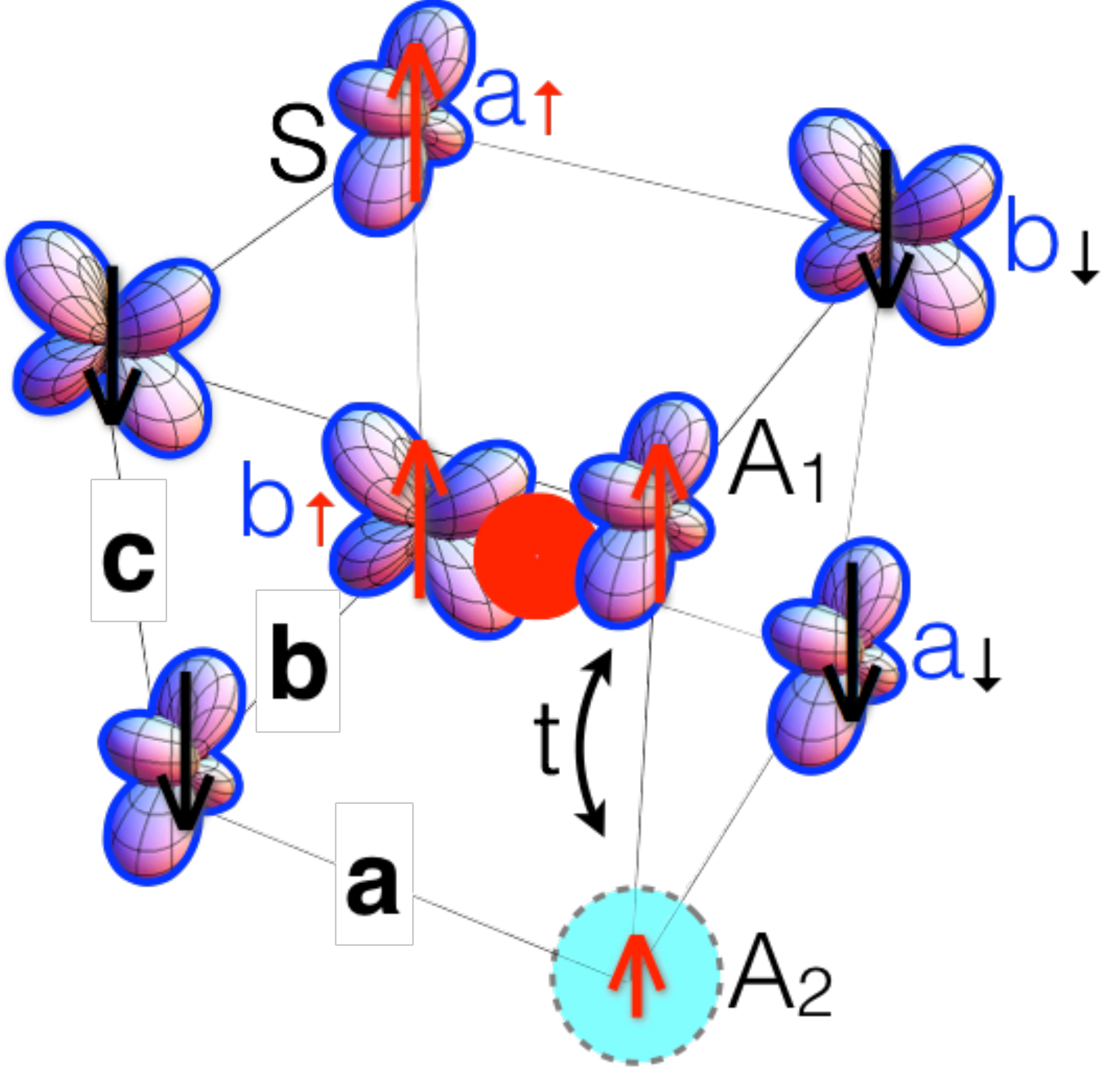} 
\caption{Occupied vanadium $t_{2g}$ orbitals of a V-cube surrounding a Ca
defect (red ball) display $G$-type orbital order with alternating
$\{a,b\}$ orbitals in all three directions (the occupied $c$ orbitals
at the V-ions are not shown). Spins $S=1$ (red/black arrows) reflect
$C$-type AF order and the strong Hund's coupling between $\{a,b\}$
and $c$ electrons at V ions. Due to the kinetic energy ($\propto t$),
the doped hole (blue circle) at site $A_{2}$, where the spin $s=1/2$
of the $c$ electron is marked, and the $a$ electron at site $A_{1}$
strongly fluctuate on the FM \emph{active} bond ($A_{1}$-$A_{2}$)
along the $c$ direction.\label{fig:cube}}
\end{figure}

For the $t_{2g}$ electrons of the vanadium perovskites, the kinetic
energy is defined as
\begin{equation}
{\cal H}_{{\rm kin}}=-t\sum_{\alpha}\sum_{\left\langle ij\right\rangle \perp\alpha}
\sum_{\sigma}\left(d_{i\alpha\sigma}^{\dagger}d_{j\alpha\sigma}+{\rm H.c.}\right),\label{kin}
\end{equation}
where we adopt a simplified notation for the $t_{2g}$ orbital basis
states \cite{Kha00}:
\begin{equation}
|a\rangle\equiv|yz\rangle,\hskip.5cm|b\rangle\equiv|zx\rangle,\hskip.5cm|c\rangle\equiv|xy\rangle.\label{abc}
\end{equation}
The orbital with flavor $\alpha\in\{a,b,c\}$ lies in the plane perpendicular
to the cubic crystal axis $\alpha$. In order to fully understand
the implications of the actual expression for the kinetic energy~(\ref{kin}),
one has to take into account that the hopping between two V ions 
occurs along $\sigma$ bonds via oxygen $p$ orbitals. Due to the
spatial symmetry of the V $t_{2g}$ and the O $p$ orbitals, the hopping:
(i) conserves the orbital flavor $\alpha$ and (ii)~is finite only
between $\alpha$ orbitals along bonds $\langle ij\rangle||\gamma$
perpendicular to the direction $\alpha$: $t_{ij}^{\alpha\beta}=t(1-\delta_{\alpha\gamma})\delta_{\alpha\beta}$
and $t>0$. Thus the hopping of $t_{2g}$ electrons is effectively
2D \cite{Kha01,Har03,Dag08,Li15}.

The most relevant Hamiltonian terms to the charge, spin and orbital
site occupations are the on-site Hubbard ($U$) and Hund's exchange
($J_{H}$) interactions (${\cal H}_{{\rm int}}$) for the triply degenerate
$t_{2g}$ orbitals \cite{Ole83,Dag10},
\begin{align}
{\cal H}_{{\rm int}} & =U\sum_{i\alpha}n_{i\alpha\uparrow}n_{i\alpha\downarrow}
-J_{H}\sum_{i}\sum_{\alpha\neq\beta}\mathbf{S}_{i\alpha}\cdot\mathbf{S}_{i\beta}
\nonumber \\
& +\frac{1}{2}\left(U-\frac{5}{2}J_{H}\right)\sum_{i}
\sum_{\alpha\neq\beta}n_{i\alpha}n_{i\beta}\nonumber \\
& +J_{H}\sum_{i}\sum_{\alpha\neq\beta}d_{i\alpha\uparrow}^{\dagger}
d_{i\alpha\downarrow}^{\dagger}d_{i\beta\downarrow}d_{i\beta\uparrow}.
\label{Hint}
\end{align}
It is the predominance of the Hubbard and of Hund's on-site interactions
over the kinetic energy that establishes the Mott insulating ground
state and the characteristic multiplet structure of states {[}see
top panel of Fig.~\ref{fig:tower}{]}, which is determined by the
hierarchy of the charge excitations. The multiplet structure can be
considered as the fingerprint of the strong correlations characterizing
these systems, and determines ultimately the magnetic properties via
the spin-orbital superexchange interactions~\cite{Ole05}.

Finally, there are additional small but nevertheless relevant terms
that control the spin-orbital states in these compounds, that is, they
determine the anisotropy in the orbital or spin sector, respectively.
The last two terms in Eq. (\ref{H3band}) reflect small deviations from
the cubic symmetry of the \emph{cubic} vanadates \cite{Kha01,Hor03,Hor07},
where
\begin{equation}
{\cal H}_{{\rm CF}}=-\Delta_{c}\sum_{i\sigma}n_{i\sigma c},
\end{equation}
is the crystal field that splits the $t_{2g}$ orbitals favoring the
$c$ orbital at each site ($\Delta_{c}>0$) \cite{Ren00}. The remaining
electron can then occupy either one of the degenerate $a$ and $b$
orbitals. The small Jahn-Teller (JT) interaction acting in the orbital
space
\begin{align}
{\cal H}_{{\rm JT}} & =\frac{1}{4}V_{ab}\sum_{\langle ij\rangle{\parallel}ab}
(n_{ia}-n_{ib})(n_{ja}-n_{jb})\nonumber \\
 & -\frac{1}{4}V_{c}\sum_{\langle ij\rangle{\parallel}c}
 (n_{ia}-n_{ib})(n_{ja}-n_{jb}).
\end{align}
favors alternating $\{a,b\}$ orbitals and AO order in the $ab$ plane
($V_{ab}>0$) and the \emph{ferro} $\{a,b\}$ orbital order along the
$c$ crystal axis ($V_{c}>0$) \cite{Kha01}.

So far, the three-band $t_{2g}$ Hamiltonian includes the interactions
with charged defects, $e$-$e$ interactions, i.e., short- and long-
range, and it provides a faithful description of the magnetic and
orbital interactions. The latter is an immediate consequence of the
proper description of the atomic multiplet structure. Hence, this model
can describe the generic magnetic and orbital ordered phases appearing
in the $R$VO$_{3}$ perovskites \cite{Fuj10}, namely two complementary
spin-orbital structures: the $C$-AF spin with $G$-AO order and the
$G$-type AF spin with $C$-type AO order, as well as a pure AO ordered
phase \cite{Fuj10}.

In the above list of interactions, for the sake of simplicity and
clarity of our multiband model, we have omitted certain terms. Among
the neglected terms one finds:
(i)~The relativistic spin-orbit interaction, although small in $3d$
transition metals like vanadium, provides additional orbital fluctuations
and contributes to orbital moment formation \cite{Hor03,Hor08}.
(ii)~The orbital polarization leads to the rotation of orbitals in
the proximity of defects and to flavor mixing \cite{Hor11,Ave13}.
Both terms influence the spin-orbital order and the metal-insulator
transition, and would be relevant for a quantitative study of the
phase diagram of doped vanadates, which is however not our intention
here.

\section{Hartree-Fock approximation\label{sec:hf}}

The main aim of this paper is to analyze the evolution of the spectral
weights in the PES and the IPES of strongly correlated spin-orbital
system with random charged defects. This is very challenging as it
is necessary to treat simultaneously and on equal footing (i) the
strong correlation problem (Hubbard) in a multi-orbital system (Hund's
off-diagonal terms and constrained hopping) in presence of strong
coupling to the lattice (Jahn-Teller) and of strong fluctuations of
all such degrees of freedom, (ii) the local perturbations introduced
by the defects into the electronic structure and the long-range nature
of their Coulomb potential, which can lead to a potential landscape
that can be tuned continuously between monopolar and dipolar, as well
as (iii) the randomness of the locations of the defects that necessarily
requires a statistical treatment. We demonstrate here that the uHF
approach provides an efficient calculation scheme that is able to
reproduce the essence of the variations in the spectral weights of
the Hubbard bands, provided one deals with a system with broken
symmetry as here happens in the spin-orbital sector.

Given the above prescriptions, the derivation of the uHF equations
is standard and we do not present it here \textit{in extenso\/};
more details can be found, for instance, in Refs.~\cite{Miz96,Miz97,Bar00,Ave13}.
The essence of the derivation is that the $e$-$e$ interactions are
replaced by the terms containing mean fields acting on the single-particle
electron densities. Following this procedure, one arrives at an effective
single-electron Hamiltonian,
\begin{eqnarray}
{\cal H}_{{\rm HF}} & = & \sum_{i\alpha\sigma}\varepsilon_{i\sigma}^{\alpha}
n_{i\alpha\sigma}+\sum_{i}\sum_{\alpha\beta\sigma}\gamma_{i\sigma}^{\alpha\beta}
d_{i\alpha\sigma}^{\dagger}d_{i\beta\sigma},\nonumber \\
 & + & \sum_{ij}\sum_{\alpha\beta\sigma}t_{ij}^{\alpha\beta}
 d_{i\alpha\sigma}^{\dagger}d_{j\beta\sigma}\,.\label{HHF}
\end{eqnarray}
This Hamiltonian can be diagonalized numerically and the mean fields
appearing in the parameters (see below) and the HF states can be determined
self-consistently within an iterative procedure. More details on the
actual calculation scheme can be found in Appendix A, together with
the treatment of the randomness in the present problem.

Following Ref.~\cite{Ave13}, we emphasize that Fock terms become
active only if their off-diagonal mean fields are finite. This happens
only if a single-particle term in the original Hamiltonian (a source
for the specific off-diagonal mean fields) induces a finite value
for them. Consequently, we adopt here only those Fock terms that couple
the same orbitals at neighboring sites (as in the kinetic energy).
The terms which couple different orbitals at the same site are inactive
in the present study (\ref{H3band}) 
as such terms do not appear in the Hamiltonian Eq. (\ref{H3band})
\cite{Ave13}.

Apart from the single-particle tight-binding terms that are treated
rigorously, the Hamiltonian (\ref{H3band}) includes the JT terms,
$e$-$e$ intersite interactions $\propto V_{ij}$, and on-site terms
$\propto U$ and $\propto J_{H}$ in Eq. (\ref{Hint}). Note that
the JT terms in ${\cal H}_{{\rm JT}}$ are effective density-density
interactions generated by the JT distortions and, as such, they should
be treated in the Hartree approximation. As these and other terms
obtained in the Hartree approximation are rather straightforward to
evaluate, they will not be listed here and we address below only the
Fock terms. The latter terms originate from the $e$-$e$ interactions
$\propto V_{ij}$ and contribute effectively to the nearest neighbor
hopping terms,
\begin{equation}
-\sum_{\left\langle ij\right\rangle}\sum_{\alpha\sigma}
V_{ij}\left\langle d_{i\alpha\sigma}^{\dagger}d_{j\alpha\sigma}\right\rangle
\left(d_{i\alpha\sigma}^{\dagger}d_{j\alpha\sigma}
+d_{j\alpha\sigma}^{\dagger}d_{i\alpha\sigma}\right).
\end{equation}
Formally, such terms appear also for pairs of more distant vanadium
ions, but the hopping is limited to nearest neighbors in the model
Hamiltonian (\ref{H3band}), so they will vanish automatically in
the self-consistent solution. Finally, the on-site Coulomb interaction
$H_{{\rm int}}$ (\ref{Hint}) generates both Hartree and Fock terms,
and one finds
\begin{align}
& n_{i\alpha\sigma}n_{i\beta\sigma}\simeq\langle n_{i\alpha\sigma}\rangle\,
n_{i\beta\sigma}+n_{i\alpha\sigma}\,\langle n_{i\beta\sigma}\rangle
-\langle n_{i\alpha\sigma}\rangle\,\langle n_{i\beta\sigma}\rangle\nonumber \\
& -\langle d_{i\alpha\sigma}^{\dagger}d_{i\beta\sigma}\rangle\,
d_{i\beta\sigma}^{\dagger}d_{i\alpha\sigma}-d_{i\alpha\sigma}^{\dagger}d_{i\beta\sigma}\,
\langle d_{i\beta\sigma}^{\dagger}d_{i\alpha\sigma}\rangle\nonumber \\
& +\langle d_{i\alpha\sigma}^{\dagger}d_{i\beta\sigma}\rangle\,
\langle d_{i\beta\sigma}^{\dagger}d_{i\alpha\sigma}\rangle,\\
& S_{i\alpha}^{+}S_{i\beta}^{-}=d_{i\alpha\uparrow}^{\dagger}d_{i\alpha\downarrow}
 d_{i\beta\downarrow}^{\dagger}d_{i\beta\uparrow}\simeq\nonumber \\
& -\langle d_{i\alpha\uparrow}^{\dagger}d_{i\beta\uparrow}\rangle\,
 d_{i\beta\downarrow}^{\dagger}d_{i\alpha\downarrow}-d_{i\alpha\uparrow}^{\dagger}
 d_{i\beta\uparrow}\,\langle d_{i\beta\downarrow}^{\dagger}d_{i\alpha\downarrow}\rangle
\nonumber \\
& +\langle d_{i\alpha\uparrow}^{\dagger}d_{i\beta\uparrow}\rangle\,
 \langle d_{i\beta\downarrow}^{\dagger}d_{i\alpha\downarrow}\rangle,\\
& d_{i\alpha\uparrow}^{\dagger}d_{i\alpha\downarrow}^{\dagger}
 d_{i\beta\downarrow}d_{i\beta\uparrow}\simeq
 \langle d_{i\alpha\uparrow}^{\dagger}d_{i\beta\uparrow}\rangle\,
 d_{i\alpha\downarrow}^{\dagger}d_{i\beta\downarrow}\nonumber \\
& +d_{i\alpha\uparrow}^{\dagger}d_{i\beta\uparrow}^ {}\,
 \langle d_{i\alpha\downarrow}^{\dagger}d_{i\beta\downarrow}\rangle
 -\langle d_{i\alpha\uparrow}^{\dagger}d_{i\beta\uparrow}\rangle\,
 \langle d_{i\alpha\downarrow}^{\dagger}d_{i\beta\downarrow}\rangle.
\end{align}
Note however that these latter Fock terms, mixing the orbitals, are
included here only to exhaust the approximate treatment of $H_{{\rm int}}$
(\ref{Hint}). They would be active, for instance, if the orbitals
were optimized locally by a finite orbital polarization interaction
\cite{Ave13}, that is not considered here though. Therefore, we do
not include them in our analysis as it suffices to decouple the individual
terms in Eq. (\ref{Hint}) using Hartree approximation.

The uHF calculations in this work are performed for three different
sets $\{A,B,C\}$ of fundamental interaction parameters $U$, $J_{H}$
and $V_{D}$ listed in Table~\ref{tab:para} for convenience. They have
been motivated by different experimental results and/or considerations
as parameters relevant for $R$VO$_{3}$. Set $A$ is deduced in this
work from PES data for the undoped LaVO$_{3}$ compound \cite{Mai00}
and gives the positions of the LHB and the UHB multiplet structure
in qualitative agreement with the PES and IPES spectra of Maiti and
Sarma \cite{Mai00}, see Fig. \ref{fig:maiti}. Set $B$ was deduced
from the optical spectroscopy and the magnetic properties of YVO$_{3}$
\cite{Kha04} and was used before to: (i)~analyze the magnetic transition
from $G$-AF to $C$-AF phase in YVO$_{3}$ \cite{Hor11}, and (ii)~study
the defects in Y$_{1-x}$Ca$_{x}$VO$_{3}$ \cite{Ave13,Ave15}. Set
$C$ is dictated by a simplified analysis of the PES and IPES data in
Ref.~\cite{Mai98}, as explained in more detail in Sec.~\ref{sec:atom}.
Otherwise, we consider $\eta=1$ in Eq.~(\ref{pots}) and vary the
doping in the range $x\in(0,0.5]$.

Sets $A$ and $C$ are rather similar concerning the most important
feature, namely the distance of the centers of LHB and UHB given by
$(U-3J_{H})=2.75$ and 3.0 eV, respectively. Set $B$ instead was
determined from a model Hamiltonian with only local Hubbard-Hund type
interactions, the missing excitonic corrections \cite{May06,Mat09}
to the optical gap require in this case a significantly smaller value
$(U-3J_{H})=2.2$ eV. An analysis of the parameter $V_{D}$ that defines
the defect potential strength as well as the $e$-$e$ interaction
will be given in Section~VIII. Although our work focuses on the role
played by $e$-$e$ interactions, we present in the following calculations
both for the \emph{realistic} set $C$ and for the set $B$, because
set $B$ displays features that are somewhat hidden in the spectra
calculated using set $C$.

\begin{table}[t!]
\caption{The three parameter sets for the interactions in Hamiltonian
(\ref{H3band}) used in the numerical uHF calculations; all parameters
are given in eV. The set $A$ refers to LaVO$_{3}$ and was used to
obtain the theoretical spectrum in Fig.~\ref{fig:maiti};
the remaining sets are for doped systems and include a finite defect
potential $V_{{\rm D}}$ (\ref{v}).}
\vskip .1cm
\begin{ruledtabular}
\begin{tabular}{ccccccc}
set  & $\Delta_{c}$ & $U$ & $J_{H}$ & $V_{ab}$ & $V_{c}$ & $V_{{\rm D}}$
\tabularnewline
\colrule $A$  & 0.1  & 4.1  & 0.45  & 0.03  & 0.05  & \textemdash{} \tabularnewline
$B$  & 0.1  & 4.0  & 0.6  & 0.03  & 0.05  & 1.0 \tabularnewline
$C$  & 0.1  & 4.5  & 0.5  & 0.03  & 0.05  & 2.0 \tabularnewline
\end{tabular}
\end{ruledtabular}

\label{tab:para}
\end{table}


\section{Doping dependence of the density of states\label{sec:dop}}


\subsection{Multiplet structure and sum rules\label{sec:multi}}

\begin{figure}[t!]
\includegraphics[width=8.2cm]{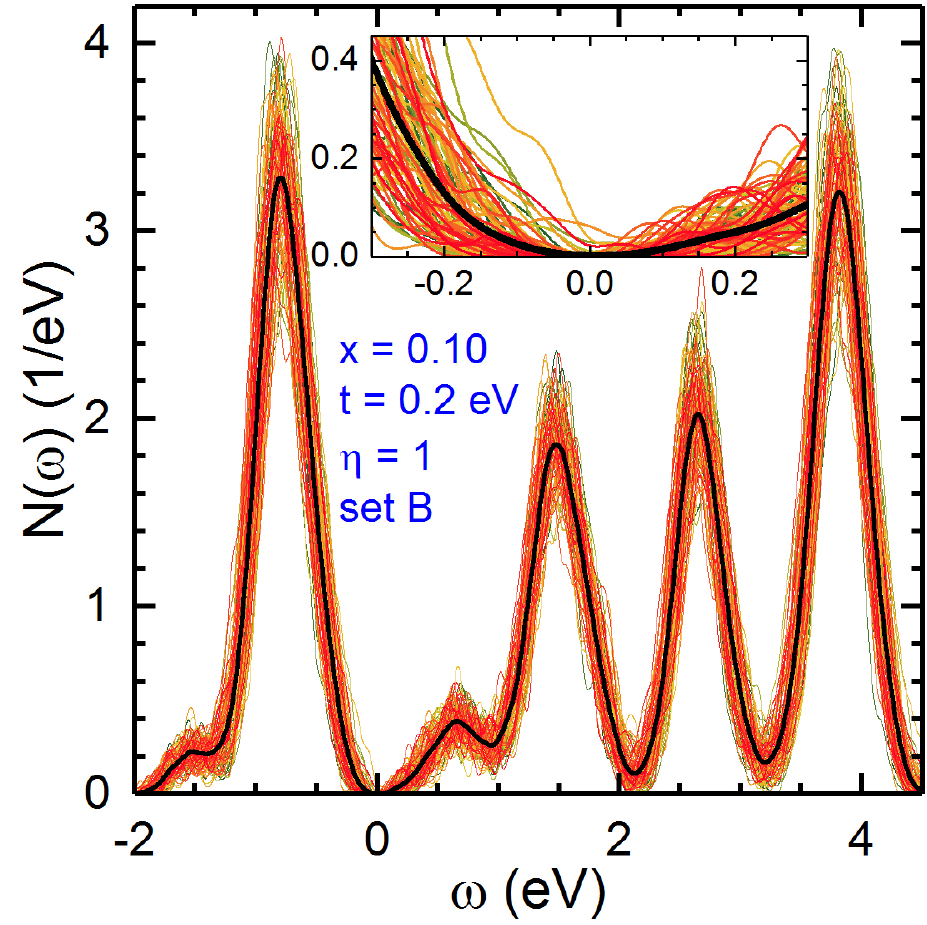} \caption{
Statistical average of $t_{2g}$ DOS $N(\omega)$ (\ref{Nw_av}) (thick
solid black line) as obtained from $M=100$ random defect realizations
(thin solid green-to-red lines) for $t=0.2$ eV and including $e$-$e$
interactions ($\eta=1$) at doping concentration $x=0.1$. Other parameters
as in set $B$ of Table~\ref{tab:para}. Inset shows the zoom of
the DOS near the Fermi energy at $\omega=0$.\label{fig_z}}
\end{figure}

In this Section, we begin our study of the doped Mott insulator with
the DOS $N(\omega)$ of the multiband model in the phase with $C$-type
spin and $G$-type orbital order (i.e., $C$-AF/$G$-AO phase). Within
the uHF method, $N(\omega)$ describes processes that correspond to
the addition of an electron for $\omega>\mu$ and to the removal of
an electron for $\omega<\mu$, where $\mu$ is the chemical potential.
Thus, $N(\omega)$ provides information relevant for the interpretation
of PES and IPES or tunneling experiments \cite{Dam03,Dam16}. The
numerical results presented in this paper are obtained for a cluster
of $N_{a}=8\times8\times8$ vanadium ions with periodic boundary conditions,
after averaging over $M=100$ randomly chosen different Ca defect
\emph{realizations} $\{s\}$ (i.e., sets of randomly chosen Ca defect
locations) at a given doping $x\in(0,0.5]$. We use as standard parameters
set $B$ of Table \ref{tab:para} and $t=0.2$ eV \cite{Kha01}. For
each defect realization $s$, we determine the $6\times N_{a}$ eigenvalues
$\{\epsilon_{s,l}\}$ and the value of the Fermi energy $\mu_{s}$.
Next, we calculate the final averaged DOS $N(\omega)$, representative
for the whole system with $N_{a}$ sites and $xN_{a}$ randomly distributed
defects, as an average over the $M$ defect realizations,
\begin{equation}
N\left(\omega\right)\equiv\frac{1}{M}\sum_{s=1}^{M}\left[
\frac{1}{N_{a}}\sum_{l=1}^{6N_{a}}\delta(\omega+\mu_{s}-\epsilon_{s,l})\right].
\label{Nw_av}
\end{equation}
The shapes of the structures arising in the LHB and the UHB are somewhat
different in each of the defect realizations $s$, but all of them
show the characteristic maxima to be discussed below, see Fig. \ref{fig_z}.

Each charged defect adds a hole to the system, thus the averaging
over the different defect realizations is performed for an electron
density $n=N_{0}/N_{a}=2-x$ per V ion, and the chemical potential
is determined as $\mu_{s}=(\epsilon_{s,N_{0}}+\epsilon_{s,N_{0}+1})/2$.
The obtained DOS satisfies the sum rule:
\begin{equation}
\int_{-\infty}^{\infty}d\omega\,N(\omega)=6,\label{sum2}
\end{equation}
and determines the total number of $t_{2g}$ electrons per site,
\begin{equation}
\int_{-\infty}^{0}d\omega\,N(\omega)=n,\label{sum1}
\end{equation}
where the overall chemical potential is at $\omega=0$. To control
the gradual change of the spectral weights found for the particular
structures observed in the DOS, it is convenient to introduce the
integrated DOS $n(\omega)$ defined as follows,
\begin{equation}
n(\omega)\equiv\int_{-\infty}^{\omega}d\omega'\,N(\omega').
\label{nw}
\end{equation}

\begin{figure}[t!]
\includegraphics[width=8.2cm]{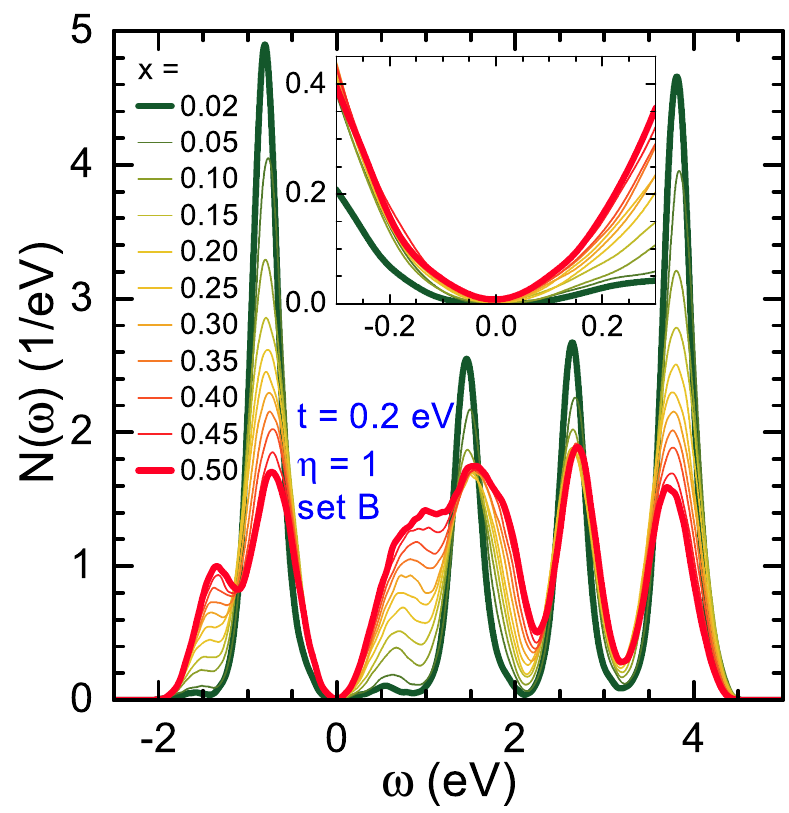} \caption{
Doping dependence of $t_{2g}$ DOS $N(\omega)$ (\ref{Nw_av}) as
obtained from $M=100$ random defect realizations for $t=0.2$ eV
and including $e$-$e$ interactions ($\eta=1$) for increasing doping
$x\in[0.02,0.50]$. Other parameters as in set $B$ of Table~\ref{tab:para}.
Inset shows the zoom of the DOS near the Fermi energy at $\omega=0$.
\label{fig:dos02}}
\end{figure}

Undoped YVO$_{3}$ (or LaVO$_{3}$) is a Mott insulator and its DOS
consists of a LHB and a UHB, separated by a wide MH gap, shown in
Fig.~\ref{fig:dos02}. The reference here are the data obtained at
the lowest value of doping $x=0.02$. The LHB is given by a single
maximum with total spectral weight $w_{{\rm LHB}}=2$
in the undoped system, which corresponds to possible one-electron
excitations in the PES, see Fig.~\ref{fig:maiti}. In contrast, the
UHB has three distinct maxima, which reflect the multiplet structure
of the excitations accessible in the IPES for the undoped system,
as discussed for the two-flavor model in Ref.~\cite{Hor11}. We observe
that the full width $W$ (at half maximum) of all these structures
and of the LHB is $W\simeq0.5$ eV at low and moderate doping. One
could argue that this broadening originates from the incoherence
expected for hole (electron) polaron motion in a system with broken
symmetry. However, one would expect for a free polaron a width
$W_{{\rm inc}}=4\sqrt{z-1}\,t$ \cite{Bri70}, where the number of
neighbors for the effective 2D hopping of $t_{2g}$ electrons is $z=4$.
This would suggest a much larger value $W_{{\rm inc}}\approx1.4$~eV.
We take this as an indication that the doped holes or polarons are
immobile and bound to defects. Hence the broadening is predominantly
due to the distribution of the local energies due to the random
defect potemtials (see Sec. \ref{sec:bond}).
We have shown before that the width of the LHB is $\sim1.0$ eV in
the absence of $e$-$e$ interactions, and is \textit{reduced} to
$\sim0.5$ eV in the presence of long-range Coulomb interactions
\cite{Ave15}, due to the screening of the defect potentials resulting
from $t_{2g}$ electrons.

A particularly exciting feature is the persistence of a soft gap right
at the chemical potential $\omega=0$ in Fig.~\ref{fig:dos02}. Although
many defect states fill into the MH gap with increasing doping $x$,
it appears that there is a mechanism at work resulting in the highest
occupied states and the lowest unoccupied states repelling each other.
This is reminiscent of the Peierls effect, but also of the Coulomb
gap mechanism. The detailed analysis of the origin of the DS gap in
our model is a central issue in this and the following Sections.

\begin{figure}[t!]
\includegraphics[width=8.2cm]{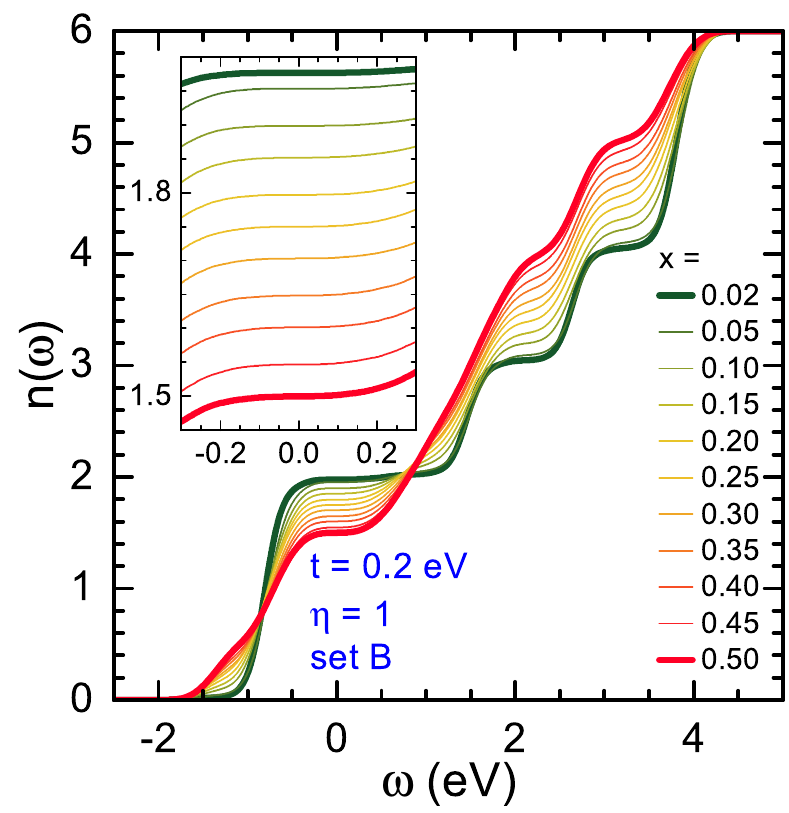}
\caption{Integrated density of states $n(\omega)$ \eqref{nw} for different
doping concentration $x\in[0.02,0.50]$ obtained for the spectra presented
in Fig. \ref{fig:dos02} for $t=0.2$ eV and including $e$-$e$ interactions
($\eta=1$). Other parameters as in set $B$ of Table \ref{tab:para}.
Inset shows the zoom of the $n(\omega)$ near the Fermi energy at
$\omega=0$.\label{fig_c} }
\end{figure}

The DOS obtained for doped $R_{1-x}$Ca$_{x}$VO$_{3}$ systems preserves
the main features seen at $x=0$: the LHB and the UHB are separated by
the MH gap, see Fig.~\ref{fig:dos02}. With increasing doping $x>0$,
the spectral weights below (above) the Fermi energy change in a
systematic way. In particular, the fundamental splitting of the MH gap
$(U-3J_H)$ persists up to surprisingly high doping. There arise also
new features due to defect states with intensities growing with doping
$x$ \textemdash{} one finds a second structure in the LHB at low energy,
see Fig.~\ref{fig:dos02}. A similar structure with a faster increase of
the spectral weight as a function of doping $x$ is observed inside the
MH gap at the low-energy edge of the first maximum in the UHB.
At the same time, the Hubbard subbands
found at $x=0$ persist, but show decreasing spectral intensities
with increasing $x$. We analyze these changes in the next Section.

It is insightful to analyze the evolution of the spectra in
Fig.~\ref{fig:dos02} by considering the integrated DOS $n(\omega)$
\eqref{nw}. At small doping $x=0.02$, $n(\omega)$ is characterized by
several steps that reflect the wide MH gap (wide plateau around
$\omega=0$) and the minima of $N(\omega)$ that separate three different
states in the UHB (two narrow plateaus), see Fig.~\ref{fig_c}. With
increasing $x$ the plateaus shrink due to the appearance of more and
more defect states inside the different gaps. The DOS integrated up to
the Fermi energy $\omega=0$ gives the total electron density, $n(0)=n$.
That is, the weight of the LHB including the low energy satellite
of the LHB is $(2-x)$. A minute increase just above $\omega=0$
indicates that the MH gap is accompanied by a small maximum, arising
at finite doping, just above the Fermi energy. For increasing $x$, the
maximum above the Fermi energy extends over a broader range,
$0.2<\omega<2.0$, suggesting that all structures arising at finite
doping are all absorbed within this broader and broader maximum.
We turn back to this discussion below in Sec.~\ref{sec:bond}.

\subsection{Atomic limit: Exact solution}
\label{sec:atom}

It is convenient to recall first the main features of the one-particle
spectra for the nondegenerate Hubbard model. One of its main successes
was to elucidate the variation of the electronic structure in a strongly
correlated electronic system with increasing/decreasing electron density
and to provide its explanation. It captures the evolution of spectral
weights within the individual Hubbard bands
\cite{Mei91,Mei93,Fuj92,Dag92,Esk94,Esk94a,deM05}.
Indeed, in a Mott insulator described by the nondegenerate Hubbard model,
the DOS consists of two Hubbard bands, the LHB and the UHB, with equal
spectral weights at half filling. These weights however change rather
fast in a hole doped insulator when the number of unoccupied
states at energies just above the Fermi energy increases \textemdash{}
these states belong to the LHB in a system without charged defects
\cite{Mei91,Mei93,Fuj92,Dag92}. A similar behavior is found for electron
doping by applying the particle-hole symmetry. The evolution of spectral
weights with increasing doping has been explained within a systematic
expansion in $(t/U)$ in the Hubbard model \cite{Esk94a}, but frequently
it is sufficient to use only the leading terms of this expansion, which
stem from the atomic limit and are written in terms of the electron
density. For the nondegenerate Hubbard model at hole doping $x$ this
approximation predicts that the LHB has altogether $(1+x)$ states, with
$(1-x)$ states below and $2x$ (unoccupied) states above the Fermi energy.

\begin{figure}[t!]
\includegraphics[width=8cm]{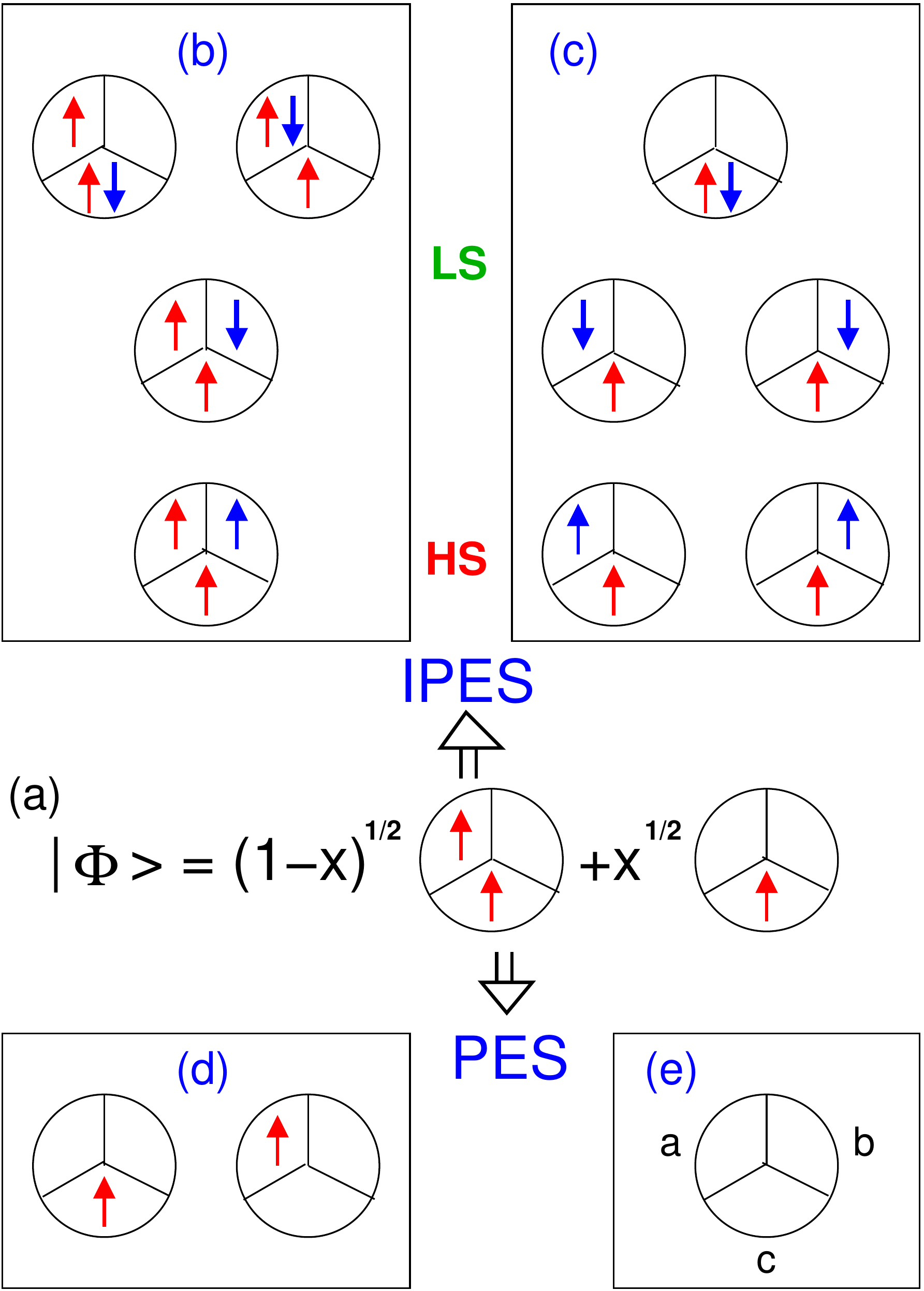}
\caption{ Artist's view of possible PES (hole doping, bottom) and IPES
(electron doping, top) excitations in doped $R_{1-x}$Ca$_{x}$VO$_{3}$
compounds, as obtained in the HF approximation. In (a), we show a
representative ground state $|\Phi\rangle$ with $\uparrow$-spin
electron(s) (red arrows) which includes both
$|c{\uparrow}a{\uparrow}\rangle$ and $|c{\uparrow}\rangle$ local
states in a doped system with amplitudes $\sqrt{1-x}$ and $\sqrt{x}\,$
\textemdash{} we select a V ion on the $\uparrow$-spin sublattice with
local undoped (doped) $|c{\uparrow}a{\uparrow}\rangle$
($|c{\uparrow}\rangle$) state. Adding an electron in IPES
(UHB) (blue arrows) generates in this ground state either:
(b) four possible $t_{2g}^{3}$ final states with probabilities $(1-x)$
each, or
(c) five possible $t_{2g}^{2}$ final states with probabilities $x$
each (with some modifications, see the main text). The states excited
in IPES are either HS or LS, with their energies listed in
Table~\ref{tab:sw}.
Removing an electron (replacing an arrow by a hole) in PES reduces
the spin by $1/2$ and gives:
(d) either the two $t_{2g}^{1}$ excited states with probabilities
$(1-x)$ in the LHB, or
(e) an empty $t_{2g}^{0}$ final state with probability $x$.
In each panel, three $t_{2g}$ orbitals are labeled as $\{a,b,c\}$
(\ref{abc}), with the convention shown in~(e).\label{fig:hub} }
\end{figure}

In the following, we adopt the atomic perspective and analyze the
$C$-AF/$G$-AO states with broken symmetry, addressing also possible
deviations resulting from the finite kinetic energy $\propto t$ and
from the uHF. For a vanadium perovskite with charged defects the holes
added to vanadium ions generate local $t_{2g}^{1}$ configurations
with probability $x$, while the $t_{2g}^{2}$ states occur with
probability $(1-x)$ in a doped system. This is schematically
represented by the ground state wave function $|\Phi\rangle$ for a
vanadium ion in Fig.~\ref{fig:hub}(a). In the case depicted in
Fig.~\ref{fig:hub} doping has removed an $|a{\uparrow}\rangle$
electron which has a higher energy than the $|c{\uparrow}\rangle$
electron due to the small crystal field $\Delta_{c}=0.1$ eV, see also
Fig.~\ref{fig:cube}. We start from these states to explain the
possible PES and IPES excitations, that result from electron removal
and addition processes, and lead to final states in the LHB and
the UHB, respectively,  but also to various defect states.

\begin{table}[t!]
\caption{ Spectral weights $w_{n}$ and atomic excitation energies
$\varepsilon_{n}$
in IPES and PES processes available in doped $R_{1-x}$Ca$_{x}$VO$_{3}$,
leading to final states shown in Fig.~\ref{fig:hub}, as found in
the uHF approximation (HF). The energies that involve a doped ion
are increased by the energy $V_{{\rm D}}$. The energies found in
HF (HF) are compared with those found in the atomic limit (exact)
where quantum fluctuations are also included; the spectral weights
$w_{n}$ are as in uHF. The HF approach reproduces
exact results for PES and for HS IPES states, where only a single
energy is shown. For convenience, the PES excitation energy for the
host is taken as a reference, $\varepsilon_{{\rm LHB}}=0$ (here,
the small crystal-field splitting $\Delta_{c}$ is ignored). }
\vskip .1cm
\begin{ruledtabular}
\begin{tabular}{cccccc}
\multirow{2}{*}{IPES} & \multirow{2}{*}{final state} & \multirow{2}{*}{spin} & \multirow{2}{*}{$w_{n}$} & \multicolumn{2}{c}{$\varepsilon_{n}$}\tabularnewline
 &  &  &  & HF  & exact \tabularnewline
\hline
$d^{2}\rightarrow d^{3}$  & $|c{\uparrow}a{\uparrow}b{\uparrow}\rangle$  & HS  & $1-x$  & \multicolumn{2}{c}{$U-3J_{H}$}\tabularnewline
 & $|c{\uparrow}a{\uparrow}b{\downarrow}\rangle$  & LS  & $1-x$  & $U-J_{H}$  & $U$ \tabularnewline
 & $|c{\uparrow}a{\uparrow}c{\downarrow}\rangle$  & LS  & $1-x$  & $U+J_{H}$  & $U$ \tabularnewline
 & $|c{\uparrow}a{\uparrow}a{\downarrow}\rangle$  & LS  & $1-x$  & $U+J_{H}$  & $U+2J_{H}$ \tabularnewline
$d^{1}\rightarrow d^{2}$  & $|c{\uparrow}a{\uparrow}\rangle$  & HS  & $x$  & \multicolumn{2}{c}{$V_{{\rm D}}$}\tabularnewline
 & $|c{\uparrow}b{\uparrow}\rangle$  & HS  & $x$  & \multicolumn{2}{c}{$V_{{\rm D}}$}\tabularnewline
 & $|c{\uparrow}b{\downarrow}\rangle$  & LS  & $x$  & $V_{{\rm D}}+J_{H}$  & $V_{{\rm D}}+2J_{H}$ \tabularnewline
 & $|c{\uparrow}a{\downarrow}\rangle$  & LS  & $x$  & $V_{{\rm D}}+J_{H}$  & $V_{{\rm D}}+2J_{H}$ \tabularnewline
 & $|c{\uparrow}c{\downarrow}\rangle$  & LS  & $x$  & $V_{{\rm D}}+3J_{H}$  & $V_{{\rm D}}+5J_{H}$ \tabularnewline
\hline
PES  & final state  & $-$  & $w_{n}$  & \multicolumn{2}{c}{$\varepsilon_{n}$}\tabularnewline
\hline
$d^{2}\rightarrow d^{1}$  & $|c{\uparrow}\rangle$  & $-$  & $1-x$  & \multicolumn{2}{c}{$0$}\tabularnewline
 & $|a{\uparrow}\rangle$  & $-$  & $1-x$  & \multicolumn{2}{c}{$0$}\tabularnewline
$d^{1}\rightarrow d^{0}$  & $|0\rangle$  & $-$  & $x$  & \multicolumn{2}{c}{$V_{{\rm D}}-(U-3J_{H}$)}\tabularnewline
\end{tabular}
\end{ruledtabular}
\label{tab:sw}
\end{table}

In IPES, one electron is added in the $d^n\rightarrow d^{n+1}$ process,
either to the $|c{\uparrow}a{\uparrow}\rangle$ initial configuration
($n=2$), see Fig.~\ref{fig:hub}(b), or at a V ion with a hole in the
$|c{\uparrow}\rangle$ initial state ($n=1$), see Fig.~\ref{fig:hub}(c).
As the doped sites are typically direct neighbors of a charge defect, the
electron energies at these sites are increased by $V_{{\rm D}}$ (\ref{v}).
Thus the HS states generated by $d^1\rightarrow d^2$ IPES transitions
appear as in-gap states above the Fermi energy. But there is also a
multiplet structure of the $d^2$ final states that overlaps partially
with the $t_{2g}^3$ multiplet states of the UHB as can be infered from
Fig.~\ref{fig:hub}. Altogether, the IPES excited states are quite
numerous, four and five for the two cases listed above, and their spin
is either increased by 1/2 in HS states, or decreased by 1/2 in LS
excited states. Their exact energies are given in Table~\ref{tab:sw}.

Taking all the states generated in IPES one finds their total weight
of $(4+x)$ which is a generalization of the $(1+x)$ found in the
nondegenerate Hubbard model \cite{Mei93} and the  $(3+x)$ for the
two-flavor model in Ref. \cite{Hor11}.
The UHB obtained for $d^2\rightarrow d^3$ excitations has three
subbands corresponding to the atomic multiplet structure
with energies $(U-3J_H)$, $U$, and $(U+2J_{H})$ as given in Table II.
They are found from the exact solution treating rigorously the
Coulomb interactions for $t_{2g}^{3}$ excited states in the atomic
limit. In a doped system, the initial $t_{2g}^2$ state occurs in the
wave function $|\Phi\rangle$ with the probability $(1-x)$, see Fig.
\ref{fig:hub}(a). Therefore, the weights of these excitations are:
$(1-x)$, $2(1-x)$, and $(1-x)$. This weight distribution is modified
when quantum fluctuations are neglected, see below.

Let us compare first the results obtained for the PES and IPES spectra
with the experimental results for undoped LaVO$_{3}$ shown in
Fig.~\ref{fig:maiti}. The spectrum consists of two distinct structures,
the LHB and the UHB separated by a broad MH gap. From the experimental
spectra one learns that the distance that separates the LHB and the HS
excitation in the UHB is $3.0$ eV. This gives the first experimental
constraint on the on-site Coulomb interaction parameters $\{U,J_{H}\}$
\textemdash{}
this energy difference obtained in the theory is $(U-3J_{H})$, see
Table~\ref{tab:sw}. Furthermore, for the undoped LaVO$_{3}$, the HF
theory predicts three peaks in the UHB, with the HF energies of
$(U-3J_{H})$, $(U-J_{H})$, and $(U+J_{H})$ for the HS, LS$_{1}$, and
LS$_{2}$ excitations, see Table II. The energetic separation of the LS$_{2}$
and the HS states is thus $4J_{H}$ that determines the lower bound for
the width of the UHB, to be still enhanced by the experimental broadening.
Note that the same result would be obtained in the electronic structure
calculations implementing the leading part of local Coulomb interactions
\cite{Vau12}.

As explained above, the highest excitation energy for the LS$_{2}$
states is doubly degenerate in HF which gives here twice larger spectral
weight than that of the other two excitations, HS and LS$_{1}$. In
contrast, the LS$_{1}$ state is doubly degenerate when quantum effects are
included (see Table~\ref{tab:sw}) and the correct weights' distribution
is 1:2:1 for the HS, LS$_{1}$, and LS$_{2}$ states.
In Fig.~\ref{fig:maiti}, we have adjusted the spectral weights of the
uHF spectra accordingly to achieve a better agreement with experiment.
Unfortunately, the individual excitations within the UHB cannot
be resolved in the data \cite{Mai00}. The whole
spectrum formed by the LHB and UHB are reproduced well by the theory,
see Fig.~\ref{fig:maiti}, after adjusting the two characteristic
parameters $\{U,J_{H}\}$ that define the position of the maximum seen
in the UHB at the energy $(U-J_{H})$ of the LS$_{1}$ state, i.e.,
relative to the energy of the LHB. The best fit was obtained with
$U=4.1$ eV and $J_{H}=0.45$ eV that defines the set $A$ of
Table~\ref{tab:para}. Apart from a somewhat decreasing overall
experimental spectral weight with increasing excitation energy $\omega$,
which cannot be reproduced without additional information about the
matrix elements, the agreement between the experiment and the theory
predictions is indeed very satisfactory
and demonstrates the presence of a multiplet structure within the~UHB.

The analysis of the IPES excitations summarized in Table~\ref{tab:sw}
elucidates additional features generated in the spectra by doping.
For the $|c{\uparrow}\rangle$ component of the ground state wave
function $|\Phi\rangle$ [Fig.~\ref{fig:hub}(a)], with a hole in the
initially occupied $a$ orbital, two HS states and three LS states may
be generated by IPES, see Fig.~\ref{fig:hub}(c). Each of these states
will have a spectral weight $x$. In contrast to the excitations
occurring in the host shown in Fig.~\ref{fig:hub}(b), the HS excitations
at a doped site include two final $t_{2g}^{2}$ states, either
$|c{\uparrow}a{\uparrow}\rangle$ or $|c{\uparrow}b{\uparrow}\rangle$,
with the same local Coulomb energy $(U-3J_{H})$ as for $t_{2g}^{2}$
host states. Thus, these excitation energies
$\varepsilon_n$ do not include the intraatomic Coulomb $U$ or exchange
$J_H$ and would appear just above the LHB in the absence of a defect
potential (i.e., at $V_{{\rm D}}=0$) \cite{Mei93}. However, in reality
the $d^{1}\rightarrow d^{2}$ excitation energy is enhanced by large
$V_{{\rm D}}$ near the charged impurity in the center of the cube
occupied by the defect (Fig.~\ref{fig:cube}) and these unoccupied
states appear deep within the MH gap. The defect potential
$V_{{\rm D}}$ acts on any local state and enhances its excitation
energy, see Table~\ref{tab:sw}. The two LS states with two different
orbitals singly occupied have the final energies of $(U-J_{H})$, so
the excitation energies are $2J_{H}$. The excitation involving double
occupancy $|c{\uparrow}c{\downarrow}\rangle$ is more subtle: This state
itself is not an eigenstate of $H_{\rm int}$ (\ref{Hint}) as a double
occupancy couples by the term $\propto J_{H}$ to
$|a{\uparrow}a{\downarrow}\rangle$ or
$|b{\uparrow}b{\downarrow}\rangle$ state.
As a result, the highest excitation energy $(V_{{\rm D}}+5J_{H})$ is
obtained only for one (fully symmetric with respect to orbital
permutations) eigenstate while the energy $(V_{{\rm D}}+2J_{H})$ for
the other two. Eventually, the configuration
$|c{\uparrow}c{\downarrow}\rangle$ shown in Fig.~\ref{fig:hub}(c) is
found with the probability of $1/3$ in each of these eigenstates.
Hence, by performing the corresponding projections, one finds that
the final exact atomic spectral weights for the energies
$(V_{{\rm D}}+2J_{H})$ and $(V_{{\rm D}}+5J_{H})$ are $8x/3$ and~$x/3$.

Here again quantum fluctuations contribute and the HF energies of two
interorbital LS states, $|c{\uparrow}a{\downarrow}\rangle$ and
$|c{\uparrow}b{\downarrow}\rangle$, are lower by $J_{H}$, i.e., by the
same amount as found for the final LS states in the case of
$d^{2}\rightarrow d^{3}$ excitations. The other states are given by
the double occupancies that are here eigenstates at energy $U$, so
the excitation energy for the accessible
$|c{\uparrow}c{\downarrow}\rangle$ state is $(V_{{\rm D}}+3J_{H})$,
see Table II. The HF spectral weights for the LS$_1$ and LS$_2$ states
are thus $2x$ and $x$.

The PES excitations are much simpler than the IPES ones as just one
electron is removed from either component of the ground state wave
function shown in Fig.~\ref{fig:hub}(a) and the spin is then reduced
by $1/2$. The $d^{2}\rightarrow d^{1}$ excitations in the host have
approximately the same excitation energy taken here as the reference,
$\varepsilon_{n}=\varepsilon_{{\rm LHB}}=0$
(we neglect again the crystal-field term $\propto\Delta_{c}$). In
contrast, a PES excitation at the hole site, $d^{1}\rightarrow d^{0}$,
starts from the $|c{\uparrow}\rangle$ state, so the energy of the
$d^{2}$ reference configuration is subtracted from $V_{{\rm D}}$
in Table~\ref{tab:sw}.

\subsection{Spectral weight distribution in the atomic limit}
\label{sec:atomw}

\begin{figure}[t!]
\includegraphics[width=8cm]{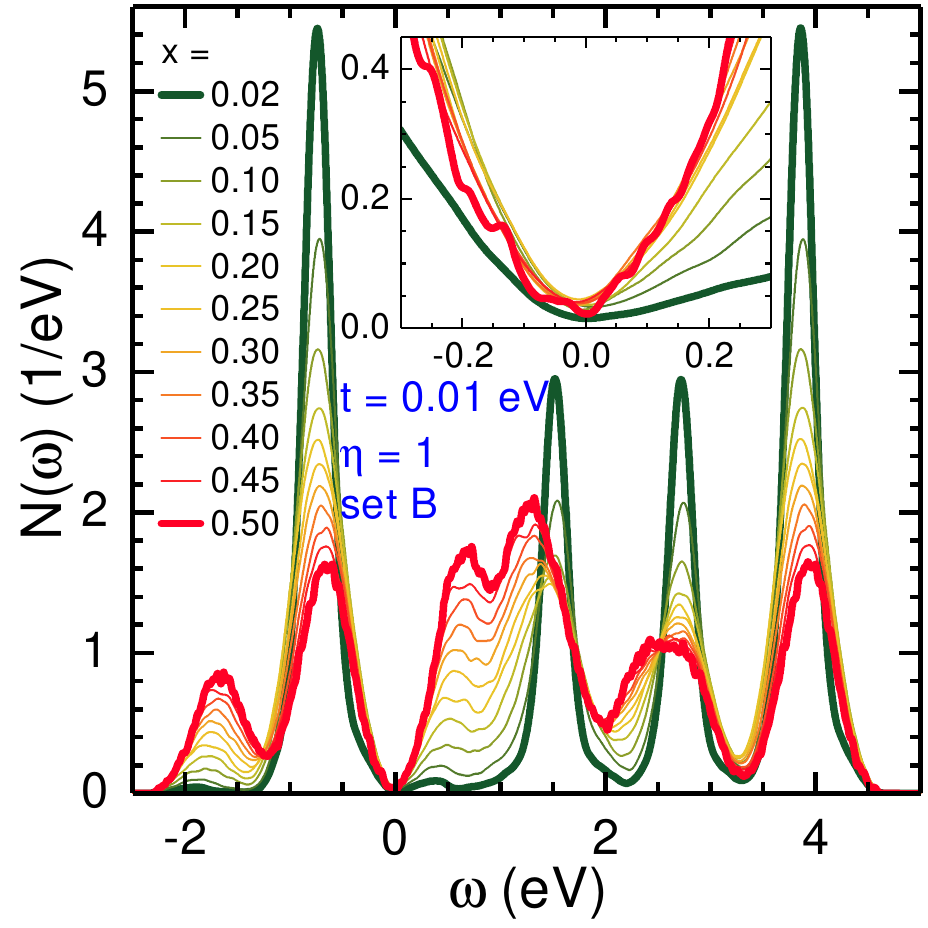}
\caption{Doping dependence of $t_{2g}$ DOS $N(\omega)$ (\ref{Nw_av}) as
obtained from $M=100$ random defect realizations for $t=0.01$ eV
and including $e$-$e$ interactions ($\eta=1$) for increasing doping
$x\in[0.02,0.50]$. Other parameters as in set $B$ in Table \ref{tab:para}.
Inset shows the zoom of the DOS near the Fermi energy at $\omega=0$.
\label{fig:dos001}}
\end{figure}

To illustrate the above theory, we determined the individual structures
in PES/IPES spectra and their evolution with increasing doping
$x\in[0.02,0.50]$ for $t=0.01$ eV. Here, the kinetic energy is chosen to
be very small indeed to generate the results representative for the
atomic limit. In Fig.~\ref{fig:dos001}, one observes that the MH gap
persists in doped systems. The LHB is well separated from the HS
excitation in the UHB by the MH gap for the entire doping regime. The
states arising within the MH gap do not close this gap but develop a
novel kinetic gap analyzed in more detail in the next Sec.~\ref{sec:gap}.

PES excitations (at $\omega<0$) are seen in the spectra as two
structures:
(i)~the LHB corresponding to individual $d^{2}\rightarrow d^{1}$
transitions at V sites that are not nearest neighbors of defects;
(ii)~a~satellite below the LHB at energy $\sim-2.0$ eV, i.e.,
originating from $d^{1}\rightarrow d^{0}$ processes at sites
occupied by doped holes, see Fig.~\ref{fig:dos001}. When the doping
$x$ increases, the spectral weight of the PES part is altogether
$\sim(2-x)$, which consists of the main peak representing the LHB
$\sim(2-2x)$ and a satellite generated by the excitations at the
hole sites with growing spectral weight $\sim x$. This agrees with
the spectral weights of the PES states listed in Table II.
The satellite moves gradually towards the LHB, reaching energy
$\sim-1.7$ eV at $x=0.45$. Excitations from undoped sites near
defects are close to the Fermi energy and can be well resolved
from the main maximum in the LHB.

The three multiplet transitions $d^{2}\rightarrow d^{3}$ that appear in
the IPES part of the uHF spectra are surprisingly well resolved even at
high doping. The spectral weights of these UHB states of the host
compound are $(1-x)$, $(1-x)$, and $2(1-x)$ for the HS, LS$_1$ and LS$_2$
states, see Fig.~\ref{fig:dos001}. The defect related features in the
IPES part ($\omega>0$) due to $d^{1}\rightarrow d^{2}$ excitations are
better resolved here than in case of $t=0.2$ eV (Fig.~\ref{fig:dos02}).
There are two distinct peaks that grow with doping within the MH gap,
at energies $\sim0.6$ eV and $\sim1.2$ eV. While the weight of the
lower maximum accumulates the weight $\sim 2x$, the weight if the
second maximum appears even somewhat larger. We recall that the weight
of the satellite below the LHB is $\sim x$, see Table~II. The third
maximum induced by doping falls near the
minimum separating the HS from the LS$_1$ excitation in the reference
multiplet structure of the UHB (for these parameters) and has a lower
weight $\sim x$. In addition, a spectacular evolution of pseudogap in
$N(\omega)$ close to $\omega=0$ is found (see inset). The highly
asymmetric spectrum changes gradually to an almost symmetric one with
increasing~$x$.

\begin{figure}[t!]
\includegraphics[width=8cm]{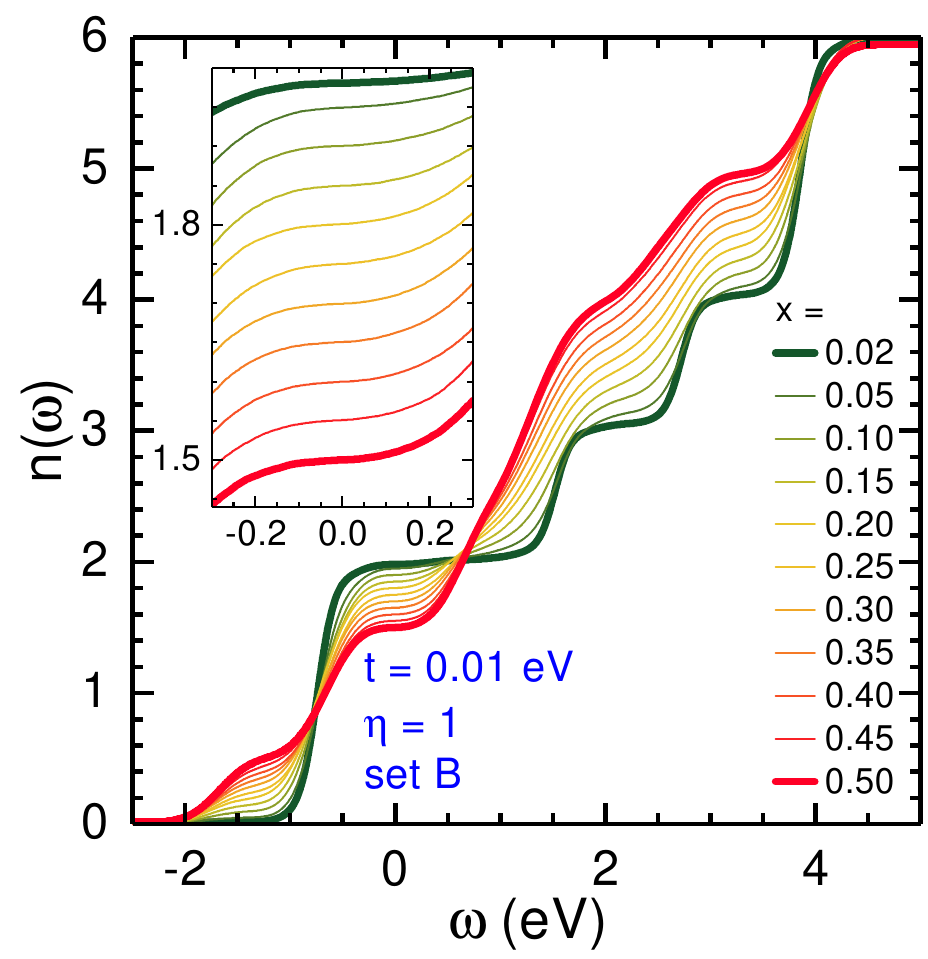} \caption{
Integrated density of states $n(\omega)$ \eqref{nw} for different
doping $x\in[0.02,0.50]$ obtained for the spectra presented in Fig.
\ref{fig:dos001} at $t=0.01$ eV and including $e$-$e$ interactions
($\eta=1$). Other parameters as in set $B$ in Table \ref{tab:para}.
Inset shows the zoom of the $n(\omega)$ near the Fermi energy at
$\omega=0$.\label{fig:nw001}}
\end{figure}

The modifications of the DOS $N(\omega)$ with increasing doping $x$
shown in Fig.~\ref{fig:dos001} can be even
better appreciated by analyzing the integrated DOS, see
Fig.~\ref{fig:nw001}. First of all, one finds an almost flat plateau
in the integrated DOS at $n(\omega)\simeq1.98$ at low doping $x=0.02$,
and the plateau at $\omega=0$ persists at higher doping. At $x=0.02$,
large steps of $(1-x)$ are found for the energies of the HS and LS$_{1}$
states in the UHB, and of $2(1-x)$ at energy of the LS$_{2}$ state.
Note that this latter excitation is well separated from the structures
generated by excitations at sites doped by holes as the former energies
are lower. Most importantly, the excitation energies are constant and
are not influenced by increasing doping, reflecting the robust local
character of both $d^{2}\rightarrow d^{3}$ and $d^{1}\rightarrow d^{2}$
processes.

At the lowest doping $x=0.02$, one is sufficiently close to the
reference undoped system characterized by the LHB at $\omega\sim-0.8$
eV, and the weak satellite feature with low intensity arises at
$\omega\simeq-2.0$ eV, identified by finite $n(\omega)$ for
$-2.0<\omega<-1.0$ eV. By removing a single electron one gains here the
energy $V_{{\rm D}}$ as this electron before the removal feels the
potential of the charged defect $V_{{\rm D}}$ in the center of the cube,
see Table II. Indeed, this latter excitation energy is lower than that
at the center of the LHB, as observed in Fig.~\ref{fig:dos001}. Faster
increase of $n\left(\omega\right)$ follows close to $\omega=-0.8$ eV and
afterwards the integrated weight grows again very slowly. Finally, the
spectral features that grow with increasing doping in
Fig.~\ref{fig:dos001} are responsible for the dramatic deformation of
a distinct step structure seen in Fig.~\ref{fig:nw001} at $x=0.02$
towards an almost steady increase of $n(\omega)$ from the onset of the
LHB to the top of the UHB, except for a quite narrow plateau near the
Fermi energy shown in the inset.

We also observe a remarkable feature in Fig.~\ref{fig:nw001}: Independently
of actual doping $x$, the integrated DOS reaches the value of two electrons
per site at $\omega\simeq 0.7$ eV. This shows that the spectral weight
missing in the PES part is compensated by the HS $d^1\rightarrow d^2$
excitations which have an approximately constant energy above the Fermi
level in the entire doping regime. A similar point is found at
$\omega\simeq-0.8$ eV and the filling of $n(\omega)=0.8$ --- here it
falls around the maximum of the LHB. These points are quite reminiscent
of the isosbestic point found in the specific heat of correlated systems
\cite{Vol97,Gre13}. Here the two isosbestic points originate from the
doping dependent spectral weight tranfers between the host Hubbard bands
and the defect states, that correspond to the final states of the
$d^1\rightarrow d^0$ and $d^1\rightarrow d^2$ transitions, respectively.

\begin{figure}[t!]
\includegraphics[width=8cm]{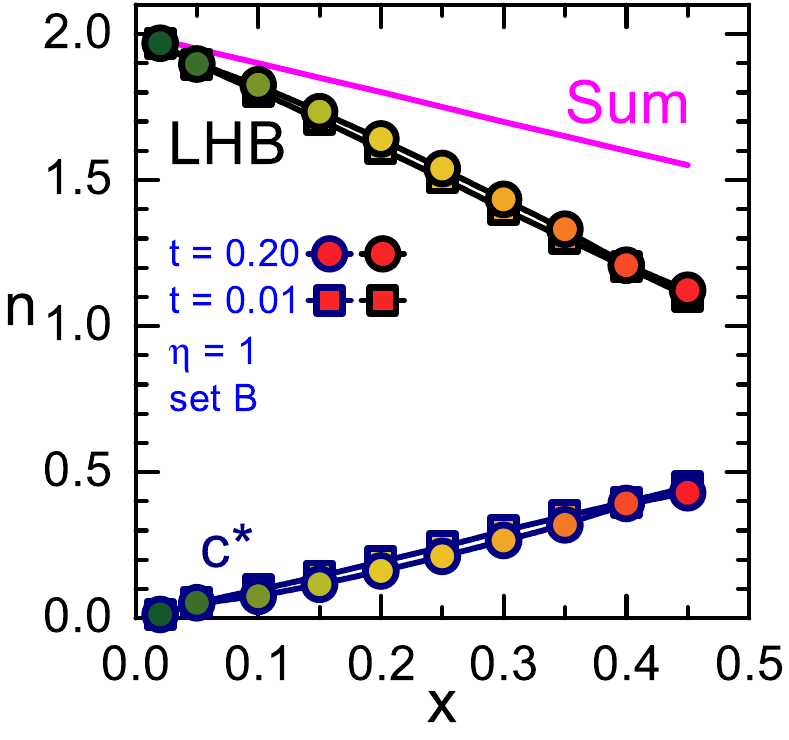} \caption{
Doping dependence of the total spectral weight below the Fermi energy
(Sum, magenta line), of the LHB (LHB, black symbols) originating from
undoped V sites and from the satellite arising due to PES excitations
at V doped ions $|c_{i\sigma}^*|0\rangle$ ($c^{*}$, blue symbols)
bound by a charged defect, as obtained near the atomic limit for
$t=0.01$ eV (squares) and at the value $t=0.2$ eV realistic for
Y$_{1-x}$Ca$_{x}$VO$_{3}$ (circles).
Other parameters as in set $B$ in Table~\ref{tab:para}.\label{fig:doping}}
\end{figure}

Summarizing, by
analyzing the data in Figs.~\ref{fig:dos02} and \ref{fig:dos001},
one concludes that the spectral weights found in the numerical
calculations both confirm those found in the atomic limit, see
Table~\ref{tab:sw}. The total spectral weight of the occupied part of
the DOS, $2-x$, reproduces the average electron density below the Fermi
energy. This weight consists of that of the LHB, represented by the
main peak in the spectra, which corresponds to $d^{2}\rightarrow d^{1}$
transitions at undoped sites with the total intensity $2(1-x)$, and
that of the satellite growing at low energy with the weight of $x$
as it reflects the $d^{1}\rightarrow d^{0}$ excitations at the V
ions with doped holes in bound states near the defects. The predicted
linear behavior $\propto x$ is indeed reproduced at $t=0.01$ eV, see
Fig.~\ref{fig:doping}, while only a small deviation from it is observed
for $t=0.2$ eV. One finds that the spectral weight in the LHB is
somewhat enhanced at the expense of the $d^{1}\rightarrow d^{0}$
satellite. This behavior is reminiscent of the kinetic spectral weight
transfer in the nondegenerate Hubbard model \cite{Mei91,Mei93}.

\subsection{Active bond and satellite of the LHB\label{sec:bond}}

\begin{figure}[t!]
\includegraphics[width=8cm]{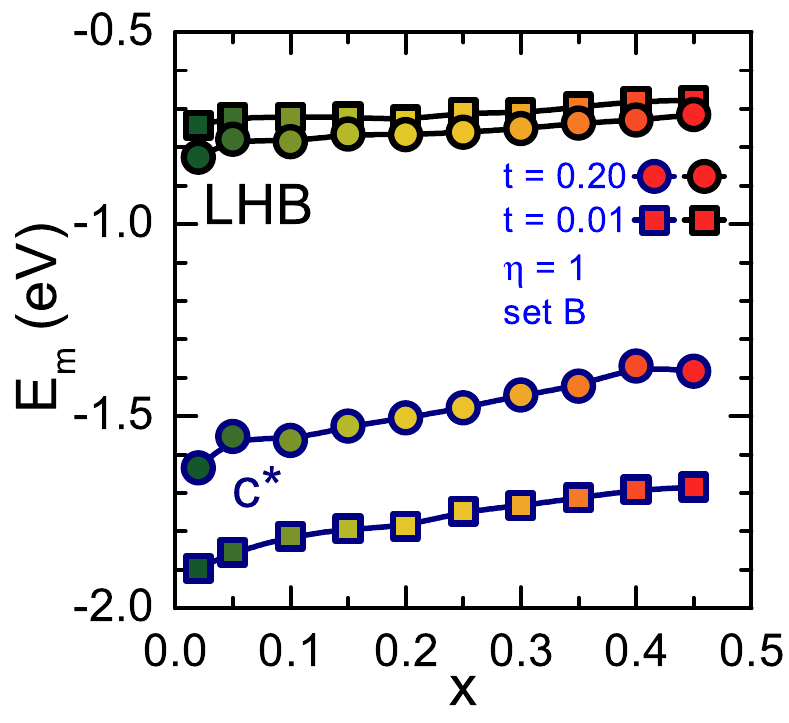} \caption{
Excitation energies $E_{m}$ assigned to the LHB $d^2\rightarrow d^1$
and the satellite $c^{*}$ attributed to $d^{1}\rightarrow d^{0}$
excitations arising from hole-like small polarons, where the doped
holes (or $d^{1}$ configurations) are localized at V-ions close to
a charged Ca defect. Data for $R_{1-x}$Ca$_{x}$VO$_{3}$ is shown for
$t=0.2$ eV (circles) and for the atomic limit $t=0.01$ eV (squares).
Other parameters as in set $B$ of Table~\ref{tab:para}.\label{fig:Em}}
\end{figure}

We emphasize that the satellite structure arising from the
$d^{1}\rightarrow d^{0}$
excitation is well separated from the remaining states in the LHB only
in a particular range of parameters and in addition when the hopping
is small, $t=0.01$ eV. For such an immobile hole the excitation energy
is practically the same as in the atomic limit, see Table~\ref{tab:sw},
\begin{equation}
E_{m}^{(0)}=E_{{\rm LHB}}+V_{{\rm D}}-(U-3J_{H}).\label{Em0}
\end{equation}
Thus, to get the excitation energies in the case of $t=0.01$ eV one
has to include the energy corresponding to the center of the LHB,
being $E_{{\rm LHB}}=-0.7$ eV in Fig.~\ref{fig:Em}, and combine it
with the energy of the $d^{1}\rightarrow d^{0}$ excitation
$(V_{{\rm D}}-U+3J_{H})$, and one arrives at the excitation energy
$E_{m}^{(0)}=-1.9$ eV shown in Fig.~\ref{fig:Em} for the low doping
$x=0.02$.

\begin{figure}[t!]
\includegraphics[width=8cm]{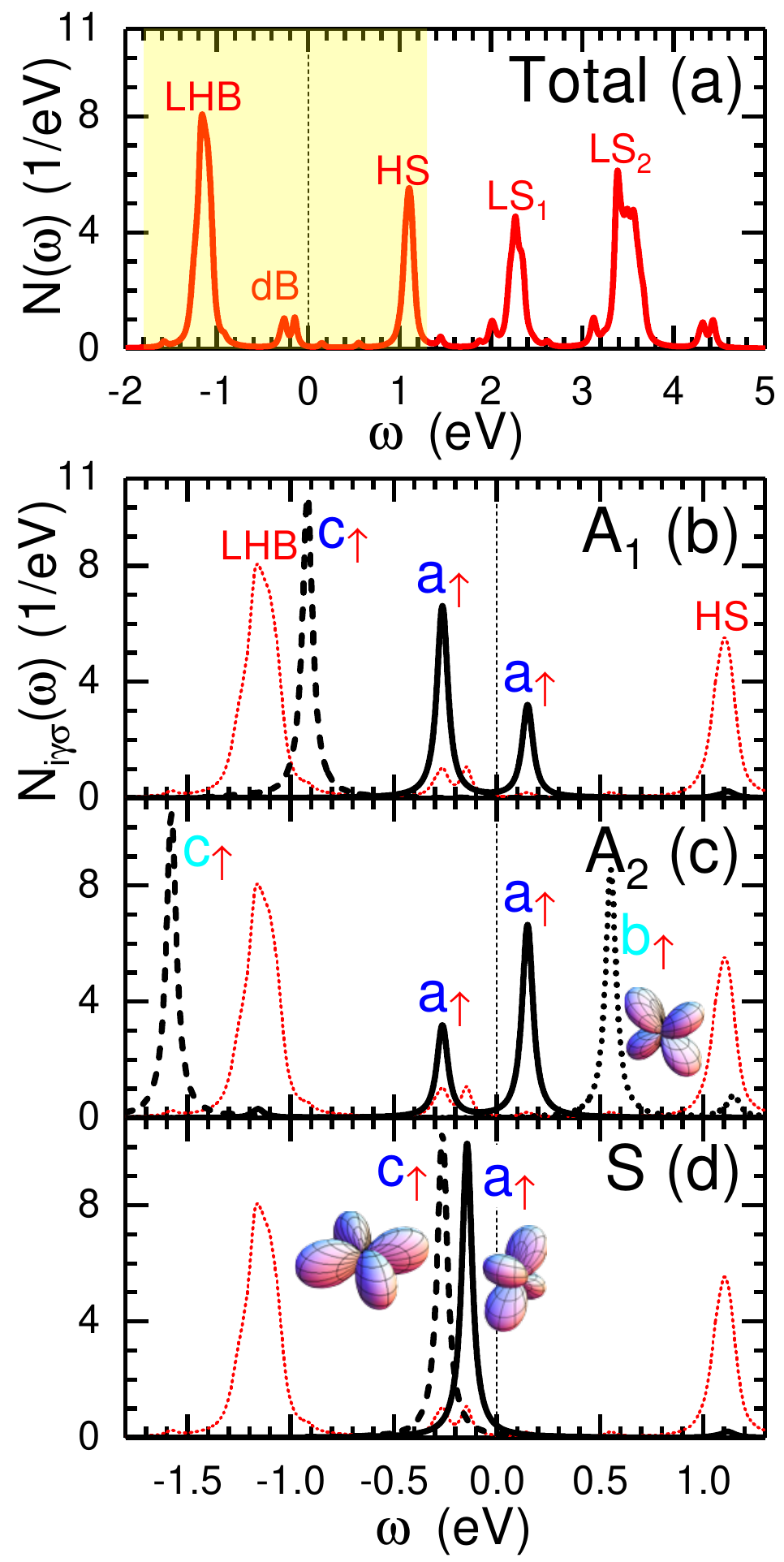}\caption{
Single particle excitations: (a) Multiplet structure (Total) in the
presence of well annealed, or equivalently short-range-potential random,
charged defects, in yellow the energy region reported in the other
panels; (b-c) Partial spectral weights $N_{i\gamma\sigma}(\omega)$
for vanadium ions $A_{1}$ and $A_{2}$, respectively, that belong
to the active bond, and (d) for an occupied spectator site $S$ as
defined in Fig.~\ref{fig:cube}. The hole is predominantly at $A_{2}$.
In (b-d), the thin dashed red curves represent the total DOS (from
panel a) and is given as reference.
Parameters as in set $B$ of Table~\ref{tab:para} and $t=0.2$ eV.
\label{fig2}}
\end{figure}

Consider now a doped site $i$ at finite and increasing $t$. The hole
as well as the other occupied V neighbors closest to a given
Ca$^{2+}$ defect (on a cube $\mathcal{C}_{m}$ surrounding a defect
at site $m$) feel the potential $V_{{\rm D}}$ \cite{Ave13}. The hole
delocalizes partly along the active FM bond $\langle ij\rangle{\parallel}c$
axis, and electronic configurations at sites $i$ and $j$ change from
$(b{\uparrow})^{0}$ and $(b{\uparrow})^{1}$ to $(b{\uparrow})^{\delta}$
and $(b{\uparrow})^{1-\delta}$ (with $\delta<0.5$ and $(c{\uparrow})^1$
electron at each site). This results in the modification of excitation
energies at both sites and one finds instead of Eq. (\ref{Em0}):
\begin{eqnarray}
E_{m}^{(i)} & = & E_{{\rm LHB}}+V_{{\rm D}}-(U-3J_{H})\,(1-\delta),\label{Emi}\\
E_{m}^{(j)} & = & E_{{\rm LHB}}+V_{{\rm D}}-(U-3J_{H})\,\delta.\label{Emj}
\end{eqnarray}
This is illustrated through a calculation for well annealed, or equivalently
short-range-potential random, charged defects shown in Fig.~\ref{fig2}.
In this figure, together with the reference overall multiplet structure
(panel a), the partial DOS $N_{i\gamma\sigma}(\omega)$ at three V
sites of the defect cube (see Fig.~\ref{fig:cube}) are displayed.
Namely, sites $A_{1}$ (panel b) and $A_{2}$ (panel c) that belong to
the FM active bond that basically accommodates the doped hole, and,
for comparison, at a site with two electrons which we call a spectator
($S$) site, see Fig.~\ref{fig:cube}(d). In the latter case, both
electrons $a$ and $c$ feel the Hubbard interaction $(U-3J_{H})$.
The remaining single $a$ electron on the active bond forms a bonding
state, whose polarity is determined by several factors:
(i) the interplay of kinetic energy and JT-fields and
(ii) in the long-range-potential case with random defects,
the random fields of the other defects. Even in absence of random
defect potentials a hole distribution polarized along the $c$ axis
is expected, favored by the JT potentials in the symmetry broken
$G$-AO state.

The energies of the occupied $c$ levels at sites $A_{1}$ and $A_{2}$
in Fig.~\ref{fig2} are very different, as they are controlled by
the distribution of the single $a$-electron on the active bond
$\langle A_1,A_2\rangle$.
In Fig.~\ref{fig2}, the hole is mainly at site $A_{2}$ and, therefore,
the $c$ electron there is at a lower energy as it does not acquire
the Hubbard interaction $(U-3J_{H})$ due to the smaller density of
$a$-electrons at site $A_{2}$. The position of the satellite $c^{*}$
coincides with the occupied $c$-level at site $A_{2}$ and therefore
monitors the polarity of the active bond and, in the disordered case,
the random fields of the further distant defects.

Hole delocalization on the active bond with increasing $t$ reduces the
energetic distance between the LHB and the satellite state $c^{*}$,
see Fig.~\ref{fig:Em}. As we have explained above, see Eq.~(\ref{Emi}),
this $\delta$-dependent energy shift by
$[V_{{\rm D}}-(U-3J_{H})(1-\delta)]$ is a fraction of the Coulomb energy
$(U-3J_{H})$ and is therefore much higher than $t$ itself. In the
considered example in Fig.~\ref{fig:Em}, one finds the value
$E_{m}^{(1)}\simeq-1.6$ eV for $t=0.2$ eV which implies a hole
delocalization on the active bond corresponding to $\delta\approx0.3$.
Thus the distance of the $c^{*}$-excitation can be used to probe the
delocalization of the doped hole on the active bond.
As the delocalization is controlled by the interplay of the kinetic
matrix element $t$ and the random defect fields, the
$d^{1}\rightarrow d^{0}$ excitation energies provide a direct measure
of the strength and the fluctuations of the random fields.

So far we have focused on the multiplet structure and on the $d^{1}$
defect state that gives rise to a satellite on the low energy side
of the LHB, and should appear as $d^{1}\rightarrow d^{0}$ transition
in PES. The observation of the satellite would yield valuable information
about the defect structure and the disorder strength. In the next
Section, we turn to further defect related transitions, namely the
$d^{2}\rightarrow d^{1}$ transitions in PES and $d^{1}\rightarrow d^{2}$
transitions in IPES. The corresponding electron removal and addition
energies fall inside the fundamental MH gap and lie below and above
the chemical potential, respectively.


\section{Mechanisms for defect states gap\label{sec:gap}}

\begin{figure}[t!]
\includegraphics[width=1\columnwidth]{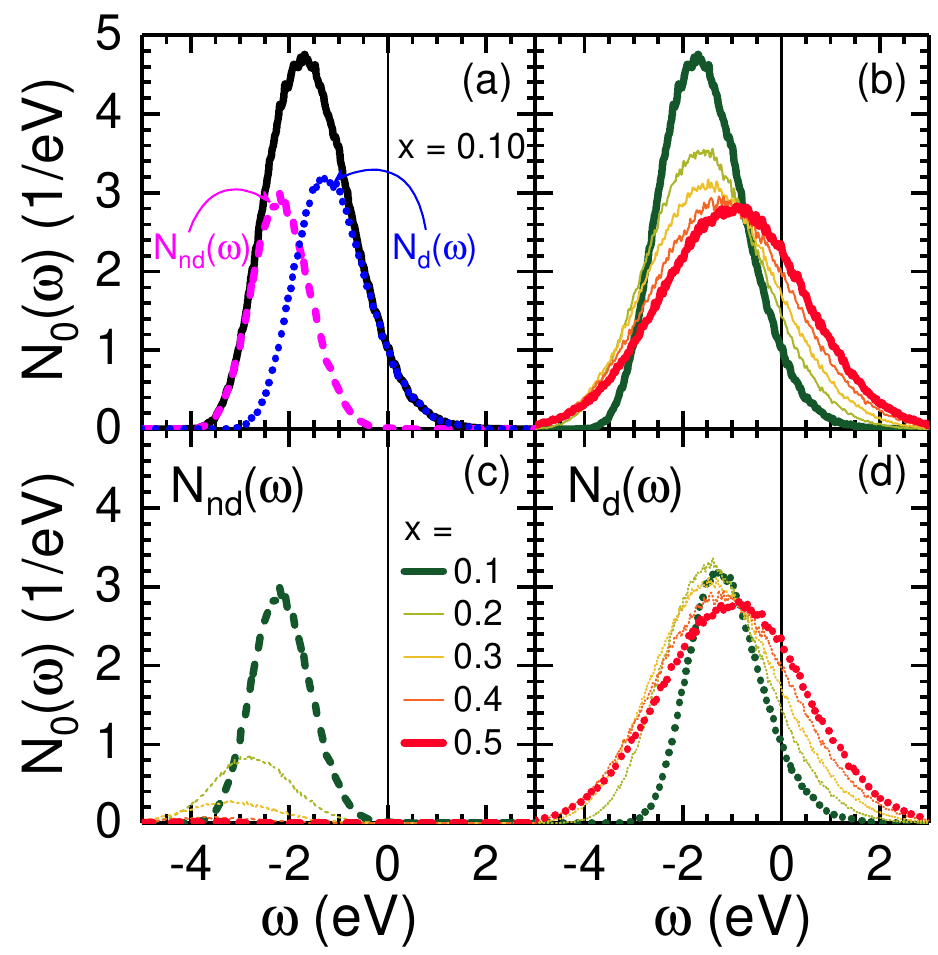} \caption{
Distribution of the local energies $h_{i}$ due to the random defect
potentials. (a) The total distribution of levels $N_{0}(\omega)$
(solid black line) for a doping concentration $x=10$\% is subdivided
in a distribution $N_{d}(\omega)$ of energy levels of V sites that
are direct neighbor of at least one defect (dotted blue line) and
a remaining distribution $N_{nd}(\omega)$ of levels related to V
sites that are not nearest neighbor of a defect (dashed magenta line).
(b) $N_{0}(\omega)$ , (c) $N_{nd}(\omega)$ and (d) $N_{d}(\omega)$
for doping concentrations $x$ from 10\% to 50\% (green to red lines).
The chemical potential $\mu$ (corresponding to $\omega=0$) lies
in the high energy tail of the level distribution $N_{d}(\omega)$
related to V sites that are nearest neighbor of a defect.\label{fig_w}}
\end{figure}

The most important features triggered by the presence of charged defects
in the system are the (defect) states that appear within the MH gap as
they determine the position of the chemical potential and the transport
properties of the material. Such states originate from the LHB and,
being repelled by effectively negative defects (for instance, Ca$^{2+}$
ions substituting La$^{3+}$ ones), are pushed upwards in energy into
the MH gap. The distribution $N_{0}(\omega)$ of the bare electron
energies $h_{i}$, as defined in Eq.~(\ref{hi}), is determined by
the Coulomb potentials of the random Ca$^{2+}$ impurities. An example
is shown for doping $x=10$\% in Fig.~\ref{fig_w}(a). The total
distribution of electron energies $N_{0}(\omega)$ {[}see Fig.~\ref{fig_w}(a){]}
can be decomposed into the distribution $N_{d}(\omega)$ of energies
of the V sites that are direct neighbors of at least one defect, i.e.,
within distance $d$, while the remaining V sites contribute to the
second distribution $N_{nd}(\omega)$ that lies at lower energy, as
shown in Figs.~\ref{fig_w}(c-d), respectively.

Figure~\ref{fig_w} highlights three important points: (i) as there
are 8 vanadium sites close to each defect, but only a single hole
per defect, the number of defect states is much larger than the number
of doped holes; (ii) hence, the chemical potential $\mu$ lies in
the high energy tail of the defect state distribution $N_{d}(\omega)$
{[}see Fig.~\ref{fig_w}(d){]}; (iii) the defect states distribution
below $\mu$ is not well separated from the states of the original
LHB. The latter point, namely that the defect states in the Hubbard
gap below $\mu$ and the LHB states cannot be easily distinguished,
is a common feature of $N_{0}(\omega)$ and of the total DOS $N(\omega)$
of the system studied in the previous Section (see Fig.~\ref{fig:dos02}).
On the contrary, above $\mu$, that is in the IPES regime, the defect
states and the HS Hubbard band states appear well separated in energy.

The opening of a soft DS gap at the chemical potential
$\mu$ in $N(\omega)$ is the most striking difference with respect
to the bare DOS $N_{0}(\omega)$. Figure~\ref{fig:dos02} displays
the pronounced depletion of the DOS $N(\omega)$ right at the chemical
potential (i.e., within the defect states band) for a system with
long-range $e$-$e$ interactions ($\eta=1$) and several doping concentrations.
One well-known mechanism for the opening of a soft gap is the combined
action of long range $e$-$e$ interactions and disorder as pioneered
by Pollak \cite{Pol70} and by Efros and Sklovskii \cite{Efr75,Efr76}
and denoted as the Coulomb gap \cite{Efr75}. The Coulomb gap arises
from a subtle optimization of the occupation of randomly distributed
localized electron states, viz. the CG model, where the
total energy is minimized by the formation of a soft gap around the
chemical potential.

\begin{figure}[t!]
\includegraphics[width=8cm]{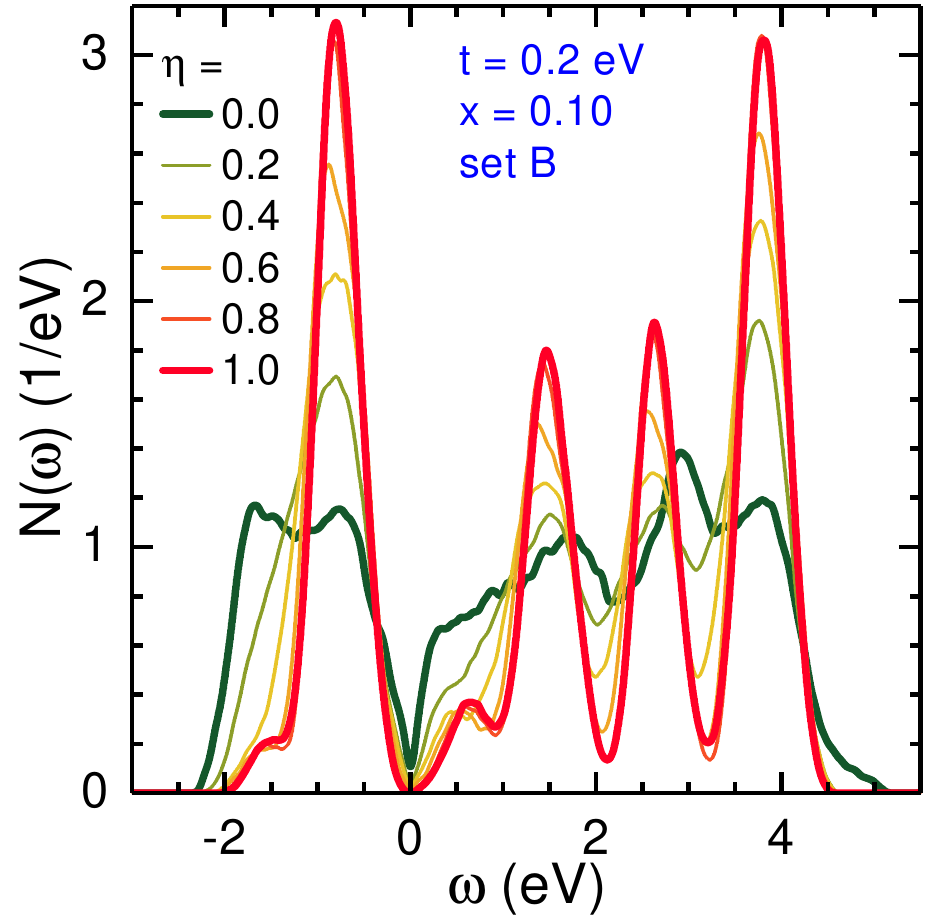} \caption{
Variation of the density of states and of the defect states gap for
parameter set $B$ of Table~\ref{tab:para} as function of the strength
of $e$-$e$ interactions $\eta=0,0.2,0.4,0.6,0.8,1.0$ and $t=0.2$
eV.\label{fig_h}}
\end{figure}

An alternative mechanism for the formation of a gap in cubic oxides
in the presence of charged defects was reported in a recent study
\cite{Ave13,Ave15}. In these systems, the gap arises because the
doped holes: (i) are bound to the defects by the defect potential
whereas the spin-orbital order greatly reduces the already constrained
(2D) mobility; (ii) are confined to one of the vertical bonds of the
defect cube hosting them by the in-plane AF spin order; (iii) can
gain kinetic energy, leading to a splitting of the topmost occupied
states, \emph{delocalizing} over that vertical bond. That is, the
system gains energy by opening a gap at the chemical potential as
in the Peierls effect. However, here the splitting does not arise
from an induced lattice distortion, but from the formation of bonding
and anti-bonding states on the FM active bond $\langle A_1,A_2\rangle$
close to a defect in the presence of a doped hole, as it is depicted
in Fig.~\ref{fig2}. The formation of a \emph{kinetic} gap between the
$|a{\uparrow}\rangle$ states in Fig.~\ref{fig2} is controlled by an
interplay of doped holes with spin and orbital degrees of freedom,
and most importantly it is controlled by the random potentials of
the defects and the $e$-$e$ interaction. Furthermore, it was found
that the kinetic gap is not destroyed by disorder if $t$ is large
enough \cite{Ave15}.

For a general case, it was pointed out that the kinetic and the
Coulomb gap mechanisms, the latter emerging from $e$-$e$ interactions,
enhance jointly the DS gap in the vanadates \cite{Ave13}.
We note that this is in contrast to the Anderson-Hubbard model with
long-range $e$-$e$ interactions where the kinetic energy suppresses
the DS gap \cite{Epp97}.

\begin{figure}[t!]
\includegraphics[width=8cm]{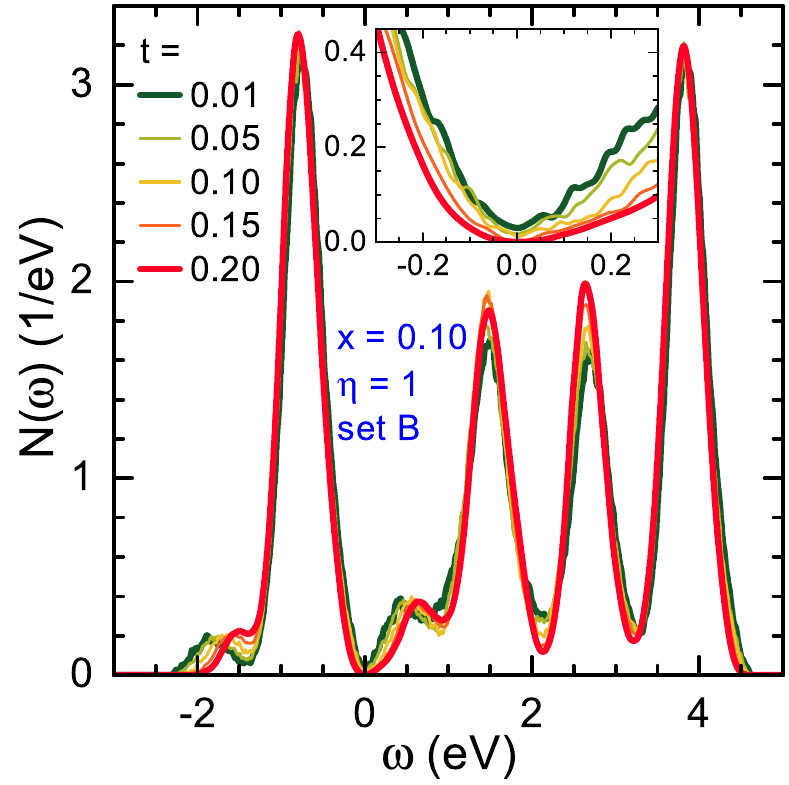} \caption{
Variation of the density of states and of the defect states gap for
parameter set $B$ of Table \ref{tab:para} for $t=0.01,0.05,0.10,0.15$,
and $0.20$ eV, and $\eta=1$. Inset shows the zoom of the DOS near
the Fermi energy at $\omega=0$.\label{fig_o}}
\end{figure}

Here, we present a systematic analysis of how the $e$-$e$ interactions
and the kinetic-gap mechanism jointly contribute to the formation of
the DS gap. In Fig.~\ref{fig_h}, for a system with doping $x=10$\%,
we investigate how the DOS $N(\omega)$ varies when the $e$-$e$
interaction is switched on while keeping the value of the hopping
integral $t=0.2$ eV fixed. This is done by changing the $\eta$ parameter
from zero to one; notice that this corresponds to the line BC in
Fig.~\ref{fig:tower}.
At $\eta=0$, the quite large width of the Hubbard bands is due
to the effective distribution of the levels generated by the Coulomb
potentials of the random defects (see Fig.~\ref{fig_w}). With increasing
$\eta$, such monopolar potentials ($\propto 1/r$) are more and more
compensated by the $e$-$e$-interaction. In the case $\eta=1$, i.e., in
the physical case, each single defect and the related bound hole act
overall as a dipole ($\propto 1/r^{2}$) and, as a consequence, the
effective distribution of the random energy levels gets narrower while
keeping the defect positions unchanged.
In Fig.~\ref{fig_h}, there is already a kinetic gap even in the
complete absence of $e$-$e$ interactions ($\eta=0$). This gap is
approximately linear in $\omega$ as we shall see in the statistical
analysis below. With increasing $\eta$, a pronounced soft gap evolves,
that can be attributed to the Coulomb gap mechanism.

In Fig.~\ref{fig_o}, we study the $t$ dependence at $\eta=1$,
i.e., in presence of $e$-$e$ interactions, for doping $x=10$\%.
This corresponds to the line DC in Fig.~\ref{fig:tower}. The DOS
changes only slightly as a function of $t$ as the width of the Hubbard
bands is essentially determined by the disorder. The evolution
of the DS gap is amplified in the inset of Fig.~\ref{fig_o}. Surprisingly,
we see here that, at small values of $t$, the Coulomb gap mechanism
in these systems is not strong enough to create a soft gap. In a certain
range $t\le t^{*}$, the data in the inset suggests that the DOS at $\mu$
is finite. As there is nevertheless a strong suppression of the DOS
we call this a pseudogap. With increasing $t$, the pseudogap changes
into a soft gap with $N(\mu)=0$. We shall see below by a rigorous
statistical analysis that for small $t$ values there is in fact a weak
singularity with $N(0)=0$ hidden in this data. As a side remark, we note
that, with increasing $t$, the satellite $c^{*}$ gradually moves upward.
The upward shift is connected with the increasing delocalization of
the hole on the active bond, as discussed above.

To analyze the behavior of the soft gap in $N(\omega)$ without
suffering from the unavoidable broadening of delta functions, we
consider the averaged integrated DOS $n(\omega)$ Eq. (\ref{nw}) that
can be studied without artificial broadening. It is worth noting
the following key features of $n(\omega)$ in the vicinity of the
Fermi energy:
(i)~at the chemical potential there is an evident gap/plateau for $t=0.2$
eV and $\eta=1$ {[}see Fig.~\ref{fig_c}{]}, but not for $t=0.01$ eV, and
(ii) on decreasing the screening $\eta\rightarrow0$, the gap/plateau
disappears even for $t=0.2$~eV.

In order to establish the statistical behavior of $N\left(\omega\right)$
at low energy in the limit of an infinite number of defect realizations
$M\to\infty$, we use that $N\left(\omega\right)$ is proportional
to the probability distribution function $P^{*}\left(\omega\right)$
that the topmost occupied state (the lowest unoccupied state) in a
generic defect realization $s$ has energy
$\omega=\Delta_{s}\left(-\Delta_{s}\right)$ relative to its Fermi energy
$\mu_{s}$. This is equivalent to the distribution of the nearest neighbor
level spacings $2\Delta_{s}$ across the chemical potential. In
Ref.~\cite{Ave15}, it has been shown that a generic defect realization
features a gap of size $E$ with a probability governed by a Weibull
probability distribution function,
\begin{equation}
P_{W}(E)=\theta\!\left(E-\zeta\right)\,\frac{k}{\lambda}\,\left(
\frac{E-\zeta}{\lambda}\right)^{k-1}
\mathrm{e}^{-\left(\frac{E-\zeta}{\lambda}\right)^{k}},\label{Wei}
\end{equation}
with shape parameter $k$, scale parameter $\lambda$ and location
parameter $\zeta$. Accordingly, if $\zeta=0,$ we have
$P^{*}\left(\omega\right)=\frac{k}{\lambda^{k}}\left|\omega\right|^{k-1}$
and $N\left(\omega\right)\propto\left|\omega\right|^{k-1}$ both for
$\left|\omega\right|\ll\lambda$.
The exponent $k$ allows to distinguish the following cases:
a soft gap for $k>2$, a linear gap for $k=2$,
a pseudogap (or singular gap) regime for $1<k<2$ and no gap for $k=1$.

Moreover, if $\zeta>0$ the Weibull probability distribution describes
a real gap, that is in this case we have $N\left(\omega\right)=0$ for
$\left|\omega\right|\leq\zeta$ and
$N\left(\omega\right)\propto\left(\left|\omega\right|-\zeta\right)^{k-1}$
for $\zeta<\left|\omega\right|\ll\lambda$. Thus, $P_{W}\left(E\right)$
results in a robust scheme to determine the behavior of
$N\left(\omega\right)$ close to the Fermi energy, that is, the presence
and type of the overall gap in the system (through $k$ and $\zeta$) as
well as the typical scale of the microscopic gap developing locally in
the system because of all microscopic mechanisms at work
(through $\lambda$).

\begin{figure}[t!]
\includegraphics[width=8.2cm]{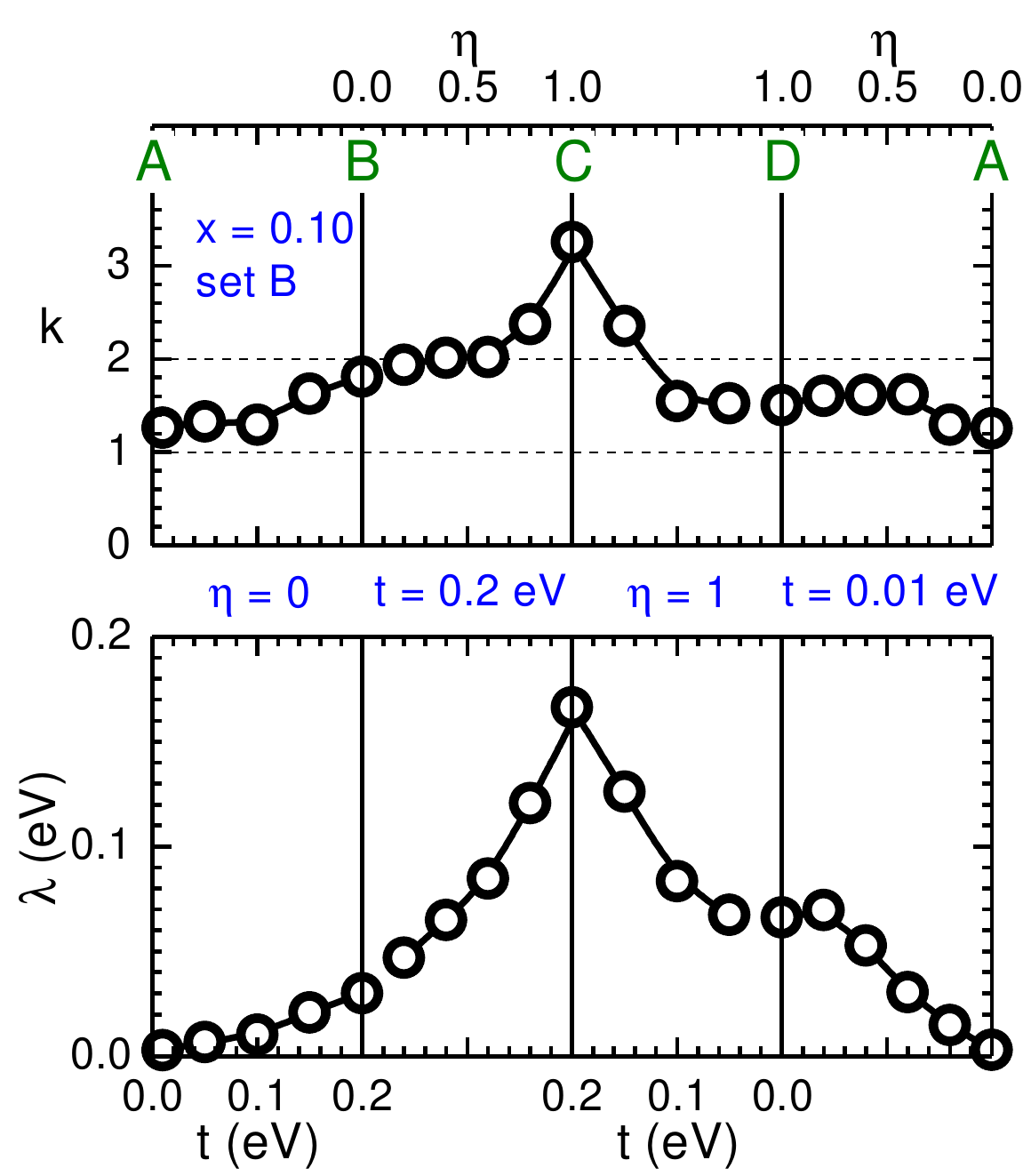} \caption{
Shape $k$ (top) and scale $\lambda$ (bottom) parameters of the Weibull
distribution (\ref{Wei}) that characterizes the soft gap at $\mu$ as
function of $e$-$e$-interaction strength $\eta$ for $t=0.2$ eV (BC)
and $t=0.01$ eV (DA), and as function of $t$ for $\eta=0$ (AB) and
$\eta=1$ (CD), respectively. For labels A-D see Fig.~\ref{fig:tower}.
\label{fig_ln}}
\end{figure}

The numerical data for the exponent $k$ and the scale parameter $\lambda$
obtained from the statistical analysis of $M=100$ defect realizations
are summarized in Figs.~\ref{fig_ln} (top panel) and \ref{fig_ln}
(bottom panel), respectively. The odd panels show the $t$ dependence
for $\eta=0$ (panel AB) and 1 (panel CD), respectively, and the even
panels --- $\eta$
dependence for $t=0.2$ eV (panel BC) and $t=0.01$ eV (panel DA),
respectively. The parameters $k$ and $\lambda$ are determined by a
statistical least-squares fit to $P_{W}(E)$ of the actual distribution
of gaps ($2\Delta_{s}$) among $100$ defect realizations and yield
for all cases a vanishing real gap, i.e., $\zeta=0$. In the top panel
of Fig.~\ref{fig_ln}, we recognize a strong variation of the exponent
$\nu=k-1$, which determines the low frequency ($\omega\ll\lambda$)
behavior of the DOS: $N(\omega)\propto\omega^{\nu}$, with the
increasing/descresing hopping or $e$-$e$ interactions, $t$ and
$\eta$. This is in striking contrast to the CG studied
by Efros and Sklovskii \cite{Efr75,Efr76}, where $\nu=d-1$ is
determined by the spatial dimension $d$ alone.

The size of the pseudogap is controlled by the scale parameter
$\lambda$ in Fig.~\ref{fig_ln} (bottom panel). In the absence of
$e$-$e$ interactions $\eta=0$ (interval AB), the disorder is very
strong indeed leading to a complete suppression of the soft gap in
the entire range of $t$ and to the vanishing of $\lambda$ scale in
the limit $t\rightarrow0$.
In panel BC, that is on increasing $e$-$e$ interactions, we see
an increase of $k$ from 2 at B, which corresponds to a gap linear
in $|\omega|$, as seen in the DOS in Fig.~\ref{fig_h}, to a soft
gap with $k\approx3$ at C. The increase of $\lambda$ from D to C,
that is on increasing $t$, displays the kinetic gap mechanism. Panel
DA ($t=0.01$ eV) shows that $e$-$e$ interactions (in the realistic
range) are not strong enough to open a soft gap when approaching the
atomic limit. When approaching the point A (at $\eta=0$) $\nu$ tends
to zero, although our result for $t=0.01$ eV is not exactly zero.

Yet as the scale parameter $\lambda\rightarrow0$ we conclude that
in the absence of $e$-$e$ interactions there is a constant DOS at
$\mu$ in the atomic limit. It is worth noting that, for a range of
$\eta$ values and $t$ values at $\eta=1$ around the D point (panels
CD and DA), we find a pseudogap with an exponent $\nu\simeq0.5-0.6$.
This Coulomb anomaly for a 3D system is a feature distinct from the
soft gap in the Efros-Sklovskii theory for the Coulomb glass and
reminiscent of the Coulomb anomalies discussed for the electron gas
by Altshuler and Aronov \cite{Alt79}.

In this Section, we have seen that even in the absence of $e$-$e$
interactions a sufficiently large hopping $t$ can open a gap in the
defect states that survives the disorder fluctuations. However, we
also found that
$e$-$e$ interaction alone may not be strong enough in the vanadium
perovskites for the emergence of a soft DS gap with a vanishing
DOS at $\mu$. The Weibull exponent $k$ is the largest when both
mechanisms, i.e., kinetic gap formation and $e$-$e$ interactions,
act together. In contrast to the Efros and Sklovskii theory, we found
here that the soft gap exponent $\nu=k-1$ of the DOS at the chemical
potential is nonuniversal but depends both on $t$ and $\eta$. Next,
we shall explore in detail the localization of electron states in the
presence of charged defects and, in particular, the role of $e$-$e$
interactions.


\section{Inverse participation number and Localization\label{sec:ipr}}


In this Section, we determine the degree of localization of the electronic
wave functions quantitatively. As useful measure of the localization of a
single particle wave function $\psi({\bf r})$, the inverse participation
number (IPN), $P^{-1}=\sum_{i}|\langle\psi|i\rangle|^{4}$, has been
introduced for models with one orbital and one spin per site $i$,
where the wave function $\psi({\bf r})$ is assumed to be normalized.
The participation number $P$ provides a measure of the number of
sites over which the single particle wave function $\psi({\bf r})$
roughly extends
\cite{Guh98}. The IPN was first considered by Bell and Dean \cite{Bel70}
in the context of the localization of lattice vibrations and subsequently
explored by Thouless \cite{Tho74}, Wegner \cite{Weg80} and others
\cite{Ono89,Epp97,Eve08} in the context of the Anderson localization
of electronic states. The IPN has also been employed in studies related
to quantum spin chains \cite{Mis16} and many-body localization \cite{Lui15}.

For systems with spin and orbital degeneracy, which are of interest
here, we shall define the IPN as,
\begin{equation}
P^{-1}=\sum_{i}\bigg(\sum_{\alpha,\sigma}
|\langle\psi|i,\alpha,\sigma\rangle|^{2}\bigg)^{2},
\end{equation}
where the internal sums in $\alpha$ and $\sigma$ are over local
orbital and spin degrees of freedom, respectively, while the remaining
sum in $i$ is over all $N$ sites in the system. The wave function
is assumed to be normalized, i.e.,
$\sum_{i,\alpha,\sigma}|\langle\psi|i,\alpha,\sigma\rangle|^{2}=1$.
For a localized state $\psi({\bf r})$ that extends over few sites,
the participation number $P$ is a small number slightly larger than
1, whereas for an (almost) completely delocalized state (e.g. a~Bloch
state), it is of order $N$, the number of sites in the system.

\begin{figure}[t!]
\includegraphics[width=\columnwidth]{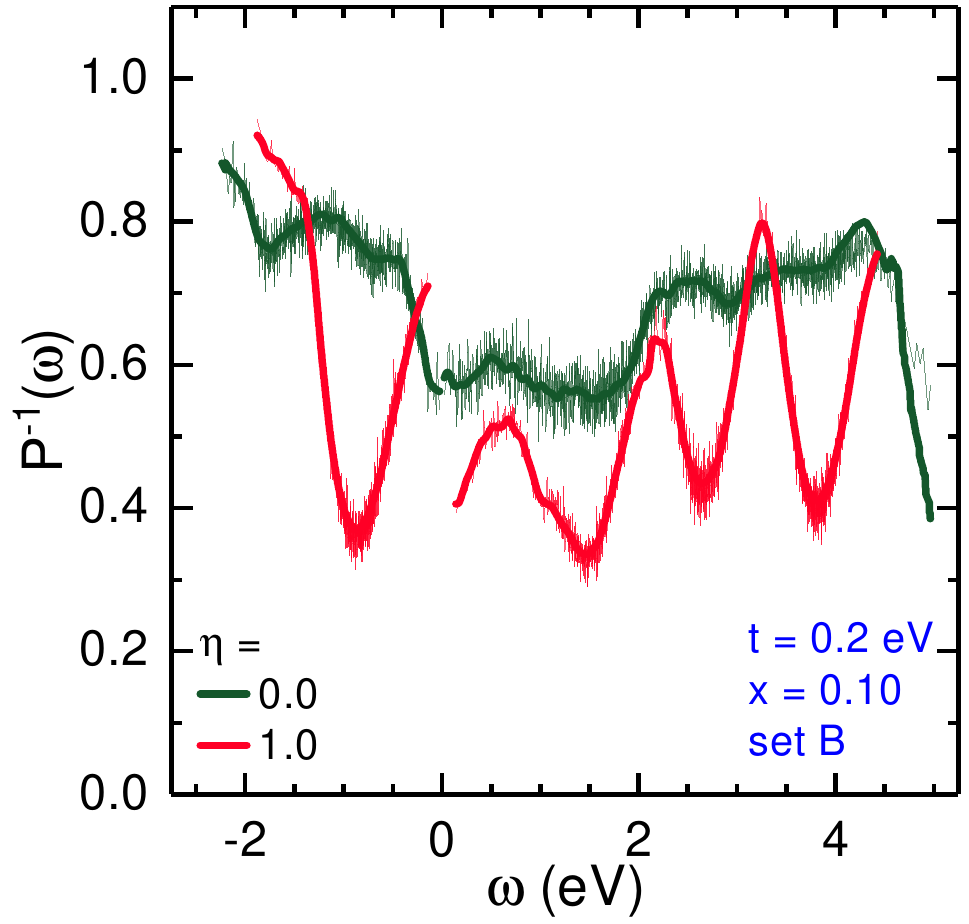} \caption{
Spectral function of the inverse participation number $P^{-1}(\omega)$
(\ref{s-IPN}) for parameter set $B$ of Table~\ref{tab:para} and
$t=0.2$ eV at $x=10$\%. The thick red curve stands for the fully
screened case $\eta=1$ and reveals the spatial extension jump at
$\mu$, whereas, in the absence of $e$-$e$ interactions ($\eta=0$,
thick green curve), the states directly below and above $\mu$ have
the same spatial extension. Thin lines stand for the average
$P_{n}^{-1}\left(\omega\right)$ vs $\omega=\langle\omega_{n}\rangle$.
\label{fig_r}}
\end{figure}

Here, we are interested in the localization of the $n$-th HF eigenstate
$\psi_{n,s}({\bf r})$ corresponding to the $n$-th HF eigenenergy
$\omega_{n,s}$ both computed for the defect realization $s$. Therefore
we shall explore the spectral function representing the statistically
averaged IPN,
\begin{equation}
P^{-1}(\omega)=\frac{1}{M}\sum_{s=1}^{M}\left[\frac{1}{N_{s}(\omega)}
\sum_{n}P_{n,s}^{-1}\delta(\omega-\omega_{n,s})\right],\label{s-IPN}
\end{equation}
where $P_{n,s}^{-1}$ is the IPN computed for $\psi_{n,s}({\bf r})$
and $N_{s}(\omega)$ is the DOS of the $s$ defect realization. It
is worth noting that the division by $N_{s}(\omega)$ is problematic
in regions with small DOS, which are of particular interest
to us. Therefore, we analyze the essentially equivalent quantity,
\begin{equation}
P_{n}^{-1}\left(\omega=\langle\omega_{n}\rangle\right)=
\frac{1}{M}\sum_{s=1}^{M}P_{n,s}^{-1},
\end{equation}
where $\langle\omega_{n}\rangle=\frac{1}{M}\sum_{s=1}^{M}\omega_{n,s}$.
This quantity has the great advantage to avoid the pathological division
by the DOS $N_{s}(\omega)$ and displays in addition the fluctuations
due to the many defect realizations in $\left\{ P_{n,s}^{-1}\right\}$
and in $\left\{ \omega_{n,s}\right\}$.

The fluctuations of the inverse participation number $P_{n}^{-1}$
for close-by energies $\langle\omega_{n,s}\rangle$ turn out surprisingly
small as can be seen in Fig.~\ref{fig_r} which displays two data sets,
one without ($\eta=0$) and another with ($\eta=1$) $e$-$e$-interactions.
At first glance, it is clear that both sets represent well localized
wave functions over the whole spectrum. Note that a value $P^{-1}\approx0.5$
corresponds to a wave function delocalized over two sites. It is important
to note here that the localization is entirely due to the disorder,
as a calculation without any defects yields $P^{-1}\approx1/L^{2}$
(not shown), where $L$ is the linear dimension of our cluster. The
$1/L^{2}$-dependence is consistent with the 2D nature of Bloch states
of the $t_{2g}$ electrons in our system.

The striking difference of the two data sets in Fig.~\ref{fig_r}
reflects the absence or presence of the screening by $t_{2g}$ electrons.
Without $e$-$e$-interactions, the monopolar defect potentials are
not screened and hence the disorder effect is stronger; this is already
manifest in the larger width of the LHB in the DOS. In the screened
case instead, the Hubbard bands are much narrower both in the DOS,
but also in the IPN distribution. The smaller values of IPN in the
center of the Hubbard bands for $\eta=1$ indicate that these states
are definitely more delocalized when the defect potentials are screened.
In the $\eta=1$ case, it is worth noting that the defect states are
much more localized, for instance in the energy region of the $c^{*}$
satellite, in comparison to the states in the center of the Hubbard
bands, which are less affected by the random defect potentials
and therefore show a smaller IPN.

The IPN of defect states inside the MH gap shows several interesting
features. In the unscreened case $\eta=0$, the inverse participation
number is approximately constant ($P^{-1}\approx0.6$) for the states
close to the chemical potential. This case corresponds to the DOS
linear in $\omega$ in Fig.~\ref{fig_h}. For the screened
case $\eta=1$, however, we observe a pronounced discontinuity in
the degree of localization of defect states right below and above
$\mu$, as seen in Fig.~\ref{fig_r}. The discontinuity in the degree of
localization of the defect states at the chemical potential reflects
the different nature of the electron removal and addition states in
presence of $e$-$e$ interactions. It is evident from Fig.~\ref{fig_r}
that the removal states are more localized than the addition states.

\begin{figure}[t!]
\includegraphics[width=\columnwidth]{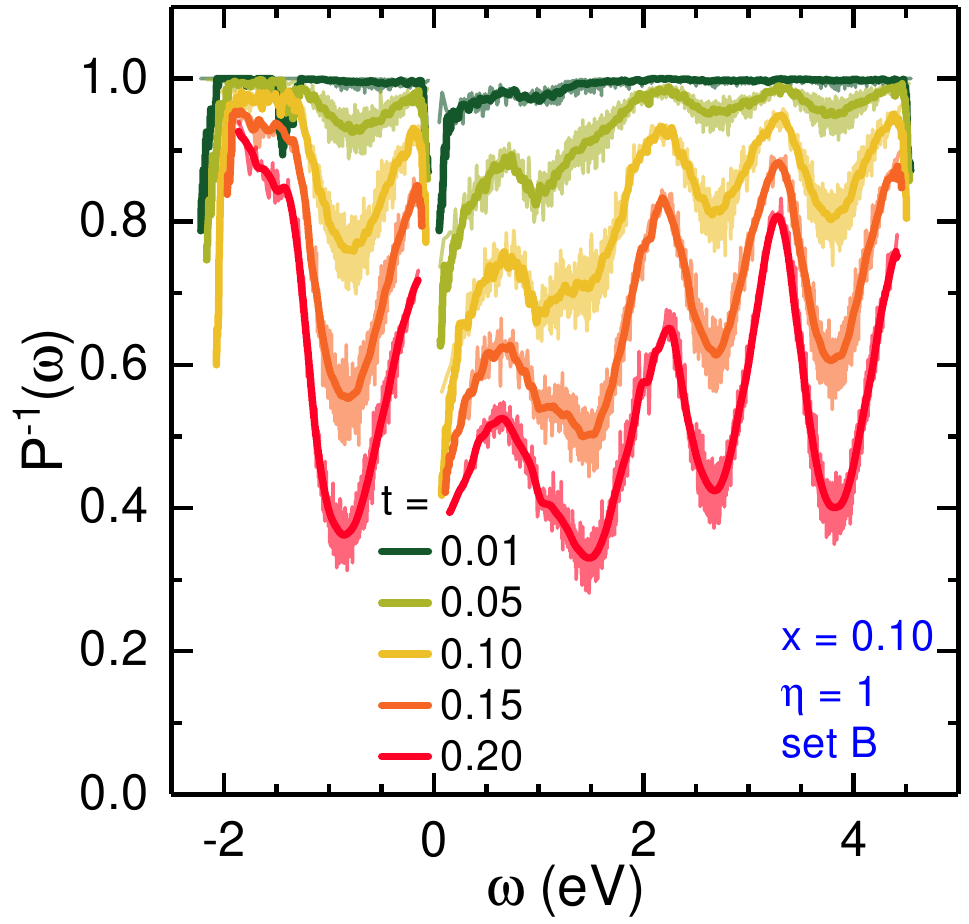} \caption{
Spectral function of the inverse participation number $P^{-1}(\omega)$
(\ref{s-IPN}) for parameter set $B$ of Table~\ref{tab:para}, $t=0.01$,
0.05, 0.10, 0.15, and 0.20 eV (green to red) and $\eta=1$ at $x=10$\%
(thick lines).
Thin lines stand for the average $P_{n}^{-1}\left(\omega\right)$
vs $\omega=\langle\omega_{n}\rangle$.\label{fig_q}}
\end{figure}


In Fig.~\ref{fig_q}, we explore the $t$ dependence of the IPN for
the case with $e$-$e$ interactions ($\eta=1$). We recognize that
in the regime of small $t$ (close to the atomic limit) $P^{-1}\approx1$.
That is the electron states are perfectly localized at a single site,
as one expects for the classical case $t=0$. The upper limit $P^{-1}=1$
shows that our definition of the IPN in the spin and orbital degenerate
case is correct. The choice of a small finite hopping $t=0.01$ eV
in Fig.~\ref{fig_q} is dictated by the dependence on $t$ of the
computation time: the smaller $t$ is the longer it takes to reach
convergence. We see that the defect states close to $\mu$ and those
at the bottom of the LHB are slightly more delocalized. For these
defect states, IPN is in the range $P^{-1}\in(0.8,1.0)$. This weak
delocalization at small $t$ is a consequence of the motion of the
holes on the active bonds and increases with increasing $t$. We also
see that the states in the center of the various Hubbard subbands
delocalize faster with increasing $t$ than the defect states in between
the Hubbard multiplets. Thus, we can conclude that all states remain
well localized even at the moderate doping concentration of $x=10$\%
and for the typical value $t=0.2$ eV of the hopping integral in these
materials.


\section{Spin-orbital polarons and reduction of magnetic and orbital order
\label{sec:polarons}}


From the study of the IPN in the previous Section, it naturally follows
that:
(i) the defect states must be visualized as localized small polaron
states that are tightly bound to defects and
(ii) the symmetry of each polaron state relative to the closest defect
must be strongly broken.
It is important to contrast here two distinct physical sources
of \textit{localization} in condensed matter physics:
\begin{itemize}
\item Wave functions \emph{localize} (i.e., their spatial extension undergoes
an extreme reduction) due to the presence of defects or of any kind
of disorder, as it happens, for instance, to the single particle states
in the Anderson localization scenario \cite{And58,Lee85}.
\item On the other hand, electrons \emph{localize} also due to strong (local)
$e$-$e$ interactions as in the half-filled Hubbard model, that is
in the absence of disorder \cite{Mot74}.
\end{itemize}
In general, the Anderson and the Mott routes to localization, as listed
above, are expected to combine in very non-trivial ways \cite{Bra15}.
In the regime of strong correlations, $t\ll U$, the ground state of the
Hubbard model at half-filling is a Mott insulator with AF spin correlations.
Added holes create spin defects while moving in the AF spin background,
and form spin polarons \cite{Mar91,Wro08}. The motion of the hole inside
the polaron cloud represents the main kinetic energy. Yet, as found for
instance in the high-$T_{c}$ materials, the entire polaron may in
addition perform a coherent motion on the magnetic energy scale
$J\propto t^{2}/U$ \cite{Mar91}.

In the case of orbital polarons in a state with AO order, the holes have
an even stronger tendency towards localization, for instance in systems
with $t_{2g}$ orbital degrees of freedom \cite{Dag08,Woh08,Wro10}.
A similar analysis of individual hopping processes as for spin polarons
\cite{Bal05,Wro08} is possible and provides a good insight both into the
optical excitations \cite{Wro12} and the spectral properties
\cite{vdB00,Dag04,Brz14,Bie16}
of orbital systems. More demanding is the theoretical description of a
charge added to a system where both spin and orbital order alternates and
excitations of both types of degrees of freedom may contribute to the
polaron cloud \cite{Zaa93,Bie17}. Results depend on the actual
symmetry of the spin and orbital order realized in the ground state.
Theoretical studies have shown that single SO polarons in systems where
both spins and $t_{2g}$ orbitals alternate, as in the $(a,b)$ planes of
the vanadium perovskites, tend to localize \cite{Woh09,Mona}.

Experimentally, such polarons were first identified in half-doped
manganites for $e_{g}$ orbital degrees of freedom \cite{Kim02,Yam13},
but exist also in doped cobaltates \cite{Dag06} and in the cubic
vanadates \cite{Ree16}. In the latter case, being of interest here,
the doped holes move in the spin-orbital ordered $C$-AF/$G$-AO state
and the emerging SO polarons perturb both spin and orbital order.
These polarons are bound to the charged defects: this eliminates
any coherent motion and the perturbation of the spin-orbital order
in the system can be described by means of a detailed analysis of
the nature and of the structure of the polaron cloud, see below. Thus,
the polaron cloud provides a many-body measure of the reduction of
the spin- and orbital- order parameter as function of doping. In this
Section, our aim is to define the SO polaron wave function and to
explore the spatial extension, or the IPN of the SO polaron.

In the spin- and orbital- ordered state, it is convenient to expand
the polaron wave function $|\psi_{n}({i})\rangle$ in a \emph{string}
basis,
\begin{equation}
|\psi_{n}({i})\rangle=\sum_{{\mathbf{\delta}},\alpha}
a_{n,\alpha}({\mathbf{\delta}})|{i+\delta};
S_{\alpha}({\mathbf{\delta}})\rangle,\label{wf}
\end{equation}
where the label $n$ distinguishes polaron states that may differ by
their spatial symmetry and their spin. However, in the following,
we shall consider only the polaron state with the minimal energy.
The wave function of the polaron $|\psi_{n}({i})\rangle$ is centered
or has its largest amplitude at the vanadium site $i$. Its kinetic
energy involves the neighboring sites $\{i+\delta\}$, on varying
${\bf \delta}=\sum_{m=1}^{n(\alpha)}{\bf \delta}_{m}$, as the hole
propagates along different paths, $S_{\alpha}({\delta})=
\{{\mathbf{\delta}}_{1},{\mathbf{\delta}}_{2},\dots,{\mathbf{\delta}}_{n(\alpha)}\}$,
where ${\bf \delta}_{m}=\pm e_{\nu}$ and $\nu=x,y,z$ are the unit
vectors of the cubic lattice. A string state,
$|{i+\delta};S_{\alpha}({\mathbf{\delta}})\rangle$ in Eq.~(\ref{wf}),
corresponds to the state of a hole after it has moved along a particular
path $S_{\alpha}(\delta)$ from site $i$ to site $i+\delta$ and,
therefore, it automatically contains information about the spin and
orbital defects created along that particular path.
Next, we proceed as in the case of the localization of single particle
wave functions and define the IPN for the many-body wave functions
of SO polarons as
\begin{equation}
P_{n}^{-1}=\sum_{{\bf \delta}}\left(
\sum_{\alpha}|a_{n,\alpha}(\delta)|^{2}\right)^{2}.
\label{iPN_polaron}
\end{equation}
As each path $S_{\alpha}({\bf \delta)}$ corresponds to a sequence
of spin-flips and/or orbital changes relatively to the underlying
spin-orbital ordered background, the energy of a string state
increases with the length of the related path and only short paths
and related string states contribute effectively as long as the
system is not close to a phase transition.

In the dilute case (i.e., small density $x=n_{h}/N_{a}$ of doped
holes), the overall reduction of the local magnetic order parameter
$m^{s}$ can be calculated from the spin deviation $m^{sP}$, with
respect to the underlying spin-orbital ordered background associated
with a single polaron,
\begin{equation}
m^{s}\simeq m_{0}^{s}-xm^{sP},\label{ms}
\end{equation}
where $m_{0}^{s}$ is the local magnetization in the undoped ground
state, as we can imagine that the interactions between polarons are
negligible. Then, $m^{sP}$ can be computed by means of the many-body
wave function of the SO polaron,
\begin{equation}
m^{sP}=\langle\psi_{0}|\hat{m}^{s}({Q})|\psi_{0}\rangle
-\langle\psi({i})|\hat{m}^{s}({Q})|\psi({i})\rangle.\label{ms-1}
\end{equation}
Here, $|\psi_{0}\rangle$ is the $N$-particle ground state of the
underlying spin-orbital ordered background and $|\psi(i)\rangle$
is a single-polaron wave function in the space of $(N-1)$ particles.

In this Section, we analyze the localization from the strong correlation
perspective. It is evident from the uHF results for the IPN in Fig.~\ref{fig_q}
that all states of the system are perfectly localized in the limit
$t\rightarrow0$. With increasing $t$ the states become gradually
more extended. Accordingly, an important probe for the delocalization
is the decrease of spin and orbital order parameters $m^{s}$ and
$m^{o}$, respectively. The staggered spin order of the $C$-AF structure,
for example, is measured by
\begin{equation}
\hat{m}_{\nu}^{s}(Q)\equiv\sum_{j}\hat{m}_{j,\nu}^{z}e^{iQ\cdot j},
\end{equation}
where $j$ and $\nu=a,b,c$ are site and orbital flavor indices, respectively.
In the following, we use the convention $\hat{m}^{z}=2\hat{S}^{z}$,
where $\hat{S}^{z}$ is the $z$-component of a spin-$1/2$ operator.
The staggered orders of $C$ and $G$ states are measured by modulation
vectors $Q_{C}=(\pi,\pi,0)$ and $Q_{G}=(\pi,\pi,\pi)$, respectively.
Furthermore, within the uHF approach, it is straightforward to distinguish
the contributions from $ab$- and $c$- electrons to the order parameters,
respectively. Thus
\begin{equation}
m^{s}=m_{ab}^{s}+m_{c}^{s},
\end{equation}
where $m_{\nu}^{s}=\langle\hat{m}_{\nu}^{s}({Q_{C}})\rangle$ for $C$-type
spin order. Similar construction applies for the $G$-type orbital order
parameter,
$m_{\nu}^{o}=\langle\hat{m}_{\nu}^{o}({Q_{G}})\rangle$.

\begin{figure}[t!]
\includegraphics[width=1\columnwidth]{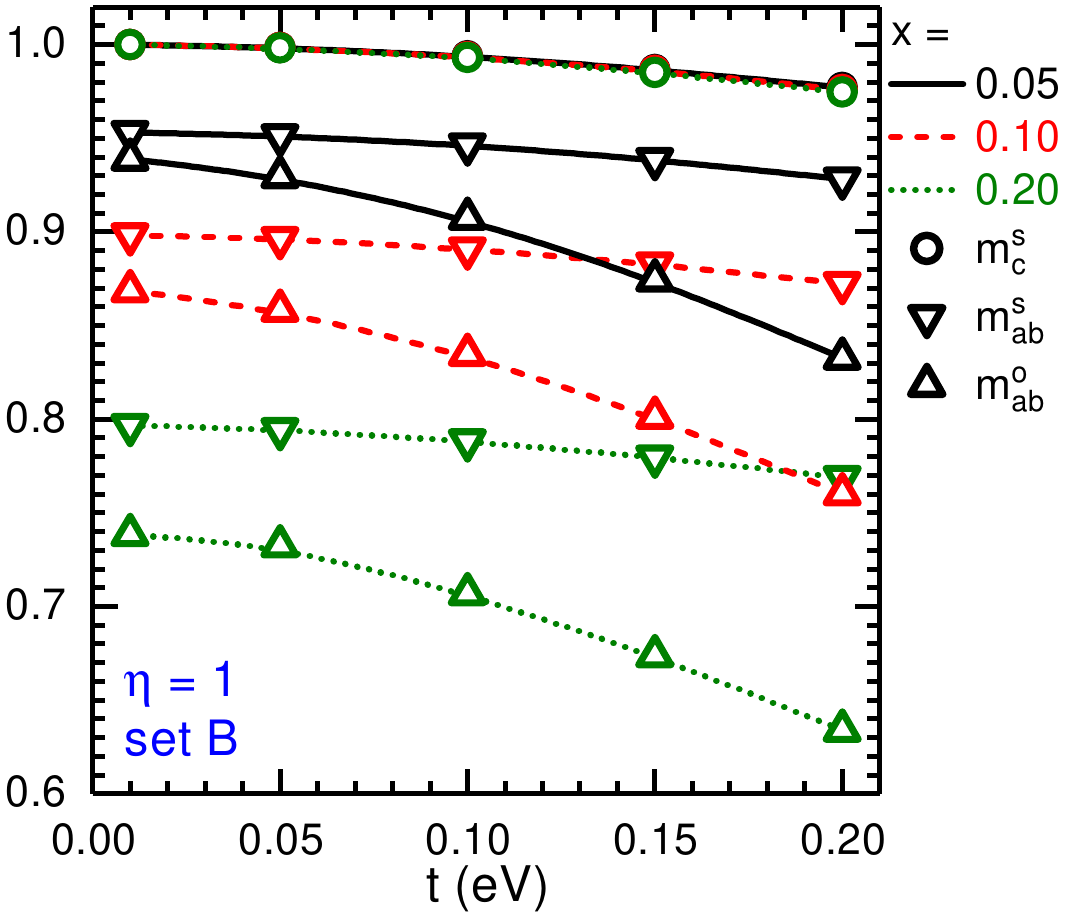} \caption{
Self-consistent uHF results for the spin-order parameters $m_{c}^{s}$
and $m_{ab}^{s}$ and the orbital-order parameter $m_{ab}^{o}$ for the
$C$-AF/$G$-AO ordered phase resulting from electrons in $a/b$ and $c$
orbitals, respectively, as functions of the hopping parameter $t$ and
for different doping concentrations $x=0.05,0.10,0.20$.
\label{fig:fig-v}}
\end{figure}


In Fig.~\ref{fig:fig-v}, the $t$ dependence of the spin-order parameters
$m_{ab}^{s}$ and $m_{c}^{s}$ and of the orbital-order parameter
$m_{ab}^{o}$ are shown. As we already explained above, hole doping
concerns the $\{a,b\}$ orbital doublet and the magnetic moment component
$m_{ab}^{s}$ is thus reduced by $x$ at $t=0$, whereas $m_{c}^{s}=1$
remains unchanged as there is no doping into $c$ orbitals. On increasing
$t$, holes move and the spin and orbital order are further reduced. This
effect is particularly small for $c$ orbitals, as their motion involves
only doubly occupied $c^{2}$ configurations and, therefore, costs the
intraorbital Hubbard repulsion $U$. The contribution of such processes
is small in comparison to that of the virtual processes involving $a$
and $b$ electrons. In Fig.~\ref{fig:fig-v}, one recognizes a strong
reduction of the orbital order $m_{ab}^{o}$, both as a function of $x$
and $t$. In fact, the kinetic processes have a stronger effect on the
$G$-type orbital order parameter $m_{ab}^{o}$ than on the $C$-type
spin order $m_{ab}^{s}$. This may be seen as a precursor of the trend
observed in the experiments performed on La$_{1-x}$Sr$_{x}$VO$_{3}$
\cite{Miy00} where the $G$-AO order melts at a lower doping than
the $C$-AF spin order.

Next, we shall use the order parameter expressions, derived in Appendix
B by assuming non-interacting small SO polarons, for the interpretation
of the uHF results. The central assumption
is the linearity in $x$ of the order parameter corrections, which
is confirmed by the present uHF results in a wide doping range, see
Appendix~B. It is important to recognize that in the insulating regime
that we are considering, each small polaron is bound to and strongly
deformed by the central charged defect. Thus, the number of relevant
paths for the doped holes are significatively reduced in comparison
to those available for a polaron not attached to a defect. Furthermore,
the polaron state is strongly influenced by the random fields resulting
from the other more-distant defects.

The expressions for the spin and orbital deviations associated with
a single polaron are derived by considering the polaron cloud of a
hole located on an active bond close to a defect. It is the FM correlations
along the $c$-direction in $C$-AF structure in combination with
the defect potential, which defines the direction of the active bond,
that yields the dominant gain of kinetic energy for the hole. The
virtual motion of holes in $a$ or $b$ direction is instead quenched
by AF correlations. Thus, the polaron wave function is characterized
by two physical parameters, namely: (i) the polarity parameter of
the active bond $\delta_{c}$, where $\delta_{c}=0$ describes a hole
localized on a single site, as for $t=0$, and $\delta_{c}=1/2$ an
equal partition between the two sites of the active bond, as in the
case of large $t$ and screened far-defect potentials ($\eta=1$);
(ii) the activation energy $e_{0}$ of the virtual string excitations.
We focus here on the spin- and orbital- order parameter due to $a/b$
orbitals that show the strongest dependence on doping:
\begin{eqnarray}
m_{ab}^{s} & \cong & 1-x\left[1+\frac{2\left(t/e_{0}\right)^{2}}
{1+\left(t/e_{0}\right)^{2}}\right],\nonumber \\
m_{ab}^{o} & \cong & 1-x\left[1+2\delta_{c}+\frac{2\left(t/e_{0}\right)^{2}}
{1+\left(t/e_{0}\right)^{2}}\right].\label{msab}
\end{eqnarray}

For an isolated single defect, the activation energy $e_{0}$ is given
by Hund's coupling $J_{H}$, whereas $\delta_{c}$ is controlled primarily
by $t$ and the Jahn-Teller coupling. In the presence of disorder,
the potentials due to the random far defects affect quite strongly
both the activation energy $e_{0}$ and the polarity parameter $\delta_{c}$.
Accordingly, these parameters are taken here as natural variational
parameters. When we apply these relations to the uHF results, obtained
as a statistical average over many defect realizations, we obtain
for $\delta_{c}$ and $e_{0}$ the values of a \emph{typical} bound
small polaron.

\begin{figure}[t!]
\includegraphics[width=1\columnwidth]{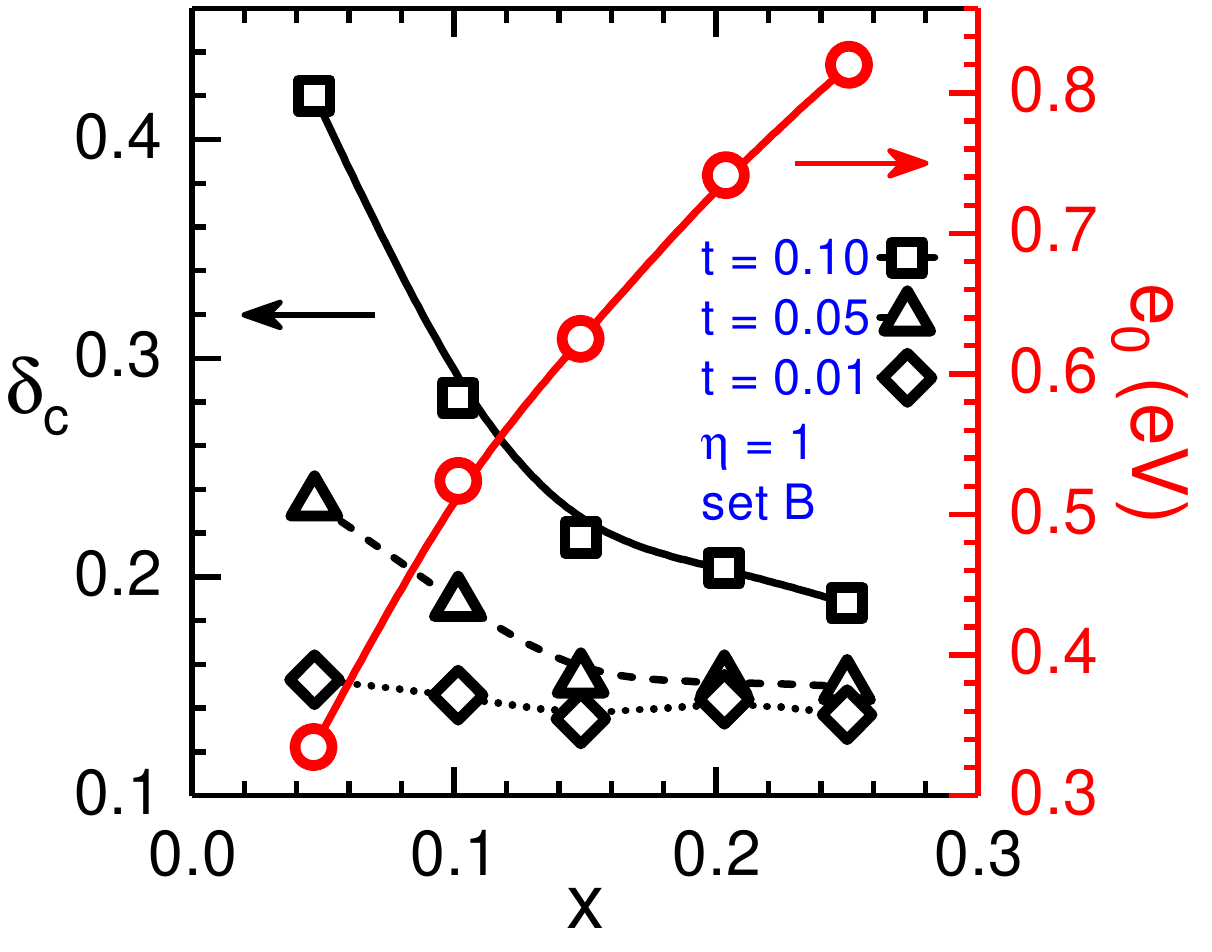} \caption{
Doping dependence of the polaron superexchange energy scale $e_{0}$
and of the polarity parameter $\delta_{c}$ of a \emph{typical} active
bond for different values of $t$ (in eV) and parameter set B of
Table~\ref{tab:para}.
The lines are simply guides to the eye.\label{fig-e0vsx}}
\end{figure}

Figure~\ref{fig-e0vsx} shows the \emph{typical} polaron parameters
obtained by fitting the uHF results shown in Fig.~\ref{fig:fig-v} for
the spin and orbital order parameters, $m_{ab}^{s}$ and $m_{ab}^{o}$,
see Eqs. \ref{msab} (see also Appendix B). We find a significant
increase of the activation energy $e_{0}$ from $0.33$ eV at $x=5$\%
to $0.82$ eV at $x=25$\%. This increase can be attributed to the
increase of the overall random potential strength on the single defect
cube when the concentration of defects becomes larger and larger.
We note that Hund's exchange, which is the energy scale $e_{0}$ for
an isolated defect as already mentioned above, is $J_{H}=0.6$ eV
for parameter set B of Table~\ref{tab:para}.

The interpretation of the data for the polarity parameter $\delta_{c}$
is more subtle. Here we determine $\delta_{c}$ directly from the
data in Fig.~\ref{fig:fig-v} via the difference of $m_{ab}^{o}$
and $m_{ab}^{s}$ using Eqs.~(\ref{msab}). The parameter $\delta_{c}$
naturally depends on $x$ and on $t$; large $t$ favors an equal
distribution on the active bond, that is $\delta_{c}=0.5$. At small
doping, this is almost realized for $t=0.1$ in Fig.~\ref{fig-e0vsx}
since the random polarization fields (i.e., the far-defect potentials)
are quite small in this case. It is worth noting that, in the low
doping regime, the small Jahn-Teller potentials, which naturally lead
to an imbalance, become also relevant.

By means of the same small-polaron wave function for a hole on an
active bond, we obtain for the IPN, with the help of the definition
in Eq.~(\ref{iPN_polaron}),
\begin{equation}
P_{a}^{-1}\cong\left(1-2\delta_{c}+2\delta_{c}^{2}\right)\left[
1-\frac{2\left(t/e_{0}\right)^{2}}{1+\left(t/e_{0}\right)^{2}}\right].
\label{Pa}
\end{equation}
The finite polarity parameter $\delta_{c}$ explains the deviation
of the IPN from $1$ in the small $t$ case for the states in the
soft gap close to the chemical potential (see Fig.~\ref{fig_q}).
The overall $t$-dependence of the IPN close to $\mu$ is determined
both by the $t$-dependence of the polarity parameter $\delta_{c}(x,t)$
and by the virtual-string contributions in the polaron wave function.
Interestingly, the nonlinear corrections driven by the random polarization
fields via the $x$-dependence of $e_{0}$ and $\delta_{c}$ are particularly
pronounced at small $x$ values (see also Appendix B).

In this Section, we have focused on the microscopic character of the
defect states that are responsible for the formation of the soft gap
and the reduction of spin and orbital order. We have shown that the
SO polaron states inside the MH gap have a reduced spatial symmetry
and are more strongly localized than the states in the center of the
Hubbard bands. The polaron states in the soft gap are controlled by
the randomness of the defect locations and by the interplay of kinetic
energy and $e$-$e$ interactions.


\section{Relation to the PES of
L\protect\lowercase{a}$_{1-x}$C\protect\lowercase{a}$_{x}$VO$_{3}$
\label{sec:pes}}


Now, we turn to the questions how the structure of the DS gap is
affected by strength of the defect potential $V_{D}$ and whether we
can identify characteristic defect-related features in the doping
dependence of existing PES data. We shall explore here the PES spectra
of La$_{1-x}$Ca$_{x}$VO$_{3}$ since for this system Maiti and Sarma
\cite{Mai98} have measured the doping dependence up to $x=0.5$,
and moreover both PES and IPES spectra exist for the parent compound
LaVO$_{3}$ \cite{Mai00}. The latter yields clear evidence of the
multiplet structure and provides an estimate for Hund's exchange $J_{H}$
(see Fig.~\ref{fig:maiti}). Moreover, these spectra give us a measure
for the principal parameter for the Mott gap $U-3J_{H}\approx2.7-3.0$
eV between the centers of the LHB and the HS multiplet of the UHB.
In the following study, we shall use $U-3J_{H}=3.0$ eV as determined
after Maiti \textit{et al.} \cite{Mai00} (set $C$ of Table~\ref{tab:para}).
Another important energy scale that follows from PES experiments is
the distance of the chemical potential from the center of the LHB,
$\mu-E_{\rm LHB}$. Surprisingly, over the whole doping range in
La$_{1-x}$Ca$_{x}$VO$_{3}$, up to $x=0.5$, this scale is approximately
constant, $\mu-E_{\rm LHB}\approx1.5$ eV. Since it is the repulsive defect
potential that pulls the $d^{2}$ defect states upward out of the LHB,
we expect also an upward shift of the chemical potential. Therefore the
distance of the chemical potential from the LHB center is expected to
scale with the parameter $V_{D}$ in the Hamiltonian Eq. (\ref{H3band}).

At this point, it is interesting to note that the pioneering work
of Mott \cite{Mot74} in the 1970s on the metallic and nonmetallic
states of strongly correlated matter was to some extent stimulated by
the experiments of Dougier and Hagenmuller \cite{Dou75}, who explored
the electrical conductivity of LaVO$_{3}$ doped with strontium. The
metal-insulator transition in this compound at $x_{c}\approx0.23$
was attributed to an Anderson type transition between localized and
delocalized states in an impurity band. The option of a Mott gap in
the defect band was ruled out \cite{Mot89}. The presence of a soft
gap with $N(\mu)=0$ in the insulating regime has not been considered.

So far, we have analyzed spectral functions in
Secs.~\ref{sec:multi}-\ref{sec:ipr} using parameter set $B$ of
Table~\ref{tab:para}. They have the advantage that satellite
structures are well resolved, representing the atomic multiplet
excitations. Interestingly, we see from spectra in Fig.~\ref{fig:dos02},
calculated with inclusion of $e$-$e$ interactions, that the LHB center
remains at a fixed distance relative to the chemical potential for all
doping concentrations, as required by the above experiments. However,
parameter set $B$ implies $U-3J_{H}=2.2$
eV and $\mu-E_{{\rm LHB}}\simeq0.8$ eV being both too small and thus
not appropriate for the La$_{1-x}$Ca$_{x}$VO$_{3}$ compounds.

\begin{figure}[t!]
\includegraphics[width=8cm]{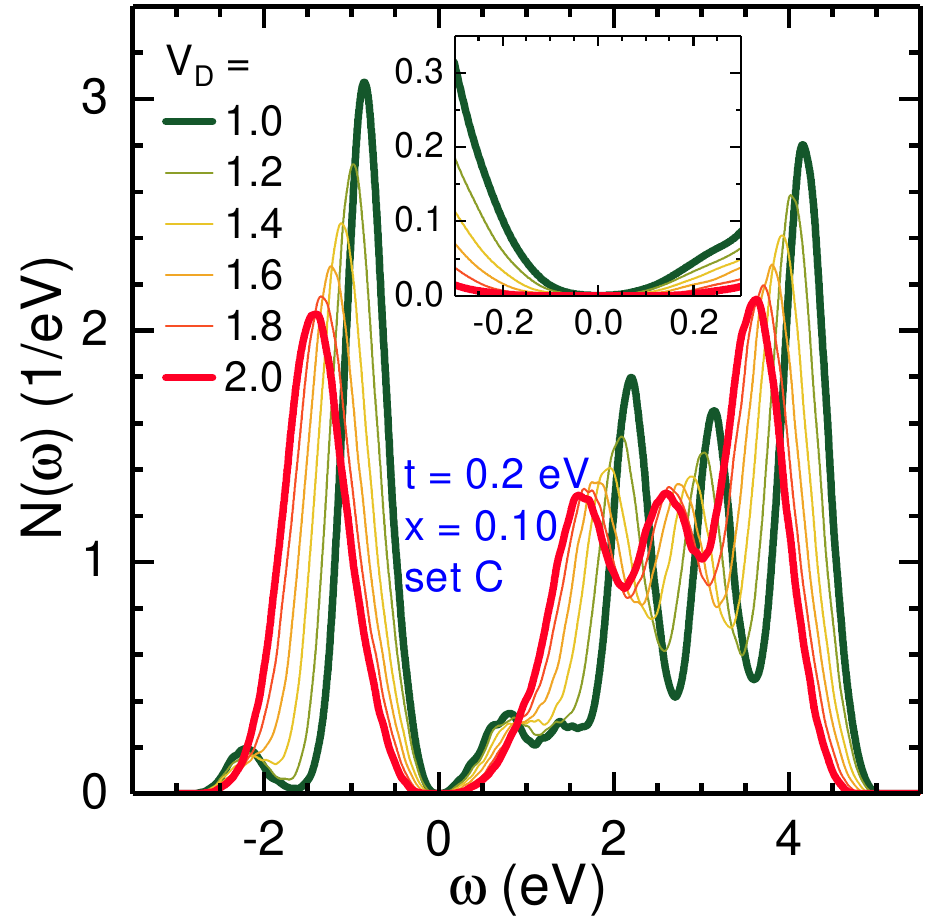} \caption{
Density of states for $x=10$\% and six different defect potentials
$V_{{\rm D}}\in[1.0,2.0]$ eV. The data shows that the distance between
the LHB and the chemical potential (at $\omega=0$) is increasing
with $V_{{\rm D}}$ while the distance between the LHB and the HS
state of the UHB stays constant. Parameters are $t=0.2$ eV and set
$C$ of Table~\ref{tab:para}. The defect potential is of Coulomb
type and $e$-$e$ interactions are included ($\eta=1$).\label{fig22}}
\end{figure}


In Fig.~\ref{fig22}, we display the $V_{D}$ dependence of the DOS
for $U-3J_{H}=3.0$ eV appropriate for La$_{1-x}$Ca$_{x}$VO$_{3}$
(set $C$ of Table~\ref{tab:para}) at $x=10$\% doping. The resulting
$N(\omega)$ spectra depend strongly on the $V_{D}$ parameter, Eq.~(\ref{v}),
which determines both the strength of the Coulomb like defect potentials
and the strength of $e$-$e$ interactions, Eq.~(\ref{pots}), guaranteeing
charge neutrality. On increasing $V_{D}$ from 1.0 to 2.0 eV, we recognize:
(i) the DS gap becomes wider and softer; (ii) the characteristic energy
scale $\mu-E_{\rm LHB}$ increases; (iii) the $d^{1}\rightarrow d^{0}$
satellite moves closer to the LHB and gets gradually absorbed. Moreover,
(iv) the defect states right above the chemical potential ($\mu=0$),
which are well separated from the HS UHB at $V_{D}=1$ eV, merge with
this latter for $V_{D}=2$ eV; and (v) the gaps between the multiplets
of the UHB get gradually filled with defect states as $V_{D}$ increases.
Finally, we see that it requires a value $V_{D}\approx2$ eV until one
reaches the experimental value $\mu-E_{\rm LHB}\approx1.5$ eV. One
also notices in Fig.~\ref{fig22} that the energy between the center
of LHB and the HS state of the UHB remains unchanged $U-3J_{H}=3.0$ eV
for the different values of $V_{D}$, as expected from Table~\ref{tab:sw}.
That is, the intra-atomic excitations and multiplet splittings are
not affected by the disorder.

\begin{figure}[t!]
\includegraphics[width=8cm]{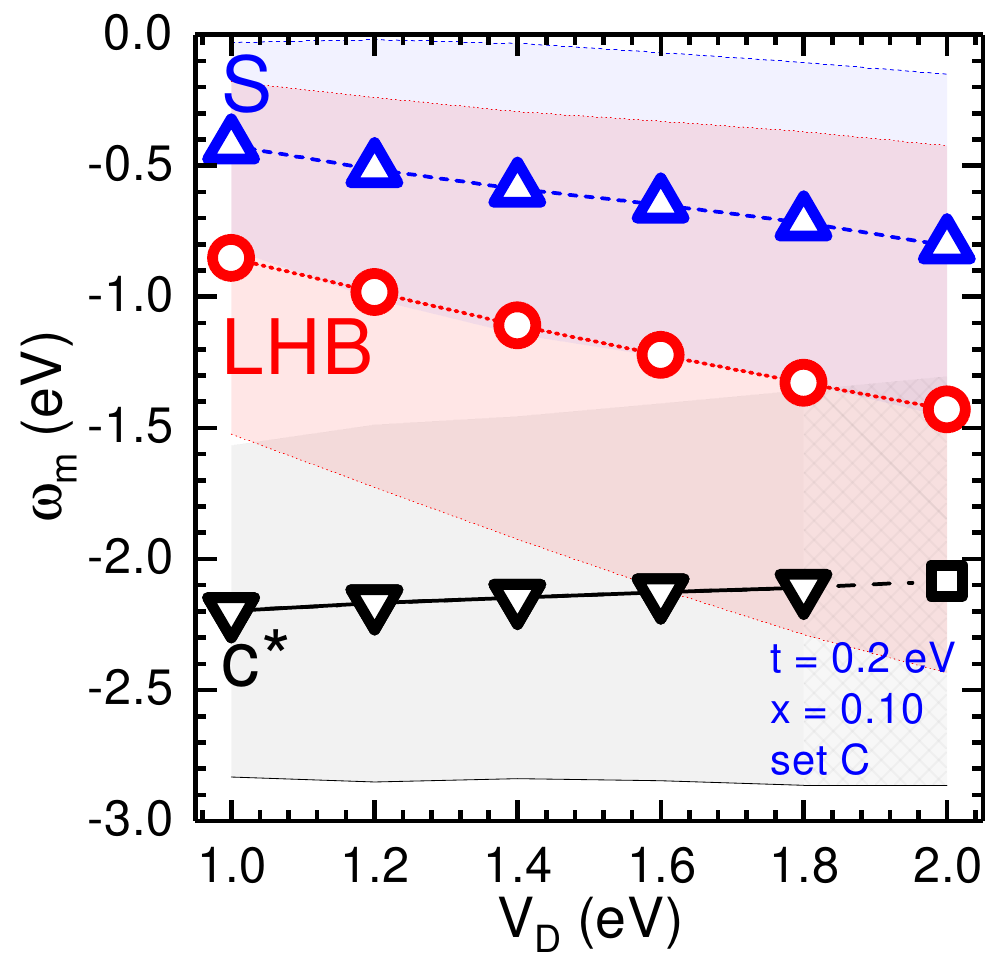} \caption{
Characteristic energies $\omega_{m}$ of the spectral features resolved
below the chemical potential $\mu$ ($\omega=0$) as functions of
the defect potential $V_{{\rm D}}\in[1.0,2.0]$ eV: the center of
gravity of the LHB (LHB, red circles), the filled defect states (S,
blue up-triangles), and the satellite due to hole states ($c^{*}$,
black down-triangles and square). The energies and the linewidths
(indicated by the shaded areas colored according to the symbol colors)
were determined by fitting the data in Fig.~\ref{fig22} with three
Gaussians, respectively. The $c*$ data for $V_{D}=2.0$ eV (black square)
has been obtained by linear fitting the data for $V_{D}<2.0$ eV as there
was no possibility to resolve such peak within the overall LHB peak.
\label{fig23}}
\end{figure}

The $V_{D}$ dependence of several characteristic single particle excitations
$\omega_{m}$ relative to the chemical potential ($\mu=0$) are plotted in
Fig.~\ref{fig23}. The characteristic energies
$\omega_{{\rm LHB}}=E_{\rm LHB}-\mu$ as well as the satellite energies,
$\omega_{c^{*}}$ and $\omega_{S}$ have been determined by a fit with three
Gaussians to the spectra in Fig.~\ref{fig22} at negative energy. The
linewidths are indicated by shading. In order to understand the variation
of $\omega_{{\rm LHB}}=E_{{\rm LHB}}-\mu$ we need an estimate for the
dependence of $E_{{\rm LHB}}$ and $\mu$ on the defect potential $V_{D}$.
The center of the LHB corresponds to a $d^{2}\rightarrow d^{1}$ transition.
In uHF, this energy is connected with the removal of the topmost occupied
$a$ (or $b$) electron of a $d^{2}$ configuration far away from a defect,
\begin{equation}
E_{{\rm LHB}}\simeq\epsilon_{a}+U-3J_{H}.
\end{equation}
The chemical potential can be approximated by a similar excitation
in the direct neighborhood of a defect, that is at a spectator site.
Hence
\begin{equation}
\mu\approx\epsilon_{a}+U-3J_{H}+V_{D},
\end{equation}
and therefore
\begin{equation}
\omega_{{\rm LHB}}=E_{{\rm LHB}}-\mu\approx-V_{D}.\label{ELHB}
\end{equation}
This simple estimate reproduces qualitatively the trend of the full
self-consistent uHF result in Fig.~\ref{fig23} that is
$\omega_{{\rm LHB}}\approx-0.8V_{D}$. The small deviation is not unexpected
since both $E_{{\rm LHB}}$ and $\mu$ will acquire upward shifts from the
long-range tails of the defect potentials leading to a small positive
correction proportional to $V_{D}$ in $E_{{\rm LHB}}-\mu$.

The satellite $c^{*}$ corresponds to a $d^{1}\rightarrow d^{0}$
transition and is a fingerprint of the spin-orbital polaron state.
As we discussed before the doped hole selects an \emph{active} FM
bond $\langle i,j\rangle$ parallel to $c$ next to a defect. The random
defect fields tend to localize the hole on a single site $i$ leading
to a configuration $|c_{i},(c_{j}a_{j})\rangle$ where the single $a$
electron is on site $j$ of the active bond. This is described by
the polarization parameter $\delta_{c}=0$. If the kinetic energy
(i.e., $t$) is strong enough, the $a$ electron delocalizes on the
active bond leading to $\delta_{c}=0.5$. The annihilation energies
of $c$-electrons in the two configurations relative to the chemical
potential $\mu$ are
\begin{eqnarray}
\omega_{c^{*}} & \simeq & \epsilon_{c}-\epsilon_{a}-(U-3J_{H})(1-\delta_{c}),\\
\omega_{S} & \simeq & \epsilon_{c}-\epsilon_{a}-(U-3J_{H})\delta_{c}.
\end{eqnarray}
Most importantly, in the expression of $\omega_{c^{*}}$, the dependence
on the polarization parameter $\delta_{c}$ contributes with a large
prefactor $U-3J_{H}$. Therefore, the delocalization of the remaining
$a$-electron in the small polaron is reflected in a large energy shift
of the $c^{*}$ satellite. That is, the state of the $a$ electron can
be probed by annihilation of $c$-electrons. Comparing the expression
$\omega_{c^{*}}$ with the data in Fig.~\ref{fig22} yields
$\delta_{c}\simeq0.3(0.33)$ at $V_{D}=1(2)$ eV, respectively.
There is no direct $V_{D}$ dependence in these expressions. The origin
of the small change of $\delta_{c}$ observed above is unclear.

As one can see in Figs.~\ref{fig22} and \ref{fig23}, the widths of both
the LHB and the UHB increase with increasing defect potential~$V_{{\rm D}}$.
Since the corresponding spectra are calculated for
fixed doping $x=0.1$ and constant kinetic energy parameter $t$,
the broadening can be directly attributed to the increasing randomness
resulting from the stronger long-range disorder potentials. Moreover,
we see that, at small $V_{{\rm D}}\simeq1.0$ eV, the $d^{1}\rightarrow d^{0}$
satellite excitations $\omega_{c^{*}}$ are well separated from the
$d^{2}\rightarrow d^{1}$ excitations at $\omega_{{\rm LHB}}$ that define
the LHB. Due to the $V_{D}$ dependence of the LHB, see Eq.~(\ref{ELHB}),
the $c^{*}$ satellite penetrates the LHB at $V_{D}\approx1.6$ eV.
In other words, the $d^{0}$ final state on the active bond becomes
unstable and changes into a $d^{1}$ state by a simultaneous $d^{2}\rightarrow d^{1}$
transition at a vanadium host site not belonging to the defect cube.
With the onset of the delocalization of holes, we come to the borderline
of the regime of well localized spin-orbital polarons. The delocalization
of holes has severe consequences for the uHF algorithm and leads to
a slowing-down of convergence of the algorithm, that is, for large
$V_{D}$ when the $c^{*}$ satellite disappears in the LHB.

\begin{figure}[t!]
\includegraphics[width=1\columnwidth]{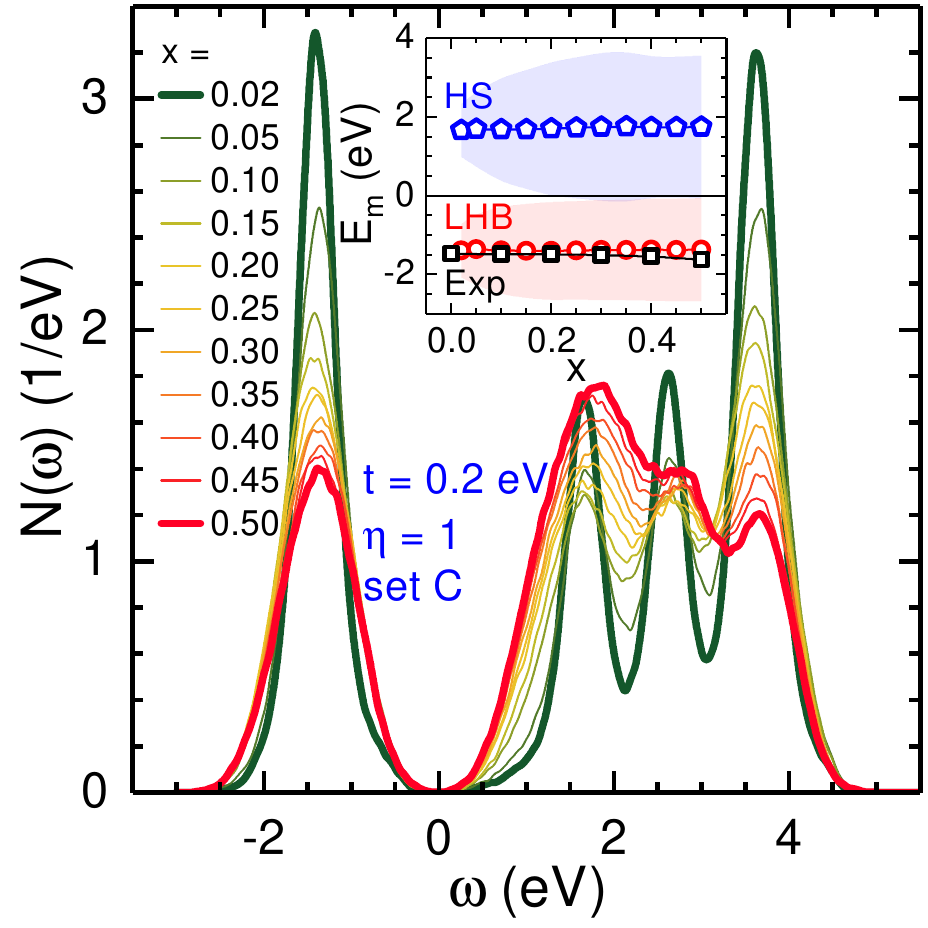} \caption{
Density of states $N(\omega)$ as a function of increasing doping
concentration $x\in[0.02,0.50]$ for set $C$ of Table~\ref{tab:para}
including $e$-$e$ interactions ($\eta=1$) and for $t=0.2$ eV.
Inset displays the small variation with $x$ of the distance of the
peak of LHB from chemical potential $\mu$ ($\omega=0$): experimental
results of Maiti and Sarma for La$_{1-x}$Ca$_{x}$VO$_{3}$ \cite{Mai98}
(black squares); theoretical data for the LHB (red circles) and the
HS state of the UHB (blue pentagons). The shaded areas colored according
to the symbol colors give a measure of the linewidths of the related
multiplet structures.\label{fig24}}
\end{figure}


The doping dependence of the uHF DOS for La$_{1-x}$Ca$_{x}$VO$_{3}$
is displayed in Fig.~\ref{fig24}. These calculations were performed
using long-range $e$-$e$ interactions ($\eta=1$) and interaction
parameters, $U-3J_{H}=3.0$ eV and $V_{D}=2.0$ eV (see parameter set
$C$ in Table I). As most important features on the PES side of the
spectra we note: (i) the persistence of a soft DS gap up
to large doping, and (ii) the essentially doping independent distance
of the chemical potential from the center of the LHB, as found in
the experimental data of Maiti and Sarma \cite{Mai98}. Unfortunately,
the existing experimental data do not allow to draw a conclusion with
any certainty concerning the existence of a $c^{*}$ satellite. Thus,
it remains as a challenge for future higher resolution experiments
to see whether this feature can be resolved. The inset highlights
the independence on doping of $\mu-E_{\rm LHB}\approx1.4$ eV and of
the fundamental Hubbard gap $(U-3J_{H})$. The decrease of the spectral
weights with doping is consistent with the sum rules discussed in
Sec.~\ref{sec:dop}.

We observe a pronounced shift of spectral weight in the UHB towards
lower energies as a function of doping, i.e., resulting in an increase
of the spectral weight in the regime of HS excited states. At large
doping, the multiplet structures of the UHB begin to merge and, as
a consequence, the UHB becomes almost featureless. The multiplet excitations
lose their distinct character as (for these parameters) these excitation
processes are degenerate with $d^{1}\rightarrow d^{2}$ excitations
at hole sites. Moreover, we note that the $d^{1}\rightarrow d^{2}$
transitions have a narrower width of their multiplet structure compared
to that for the $d^{2}\rightarrow d^{3}$ transitions of the host.

In Fig.~\ref{fig24}, we recognize a moderate increase of the width
of the LHB with increasing defect and doping concentration. This one
may have expected since there are two prevailing mechanism that contribute
to an additional broadening of the LHB when the doping concentration
is increased. Namely, (i) there are the growing satellites $c^{*}$
and $S$ that contribute to the increase overall width of the LHB,
and (ii) due to the increasing doping concentration, the defects get
closer and thereby the disorder fluctuations are expected to increase.
There is, however, also a counteracting effect due to an increased
screening resulting from the defect states and the narrowing of the
soft gap. We note that this screening, which is still non-metallic,
is fully included in our calculation, due to the finite field method
we are using. We also remark that, in terms of an alternative random
phase many-body theory, our treatment of $e$-$e$ interactions for
the $t_{2g}$ subsystem would correspond to the calculation of the
full dielectric matrix \cite{Hor88} for an inhomogeneous and random
system. The complex dielectric matrix would then yield a quite similar
or even the same polarization charge density induced by the defect
potentials as in our uHF calculations. This perspective highlights
that the dielectric screening in systems with defect states and soft
gaps is rather involved and, to the best of our knowledge, it has
not been addressed yet by any many-body theoretical approach.

\section{Discussion and summary\label{sec:summa}}

This work elucidates the inherent complexity of charged defect states
in orbitally-degenerate Mott insulators. We have studied a generic
three-band model representing the $t_{2g}$ electrons of the vanadium
ions that guarantees a faithful description of the magnetic and orbital
ordered phases of the vanadium perovskites such as LaVO$_{3}$ or
YVO$_{3}$. The model developed here includes the local Hubbard-Hund
interactions and the long-range $e$-$e$ interactions. Therefore,
it describes the dielectric screening of the Coulomb potentials of
the random charged defects due to the $t_{2g}$ electrons explicitly
in the regime of strong electron correlations. Nevertheless, there
are relevant interactions that we have omitted, such as spin-orbit
and orbital polarization interactions, as discussed in
Sec.~\ref{sec:model}. We have neglected such terms to achieve a closer
correspondence to the generic models used in studies of disorder such
as the quantum Coulomb glass or the Anderson-Hubbard model. Our model
may be thus seen as multiflavor generalization of these models.

The calculation of the one-particle excitation spectra of a Mott insulator
with orbital degeneracy and random defects is a complex task that
was only possible to undertake within the unrestricted Hartree-Fock
approximation. We stress that this is the only scheme able to: (i)
control simultaneously the strong correlations, the Coulomb interactions
and the randomness of defects even at high doping and (ii) reproduce
the essence of Mott physics, i.e., the atomic multiplet structure
and the variation of the spectral weights in the Hubbard bands.
A more detailed description of the algorithm used for the treatment
of disorder is given in Appendix A. There, we also display a subtle
convergency test for our algorithm and show that the gradual switching
on and off of $e$-$e$ interactions leads basically to the same statistical
average for the density of states $N(\omega)$ as for the initial
state.

An important feature of the vanadium perovskites is the robustness
of the Mott-Hubbard gap and of the spin and orbital order up to high
doping concentrations. Our unrestricted Hartree-Fock results for the
single particle density of states $N(\omega)$ reveal the persistence
of the atomic multiplet structure of V-ions up to doping concentrations
as large as 50\% in photoemission and inverse photoemission spectra.
At the same time, there is a strong spectral weight transfer from the
Hubbard bands into the Mott-Hubbard gap and into satellite structures
of the lower Hubbard band and the multiplets of the upper Hubbard
band. We have analyzed the energetics and the doping dependence of
spectral weights of the different excitations in the atomic limit
that provides valuable information for the insightful interpretation
of photoemission spectra.

The opening of a gap in the defect states at the chemical potential
$\mu$ inside the Mott gap is a subtle feature in doped Mott insulators.
We have presented a very detailed analysis how $e$-$e$ interactions
and the kinetic gap mechanism jointly contribute to the formation of
the gap in defect states. As we have shown, the states in the defect
states gap can be understood as small spin-orbital polaron states bound
to defects. Note that, in the presence of $e$-$e$ interactions, these
objects are electric dipoles that interact effectively via dipole-dipole
interactions. The kinetic energy gain of the doped hole moving within
the spin-orbital polaron on a ferromagnetic active bond is the origin
of a level splitting at $\mu$ and thereby of the kinetic mechanism
for the defect states gap formation. We have demonstrated that the
defect states gap survives the disorder average and typically yields
a soft gap.

Central role for the quantitative description of the insulating state
is played here by a statistical analysis of the gap using a Weibull
distribution as introduced in Ref.~\cite{Ave15}. Using this tool we
have shown that the exponent $\nu$ of the soft gap in
$N(\omega)\propto|\omega|^{\nu}$ at the chemical potential is
nonuniversal, that is, it depends sensitively on:
(i) the strength of $e$-$e$ interactions and
(ii) on the hopping parameter $t$ which controls the kinetic energy.
In particular, we note here that in the atomic limit ($t=0$), $e$-$e$
interactions are not strong enough to produce a Coulomb gap in the
vanadium perovskites, as one might expect on the basis of
Efros-Shlovskii theory. Instead, we find in this regime exponents that
rather correspond to a Coulomb or zero-bias anomaly $0<\nu<1$.

We have investigated the degree of localization of the unrestricted
Hartree-Fock wave functions quantitatively. We used here as a measure
the inverse participation number, which we have generalized for systems
with spin and orbital flavors. In particular, we have found that the
defect states in the soft gap are typically localized on one up to two
sites, and they are more strongly localized than the states inside the
Hubbard bands. This feature is surprising and the inverse participation
number provides an independent proof for \textit{small} spin-orbital
polarons bound to defects. Interestingly, we observed a strong
discontinuity in the localization of states right below and above $\mu$
in the presence of $e$-$e$ interactions, whereas in their absence no
discontinuity could be found.

Using the unrestricted Hartree-Fock, we have calculated the doping
dependence of spin- and orbital- order parameters in the $C$-AF/$G$-AO
phase. Our results reveal a faster decrease of the orbital order as
compared to the spin order. The ordered state persists beyond $x=0.5$
in our calculations. This robustness is an independent sign that doped
holes form small polarons. Starting from the atomic limit, we have
constructed the many-body wave function of a spin-orbital polaron bound
to a defect. We have then established that the reduction of spin and
orbital order obtained by unrestricted Hartree-Fock is consistent with
that calculated with help of the many-body spin-orbital polaron state.
We note that, for a quantitative comparison with existing experimental
data for the reduction of spin and orbital order and the insulator
to metal transition \cite{Fuj05,Ree16}, the inclusion of spin-orbit
interaction and orbital polarization terms \cite{Ave13} is required.

In photoemission experiments of gapped systems, the position of the
chemical potential $\mu$ is determined by defects. For the doped cubic
vanadates we find that $\mu$ lies in the center of the defect states
gap that forms inside the Mott-Hubbard gap. The defect potential
$V_{{\rm D}}$, which basically confines doped holes to a cube formed
by the V ions being the nearest neighbors of a charged defect, provides
a measure of the upward shift of the in-gap defect states relative
to the lower Hubbard band \cite{Hor11}. We found here that the distance
of $\mu$ from the center of the lower Hubbard band scales linearly with
$V_{D}$. Moreover, we found that this distance is basically unchanged
by doping up to 50\%, i.e., consistent with the photoemission study by
Maiti and Sarma \cite{Mai98,Mai00} for La$_{1-x}$Ca$_{x}$VO$_{3}$. We
interpret this as a manifestation of small spin-orbital polaron physics.
Furthermore, we have shown that the intensity of the $d^{1}\rightarrow d^{0}$
satellite of the lower Hubbard band, that grows proportional to the
doping concentration $x$, provides a direct fingerprint of the trapped
spin-orbital polarons. The position in energy of this satellite relative
to the lower Hubbard band and its width yields detailed information
about the polarization of the SO polaron and the strength of random
defect fields. For the La$_{1-x}$Ca$_{x}$VO$_{3}$ system, we found
that the satellite is not well separated from the lower Hubbard band
and thus has not been resolved in the existing PES experiments so
far. Certainly, it would be very interesting to have high resolution
photoemission or tunneling experiments carried out on this compound.

It is important to stress that our model, which includes simultaneously
the defect Coulomb potential and the $e$-$e$ interactions, is
qualitatively different from the Coulomb glass type of models not just
because it contains orbital and spin degrees of freedom. The Coulomb
glass model contains only a random distribution of defect state energies,
thus the dipole (negative defect charge and positive bound hole) formation
cannot take place. It is just this latter feature that makes an important
difference between the results obtained by these two types of defect
modelization (ours being definitely more realistic) and that leads
to a weakening of the disorder with increasing $e$-$e$ interactions.
Instead, for Coulomb glass type of models, $e$-$e$ interactions
enhance localization \cite{Pol12}.

Summarizing, we have demonstrated a systematic redistribution of the
spectral weights with increasing doping by charged defects in Mott
insulators using the example of $R_{1-x}$Ca$_{x}$VO$_{3}$. Even when
doping is rather high, these systems remain insulating and the large
Mott-Hubbard gap accommodate the spectral weight due to defect states,
with the filled and unfilled defect states separated by a kinetic gap.
We have also shown that the treatment of defects in strongly correlated
materials with orbital degrees of freedom requires important extensions
of the Hubbard-Anderson model. We designed a well motivated model for
systems with orbital degrees of freedom to uncover the origin of the
\emph{satellite} structures that appear in doped systems and to explain
the corresponding sum rules. In this way, the interpretation of
photoemission and inverse photoemission experiments on Mott insulators
doped with charged defects becomes possible. We have also explained why
the spectral features, which arise from different excitation processes,
can be energetically very close or overlap with one another, making
them unresolvable in the experimental spectra.

Finally, one may ask whether there is any fundamental difference between
doping into the vanadadium perovskites and the cuprate high-$T_{c}$
transition metal oxides. In both cases the formal charge of the defects
and the distance to the relevant electrons to the Cu or V ions,
respectively, are similar. Nevertheless, in the cuprates, defects form
shallow impurity states, whereas in the vanadates they are deep inside
the gap. Our analysis shows that the additional orbital degree of freedom
makes an important difference. The confinement of spin-orbital polarons
in vanadates is \emph{typically} on a single active bond, and thus is
much stronger than the localization of spin polarons in cuprates. This
also implies that the screening of defect potentials at low energy due
to the defect states is weaker in vanadates, effectively leading to
stronger binding of carriers and to a shift of the insulator-to-metal
transition to much higher doping concentrations than in high-$T_{c}$
materials.

\begin{figure*}[t!]
\includegraphics[width=16cm]{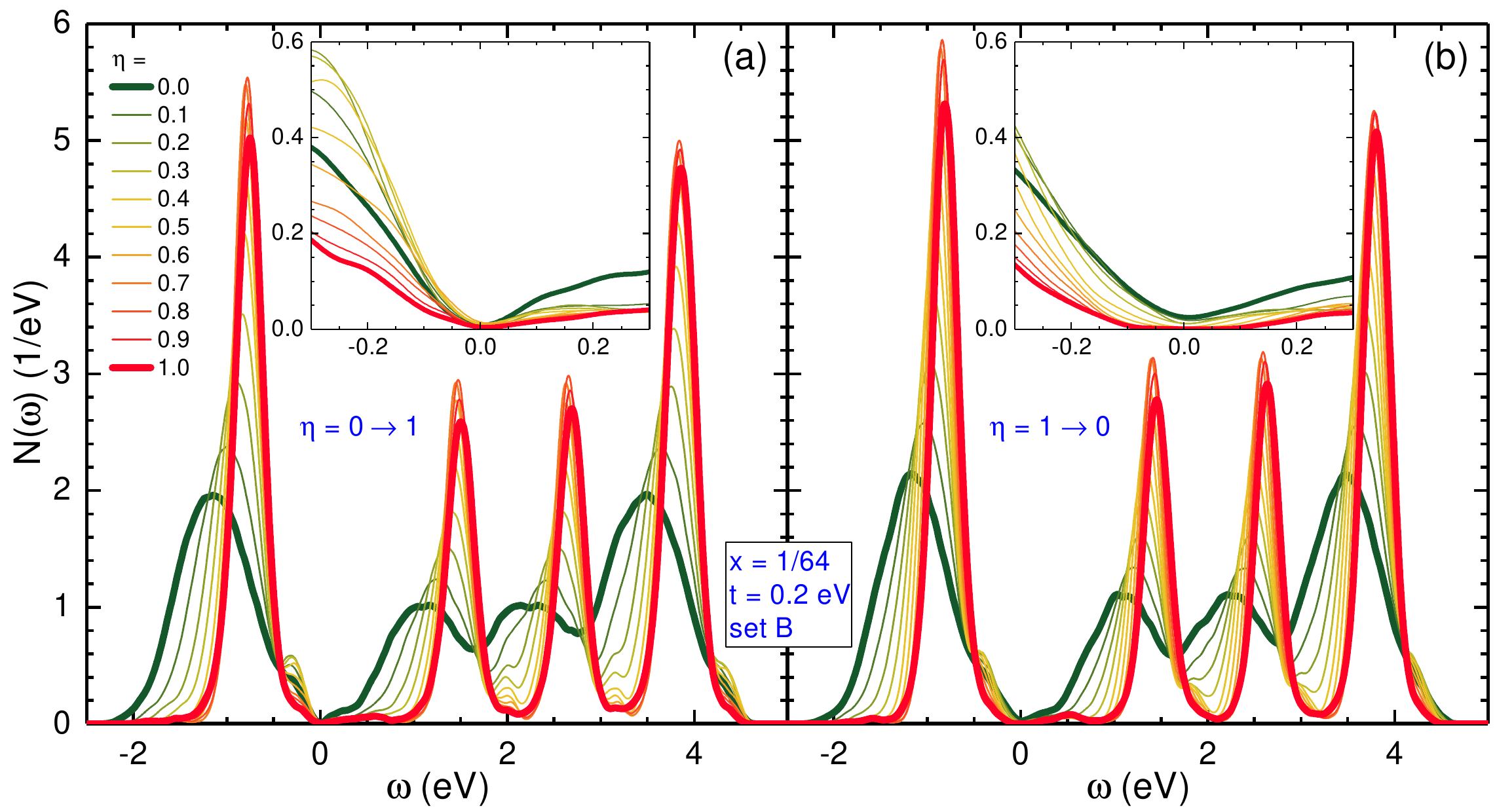} \caption{$t_{2g}$
DOS $N(\omega)$ as obtained for $t_{2g}$ states from $M=100$ random
defect realizations for $t=0.2$ eV at doping $x=1/64$, and for:
(a) increasing value of $e$-$e$ interaction (\ref{pots}) in steps of
$0.1$ from $\eta=0$ (thick green line) to $\eta=1.0$ (thick red line), and
(b) decreasing value of $e$-$e$ interaction (\ref{pots}) from $\eta=1.0$
(thick red line) to $\eta=0$ (thick green line). Insets show zooms
of $N(\omega)$ near the Fermi energy at $\omega=0$. Other parameters
as in set $B$ of Table~\ref{tab:para}.\label{fig_k}}
\end{figure*}

\acknowledgments

We thank D.~D.~Sarma for insightful discussions. A.~M.~O. kindly
acknowledges support by Narodowe Centrum Nauki (NCN, National Science
Centre, Poland) under Project No.~2012/04/A/ST3/00331.

\appendix

\section{The algorithm to determine stable configurations with random defects\label{appa}}

As we described in Sec.~\ref{sec:multi}, we present results obtained
by averaging over $100$ randomly chosen defect realizations. Any
\emph{random defect realization} is a configuration of defects defined
by a set of randomly-chosen positions within the $R$ lattice of doped
$R_{1-x}$Ca$_{x}$VO$_{3}$ that host the Ca defects. Such positions,
in terms of Cartesian coordinates, are obtained by subsequent calls
to a pseudo-random number generator. No constraint is imposed (e.g.,
no minimal distance between the defects is required) except for the
obvious hard-core constraint that any site is singly occupied, either
by an $R$ ion or by a Ca defect ion. The number of positions/defects
in the set defines the degree of doping of the system $x$.

The employed Hartree-Fock self-consistency procedure is based on recursion:
the values of all single-particle correlation functions appearing
in the Hamiltonian are computed at each iteration after diagonalizing
the Hamiltonian where the values computed at the previous iteration
have been plugged in. The first iteration obviously requires
externally-provided
starting values (initialization), which are supplied on the basis
of an educated guess. The recursion stops after a finite number of
steps when all values in two subsequent iterations differ (actually
the relative difference is used: $|x_{n-1}-x_{n}|/|x_{n-1}|$) at
most by a chosen amount (i.e., $10^{-3}$). In order to speed up the
convergence and to avoid long cycles involving states with almost
identical energies and slightly different values of the single-particle
correlation functions characterizing them, the actual new values at
each iteration are built as a linear combination of the old ones and
the ones effectively obtained by diagonalizing the Hamiltonian. The
proportion among the two components starts quite low in favor of the
old values (0.1) and increases steadily with the iterations up to
exclude the old values from the combination within 200 steps.

The size of the system ($N_{a}=8\times8\times8$), the number of orbital
and spin degrees of freedom per site ($6=3\times2$), and the number
of random defect realizations to statistically average upon (100)
make the numerical calculations very demanding and unavoidable to
parallelize the code in order to keep the convergence for a set of
Hamiltonian parameters within one day. In the finite system of $N_{a}$
sites (atoms) considered here, we can change the defect concentration
in steps of 1/512. For instance, the low doping of $x=1/64$ used
in Fig.~\ref{fig_k} corresponds to 8 random defects. Each of these
defects defines a cube of V ions around it and, at the very beginning,
one hole is doped to each cube. \textit{A priori,} we have no information
about the ultimate doped holes' configuration for a given distribution
of random defects, but the lowest energy configuration is self-consistently
chosen to minimize the random fields acting on each hole.

The \emph{procedure} developed to find this configuration is the following
one: (i) we start from a \emph{neutral} configuration (homogeneous
doping of 1/8) at each corner of any cube surrounding a defect at
$\eta=0$; (ii) increasing $e$-$e$ interactions we reach $\eta=1.0$
with full convergence at each value of $\eta$ along the path in the
parameter space for each starting configuration of random defects;
(iii) starting from the final point of (ii) {[}and, therefore, exactly
for all the same random defect configurations{]} we reach $\eta=0$
with again full convergence at each value of $\eta$ along this path.
But, in (ii) and/or in (iii), if the convergence was lost along the
way {[}for a value of $\eta\in(0.0\to1.0\to0.0)${]}, a new random
configuration was used. Accordingly, if one looks carefully and, in
particular, in the zoomed inset, one will see that also $\eta=1.0$
is slightly different between the two panels (i.e., the two pathways)
because of the last requirement.

The rebooting procedure is necessary to stabilize the final results:
When a set of values for a given Hamiltonian parameter is explored,
if the convergence is not reached from the initial value of the set
to the final value of it, we generated another random defect realization
and reboot the search in order to get a smooth progression per each
random defect realization. This rebooting procedure allows one to
avoid taking into account random defect realizations with extremely
weird relative positions of the defects that would hinder the convergence
and that would be anyway discarded by \emph{nature}.

The algorithm described above is unbiased, but rather involved. Therefore,
we present here a test of convergence carried out by performing the
calculations at low doping, $x=1/64$ , and $t=0.2$ eV in two ways:
(i) starting with the case of absent $e$-$e$ interactions ($\eta=0$)
and gradual switching on the screening interactions up to $\eta=1.0$,
and (ii) starting from the screened $e$-$e$ interactions at $\eta=1.0$
and reducing the screening down to $\eta=0$. It is worth noting that
the calculation performed starting from monopole interactions ($\eta=0$)
and reaching the state with screened $e$-$e$ interactions ($\eta=1.0$)
leads to a physically very different state controlled effectively
by dipolar interactions between defect centers, see Fig.~\ref{fig_k}(a).
It is remarkable that the subsequent switching off of $e$-$e$ interactions
leads back to essentially the same initial state, see Fig.~\ref{fig_k}(b).
For such a complex landscape we are dealing with here, this is indeed
a great result.

In fact, one finds very similar results independently of whether the
$e$-$e$ interactions are increased or reduced, see Figs.~\ref{fig_k}(a)
and \ref{fig_k}(b). In both cases, the system is an insulator, with
distinct structures representing the LHB and UHB. If $e$-$e$ interactions
are absent ($\eta=0$), the maxima corresponding to the LHB and the
HS/LS excitations in the UHB are wide and the states that correspond
to the excitations at sites occupied by holes cannot be resolved from
these peaks. When the $e$-$e$ interactions increase towards fully
screened ones ($\eta=1.0$), the defect states appear instead at the
lower edges of both the LHB and the HS subband in the UHB. Both figures
are remarkably similar even for the zoomed DOS near the Fermi energy,
where a broader gap is visible at $\eta=1.0$. This test confirms
that the entire procedure, which includes the averaging over $M=100$
random defect realizations, is reliable and gives well converged and
reproducible results.


\section{Bound spin-orbital polarons\label{appc}}


In this Appendix, we shall analyze the reduction of the spin and orbital
order parameters as functions of the doping $x$ and of the kinetic
energy parameter $t$ in the frame of the spin-polaron theory. This
strong coupling approach rests on the idea that, in the dilute limit,
polaron-polaron interactions should be negligible due to the local
nature of the polarons. This implies that the overall reduction of
the order parameters can be determined in terms of the reduction induced
by the polaron cloud of a single polaron, whose size scales with the
kinetic energy parameter $t$. Accordingly, the overall reduction
of the orders scales linearly with $x$, as in Eq.~(\ref{ms}). Here,
we shall first test up to which concentration $x$ and to which kinetic
energy parameter $t$, the independent polaron picture holds by inspecting
the uHF results for the respective order parameters. Next, we shall
derive the wave function for a spin polaron bound to a defect using
the string basis introduced in Sec.~\ref{sec:polarons}. Finally,
we shall compare the $t$-dependence of the spin-polaron cloud with
the order parameters calculated by the uHF method including the disorder.
This will provide the basis for a qualitative and quantitative interpretation
of the uHF results in the frame of the spin-orbital polaron theory
emerging from the strong correlation picture.

\begin{figure}[t!]
\includegraphics[width=8cm]{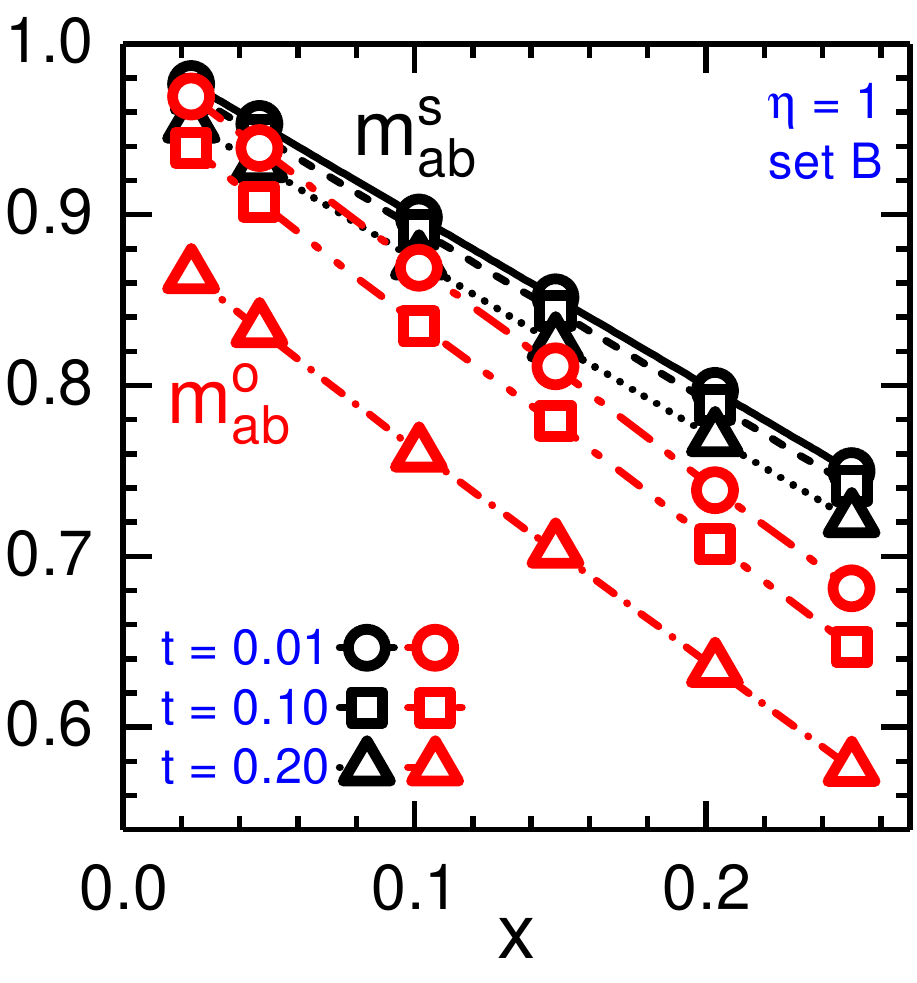}\caption{
Contributions from electrons in $a/b$ orbitals to spin and orbital
order parameters $m_{ab}^{s}$ and $m_{ab}^{o}$ , respectively, in
the $C$-AF/$G$-AO ordered phase. The results from uHF theory are
plotted as function of doping $x$ for different hopping parameters
$t=0.01$, 0.10 and $0.20$ eV. Parameters as in Fig.~\ref{fig:fig-v}.
\label{fig-mabvsx}}
\end{figure}


Figure~\ref{fig-mabvsx} displays the doping dependence of spin and
orbital order parameters $m_{ab}^{s}$ and $m_{ab}^{o}$, respectively,
that are due to the electrons in $a$ and $b$ orbitals. The figure
highlights an approximate linearity in $x$ of the uHF results up to high
doping concentrations. This is a clear evidence that the small-polaron
picture applies for the values of $t$ considered in Fig.~\ref{fig-mabvsx}
and suggests that the interactions among polarons are of minor importance
and, therefore, negligible in first approximation. Moreover, the uHF
results show that the decrease of the spin moments $m_{ab}^{s}$ with
$t$ is much slower as compared to that of the orbital-order parameter
$m_{ab}^{o}$. This confirms that the spin order in doped systems
is more robust \cite{Miy00} as the FM polarization along the active
bonds is not affected when the orbital order is locally disturbed.
Nevertheless, in Fig.~\ref{fig-mabvsx}, one notices small nonlinearities
in $m_{ab}^{s}$ at small values of $x$, which we attribute to the
random far-defect potentials.

The spatial symmetry of the polaron depends on the magnetic and the
orbital order. For instance, a hole in the AF CuO$_{2}$ planes of
high-$Tc$ superconductors takes the form of a Zhang-Rice singlet,
that is, a combination of the Cu $d_{x^{2}-y^{2}}$ orbital and of
a $d$-wave admixture of the $2p$ states from the four oxygen neighbors.
The symmetry of the wave function is $C_{2}$ relative to the $c$-axis
(but $C_{4}$ for the probability distribution). A spin polaron in
the $C$-AF/$G$-AO state has a $C_{2}$ symmetry for both the wave
function and its probability distribution. Since these SO polarons
are bound to the defect, their symmetry is further reduced due to
the strong Coulomb attraction of the charged defect.

\begin{figure}[t!]
\includegraphics[width=\columnwidth]{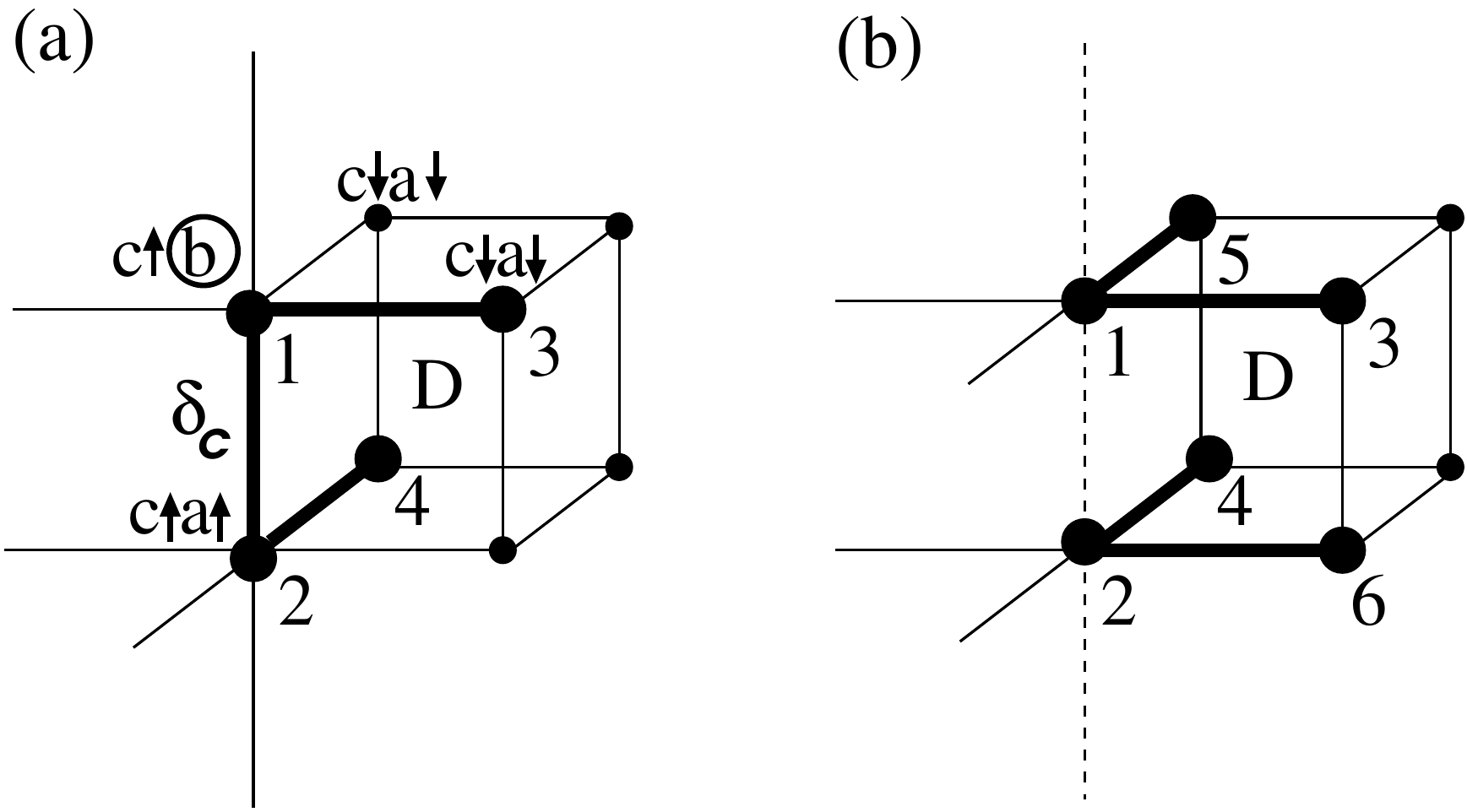} \caption{
String representation of the dominant contributions to the polaron
wave function for a hole created by the annihilation of an occupied
$b$ orbital at site 1 close to the defect D. The hole delocalizes
differently depending on the electron hopping in:
(a) $a$ or $b$ orbitals and
(b) $c$ orbitals.
The polarity $\delta_{c}$ of the active FM bond $\langle1,2\rangle$ is
mainly controlled by an interplay of the hopping amplitude $t$ on that
bond and the difference of random potentials between the sites 1~and~2.
\label{fig-strings}}
\end{figure}


Figure~\ref{fig-strings} shows the most important string states
that contribute to the bound polaron wave function for a hole created
at the V site 1 by annihilation of a $b$ electron with spin up in
the undoped ground state $|\Phi_{0}\rangle$:
\begin{equation}
|\Phi_{0}\rangle=|c{\uparrow},b{\uparrow};c{\uparrow},a{\uparrow};
c{\downarrow},a{\downarrow};c{\downarrow},b{\downarrow}\rangle,
\label{Phi0}
\end{equation}
The vacuum state $|\Phi_{0}\rangle$ indicates the occupied orbitals
and spin orientations consistent with the $C$-AF/$G$-AO order for
the ions sited at sites 1, 2, 3 and 4 [see Fig.~\ref{fig-strings}(a)].
Due to the strong attraction of the defect D the relevant string states
1, 2, 3 and 4 are all on the defect cube. Moreover, the selected strings
reflect the constraints due to the spin- and orbital- order and the flavor
conserving constrained hopping processes. Figures~\ref{fig-strings}(a) and
\ref{fig-strings}(b) distinguish strings for electrons in $a/b$ and $c$
orbitals, respectively. We shall discuss here only the contributions that stem
from $a$ and $b$ orbitals that have the largest effect on the order parameters.
The relevant string states are:
\begin{eqnarray}
|\Phi_{1}\rangle & = & |c{\uparrow},0;c{\uparrow},a{\uparrow};
c{\downarrow},a{\downarrow};c{\downarrow},b{\downarrow}\rangle,\nonumber \\
|\Phi_{2}\rangle & = & |c{\uparrow},a{\uparrow};c{\uparrow},0;
c{\downarrow},a{\downarrow};c{\downarrow},b{\downarrow}\rangle,\nonumber \\
|\Phi_{3}\rangle & = & |c{\uparrow},a{\downarrow};c{\uparrow},a{\uparrow};
c{\downarrow},0;c{\downarrow},b{\downarrow}\rangle,\nonumber \\
|\Phi_{4}\rangle & = & |c{\uparrow},a{\uparrow};c{\uparrow},b{\downarrow};
c{\downarrow},a{\downarrow};c{\downarrow},0\rangle.\label{Phi}
\end{eqnarray}

Here $|\Phi_{1}\rangle$ is the original configuration arising from
the annihilation of an up-spin electron in a $b$ orbital at the V-site
1. In state $|\Phi_{2}\rangle$, the $a$ electron has moved along
the FM bond from site 2 to site 1. This interchange results in an
orbital excitation at site 1. Since the Jahn-Teller energy is small,
the kinetic energy will mix these two states. However, the active
bond 1-2 is further controlled by the random far-defect potentials.
Therefore, it is useful to characterize the distribution of electrons
(or holes) on the active bond by the polarity parameter~$\delta_{c}$:
\begin{equation}
|\Psi_{1}(\delta_{c})\rangle=\sqrt{1-\delta_{c}}|\Phi'_{1}\rangle
+\sqrt{\delta_{c}}|\Phi'_{2}\rangle,\label{Psi1}
\end{equation}
where for $n=1$ and 2:
\begin{equation}
|\Phi'_{n}\rangle=\frac{|\Phi_{n}\rangle+\frac{t}{e_{0}}
|\Phi_{n+2}\rangle}{\sqrt{1+\left(\frac{t}{e_{0}}\right)^{2}}}.\label{Phiprime}
\end{equation}
These states contain admixtures from strings 1 and 2, and 3 and 4,
respectively, which involve low-spin excitations with excitation energy
$e_{0}=J_{H}$. According to Eq. (\ref{Psi1}), the polaron state for
$\delta_{c}=0$ corresponds to a hole at site 1 and for $\delta_{c}=1$
to a hole at site 2, in combination with an orbital excitation at
site 1. Increasing the kinetic energy parameter $t$ will eventually
push $\delta_{c}$ towards $1/2$.

\begin{figure}[b!]
\includegraphics[width=8cm]{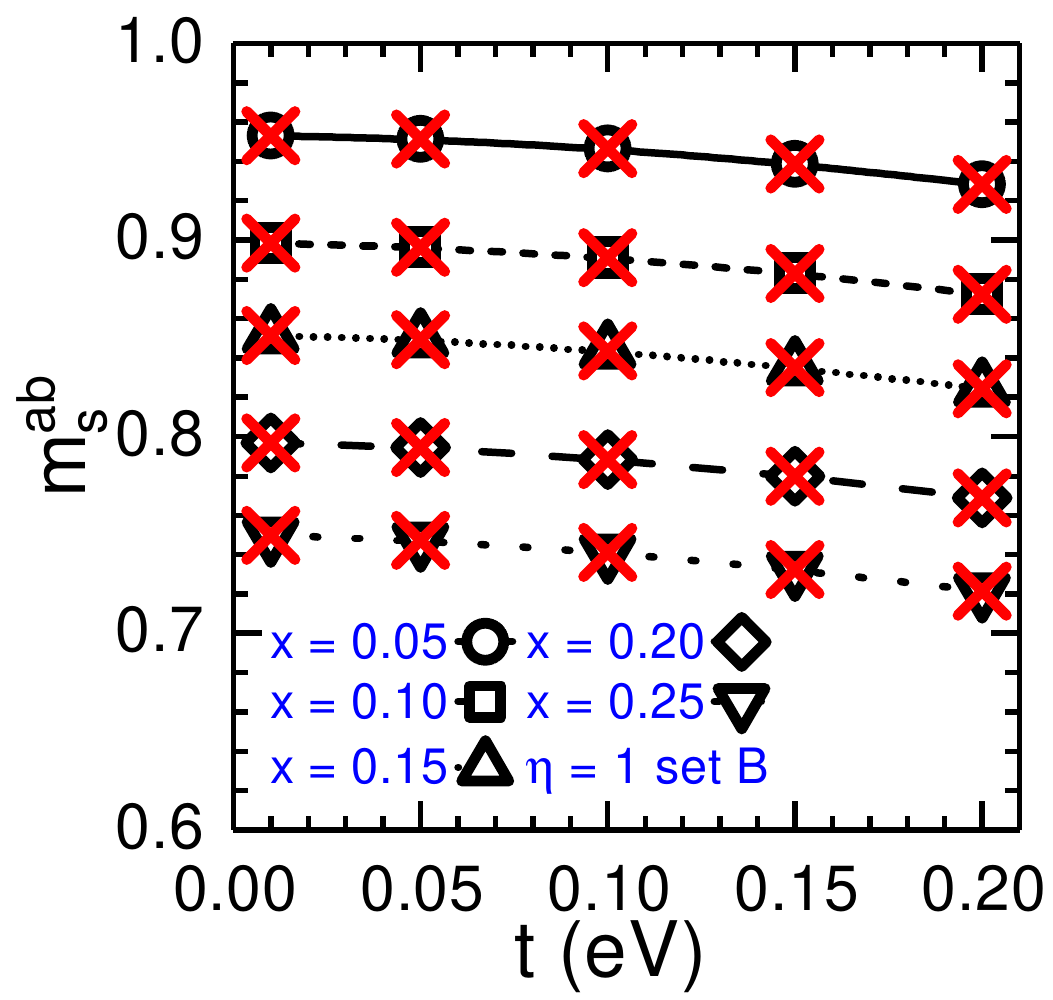} \caption{
Contributions from electrons in $a/b$ orbitals to spin order parameter
$m_{ab}^{s}$ as function of $t$ as calculated by the uHF method
(black symbols). Parameters as in Fig.~\ref{fig:fig-v}. A fit of
$m_{ab}^{s}$ to the polaron theory Eq.~(\ref{msabA}) for doping
concentrations $x=5,10,15,20$ and $25$\% is marked by red crosses.
\label{fig-msvst}}
\end{figure}


By means of the spin-polaron wave function $|\Psi_{1}(\delta_{c})\rangle$
we can evaluate the reduction of the spin and orbital order due to
a single polaron (see Eq. (7.4))
\begin{eqnarray}
m_{ab}^{sP} & \cong & 1+\frac{2\left(t/e_{0}\right)^{2}}
{1+\left(t/e_{0}\right)^{2}},\nonumber \\
m_{ab}^{oP} & \cong & 1+2\delta_{c}+\frac{2\left(t/e_{0}\right)^{2}}
{1+\left(t/e_{0}\right)^{2}}.\label{mPab}
\end{eqnarray}
We note that the potential of the random far defects will modify the
excitation energy $e_{0}$. Moreover, the polarity parameter $\delta_{c}$
too is determined by the interplay of the kinetic energy parameter
$t$ and the same far-defect random potentials.

The total reduction of spin- and orbital- order associated with $a/b$
electrons is then (Eq.(7.3)):
\begin{eqnarray}
m_{ab}^{s} & \cong & 1-x\left[1+\frac{2\left(t/e_{0}\right)^{2}}
{1+\left(t/e_{0}\right)^{2}}\right],\nonumber \\
m_{ab}^{o} & \cong & 1-x\left[1+2\delta_{c}+\frac{2\left(t/e_{0}\right)^{2}}
{1+\left(t/e_{0}\right)^{2}}\right].\label{msabA}
\end{eqnarray}
In Fig.~\ref{fig-msvst}, we report a least-square fit of the uHF
results for the spin order parameter $m_{ab}^{s}$ to the above polaron
expression with a single variational parameter, namely the activation
energy $e_{0}$. Since the uHF results represent a statistical average
over many defect realizations, the resulting activation energy $e_{0}(x)$
characterizes a \emph{typical} spin-orbital polaron bound state. It
is worth noting that $m_{ab}^{s}$ does not depend on the active bond
parameter $\delta_{c}$. This is distinct from the orbital order since
the motion of the hole on the active bond creates an orbital defect
yet no spin defect (see discussion in Sec.~\ref{sec:polarons}).

Our results show that even up to a doping concentration of $x=25$\%,
the reduction of both order parameters is approximately linear, this
holds true even when $t=0.2$ eV, i.e., for a typical value of the
kinetic energy parameter for the vanadate compounds. It is worth noting
that the uHF is able to describe also the interaction between polarons,
which appears however irrelevant in the data range shown in
Fig.~\ref{fig-msvst}.
That the independent polaron ansatz works so well for such huge doping
concentrations may appear really surprising. On the other hand, one
may have expected this given the extreme localization of the defect
states we have inferred by exploring the inverse participation number
in Sec.~\ref{sec:ipr}.

\bibliographystyle{apsrev4-1-prx}

%

\end{document}